\documentclass[12pt,modern,trackchanges]{aastex62}
\usepackage{amsmath,amssymb,color,verbatim}
\newcommand{\etal}{et~al.\/}
\newcommand{\ie}{i.e.\/}
\newcommand{\eg}{e.g.\/}

\newcommand{\msun}{M$_\odot$}

\shorttitle{Fire in the Heart of NGC\,253}
\shortauthors{Mangum \etal}

\begin{document}
\title{Fire in the Heart: A Characterization of the High Kinetic
  Temperatures and Heating Sources in the Nucleus of NGC\,253}

\author[0000-0003-1183-9293]{Jeffrey G.~Mangum}
\affiliation{National Radio Astronomy Observatory, 520 Edgemont Road,
  Charlottesville, VA  22903-2475, USA}

\author[0000-0001-6431-9633]{Adam G.~Ginsburg}
\altaffiliation{National Radio Astronomy Observatory Jansky Fellow}
\affiliation{National Radio Astronomy Observatory, P.O. Box O, 1003
  Lopezville Road, Socorro, NM 87801-0387, USA}

\author[0000-0002-7495-4005]{Christian Henkel}
\affiliation{Max-Planck-Institut f\"ur Radioastronomie, Auf dem H\"ugel
  69, 53121 Bonn, Germany}
\affiliation{Astronomy Department, Faculty of Science, King Abdulaziz
  University, P.~O.~Box 80203, Jeddah, Saudi Arabia}

\author[0000-0001-6459-0669]{Karl M.~Menten}
\affiliation{Max-Planck-Institut f\"ur Radioastronomie, Auf dem H\"ugel
  69, 53121 Bonn, Germany}

\author{Susanne Aalto}
\affiliation{Department of Earth and Space Sciences, Chalmers
  University of Technology, Onsala Observatory, SE-439 92 Onsala,
  Sweden}

\author[0000-0001-5434-5942]{Paul van der Werf}
\affiliation{Leiden Observatory, Leiden University, 2300 RA, Leiden,
  The Netherlands}

\correspondingauthor{Jeff Mangum}
\email{jmangum@nrao.edu}
\submitjournal{The Astrophysical Journal}

\begin{abstract}
The nuclear starburst within the central $\sim 15^{\prime\prime}$ ($\sim
250$\,pc; $1^{\prime\prime} \simeq 17$\,pc) of NGC\,253 has been extensively
studied as a prototype for the starburst phase in galactic evolution.
Atacama Large Millimeter/submillimeter Array (ALMA) imaging within
receiver Bands 6 and 7 have been used to investigate the dense gas
structure, kinetic temperature, and heating processes that drive the
NGC\,253 starburst.  A total of 29 transitions from 15
molecular species/isotopologues have been identified and imaged at
$1.^{\prime\prime}5$--$0.^{\prime\prime}4$ resolution, allowing for
the identification of five of the previously studied giant molecular
clouds (GMCs) within the central molecular zone (CMZ) of NGC\,253.  Ten 
transitions from the formaldehyde (H$_2$CO) molecule have been used to
derive the kinetic temperature within the $\sim
0.^{\prime\prime}5$--$5^{\prime\prime}$ dense gas structures imaged.  On $\sim
5^{\prime\prime}$ scales we measure $T_K \gtrsim 50$\,K, while on size
scales $\lesssim 1^{\prime\prime}$ we measure $T_K \gtrsim 300$\,K.
These kinetic temperature measurements further delineate the
association between potential sources of dense gas 
heating.  We have investigated potential heating sources by
comparing our measurements to models that predict the physical
conditions associated with dense molecular clouds that possess a
variety of heating mechanisms.  This comparison has been supplemented with
tracers of recently formed massive stars
(Br$\gamma$) and shocks ([FeII]).  Derived molecular column 
densities point to a radially decreasing
abundance of molecules with sensitivity to cosmic-ray and mechanical
heating within the NGC\,253 CMZ.  These measurements are consistent
with radio spectral index calculations that suggest a higher
concentration of cosmic-ray-producing supernova remnants within the
central 10\,pc of NGC\,253.
\end{abstract}


\section{Introduction}
\label{Intro}

The comparison between the properties of the star formation process in
our Galaxy and that found in galaxies that appear to be producing a
plethora of stars over a relatively short time period is dramatic.
Taking NGC\,253 as the prototype for a starburst galaxy, the giant
molecular clouds (GMCs) are
$\sim 50$\% larger, are $\sim 100$ times more massive, have velocity
dispersions that are $\sim 10$ times larger, and have freefall times $\sim
3$ times shorter than GMCs in the Milky Way disk\footnote{See Table~4 in
  \cite{Leroy2015} for an excellent comparison of physical properties
  in Milky Way disk and NGC\,253 GMCs.}.  With its relative proximity \cite[$d =
3.5$\,Mpc;][]{Rekola2005} and optimal disk orientation of $\sim
76^\circ$ \citep{McCormick2013}, which presents disk velocity
excursions running from $\sim 180$ to $\sim 300$\,km/s, NGC\,253
provides an excellent perspective to earthly observers of the
extragalactic star formation process.  As higher-resolution and more
sensitive infrared through millimeter measurements have become
available, the spectral and structural complexity of the central
kiloparsec of the NGC\,253 molecular disk has become more apparent.
Structures that are reminiscent of Milky Way massive star formation
regions with spectral richness rivaling those measured toward hot core
sources and our own Galactic center region can now be measured and analyzed,
providing valuable clues to the burst mode of star formation in
external galaxies.

Using the millimeter/submillimeter spatial and spectral properties
measured toward NGC\,253, we have endeavored to understand some basic
properties of the star formation process in this galaxy.
Specifically, what is the kinetic temperature within the dense gas
that is in the process of forming stars, and what are the heating
processes that drive those kinetic temperatures?
Section~\ref{Observations} presents our Atacama Large
Millimeter/submillimeter Array (ALMA) frequency Band
6 and 7 spectral line emission measurements, from which the spatial
(Section~\ref{CompID}) and spectral (Section~\ref{LineID}) properties
have been derived.  Molecular spectral line integrated intensities have been
extracted using a python-based script (Section~\ref{Extraction}), from
which molecular column densities have been derived
(Section~\ref{Colden}).  Section~\ref{H2COTk} presents our analysis of
the imaged transitions from the formaldehyde (H$_2$CO) molecule that
have been used to derive the kinetic temperature within the identified
GMCs that inhabit the NGC\,253 nucleus.  With this information, we
then use 
the molecular abundances inferred from our measurements, in
corporation with infrared through radio studies of the NGC\,253
nuclear disk, to constrain chemical models that trace the influence
of photon-dominated region (PDR), X-ray-dominated region (XDR),
cosmic-ray-dominated (CRDR), and mechanical heating within the dense
molecular gas
(Section~\ref{HighTK}).  In Section~\ref{vibHNCandCH3OH} we discuss
the anomalous spatial distributions presented by our measured CH$_3$OH
and vibrationally excited HC$_3$N and HNC transitions and their
association with infrared and radio emission sources.  We conclude
with a discussion of the connection between potential sources of
heating and the measured molecular abundances in the NGC\,253 CMZ.

\section{Observations}
\label{Observations}

A single field was observed toward NGC\,253 using the Atacama Large
Millimeter Array (ALMA) 12m Array, Atacama Compact Array (ACA), and
Total Power (TP) antennas at Bands 6 and 7 (ALMA projects
2013.1.00099.S and 2015.1.00476.S).  The phase center was at
RA(J2000)=00:47:33.1339, Dec(J2000)=$-25$:17:19.68,
V$_{hel}$=258.8\,km/s.  ALMA standard observing routines
were used for all three types of measurements, which included
pointing, flux, and phase calibration measurements.  The TP
measurements have proven to be unusable owing to incomplete telluric
line removal encountered during processing in the Common Astronomy
Software Applications (CASA) data reduction package.  As we have not
included these measurements in our analysis, we provide no further
information on them. 

\begin{deluxetable*}{llllll}
\tablewidth{0pt}
\tablecolumns{6}
\tablecaption{Observations Summary\label{tab:obssum}}
\tablehead{
\colhead{Band} &
\colhead{Array} &
\colhead{Obs Date/Start Time} &
\colhead{t$_{on}$ (minutes)} &
\colhead{N$_{ant}$} &
\colhead{Baselines} \\
&&&&& (Min,Max) (m)
}
\startdata
\multicolumn{6}{l}{NGC\,253, RA(J2000)=00:47:33.1339,
  Dec(J2000)=$-25$:17:19.68, V$_{hel}$=258.8\,km/s} \\
\tableline
6 & 12m & 2014-12-28 23:57:24 & 6.8 & 38 & (15,349) \\
6 & 12m & 2015-05-02 16:18:28 & 38.1 & 34 & (15,349) \\
6 & ACA & 2014-06-04 09:12:16 & 24.2 & 9 & (9,49) \\
6 & ACA & 2014-06-04 10:17:53 & 24.2 & 9 & (9,49) \\
6 & ACA & 2014-06-04 11:25:14 & 24.2 & 9 & (9,49) \\
7 & 12m & 2014-05-19 09:15:59 & 34.3 & 34 & (21,650) \\
7 & ACA & 2014-05-19 10:20:02 & 32.8 & 8 & (9,49) \\
7 & ACA & 2014-06-08 09:49:14 & 32.8 & 10 & (9,49) \\
\enddata
\end{deluxetable*}

\begin{deluxetable*}{llll}
\tablewidth{0pt}
\tablecolumns{4}
\tablecaption{Spectral Setup\label{tab:spectralsetup}}
\tablehead{
\colhead{Rest Frequency} &
\colhead{Bandwidth} &
\colhead{N$_{chan}$} &
\colhead{Channel Width} \\
\colhead{(GHz)} & \colhead{(GHz)} && \colhead{(MHz
  and km/s)}
}
\startdata
\multicolumn{4}{l}{ALMA Band 6 12m Array Correlator} \\
218.222192 & 1.875 & 960 & 1.953/2.686 \\
219.908525 & 1.875 & 960 & 1.953/2.665 \\
234.700 & 2.000 & 122 & 15.625/19.978 \\
\tableline
\multicolumn{4}{l}{ALMA Band 6 ACA Correlator} \\
218.222192 & 1.992 & 1024 & 1.992/2.67 \\
219.908525 & 1.992 & 1024 & 1.992/2.65 \\
234.700 & 1.938 & 128 & 15.625/20.05 \\
\tableline
\multicolumn{4}{l}{ALMA Band 7 12m Array Correlator} \\
351.768645 & 1.875 & 480 & 3.906/3.33 \\
363.419649 & 1.875 & 480 & 3.906/3.33 \\
364.819285 & 1.875 & 480 & 3.906/3.33 \\
\tableline
\multicolumn{4}{l}{ALMA Band 7 ACA Correlator} \\
351.768645 & 1.992 & 510 & 3.906/3.33 \\
363.419649 & 1.992 & 510 & 3.906/3.33 \\
364.819285 & 1.992 & 510 & 3.906/3.33 \\
\enddata
\end{deluxetable*}

Tables~\ref{tab:obssum} and \ref{tab:spectralsetup} summarize the general
observational characteristics of the ALMA Band 6 and 7 measurements of
NGC\,253 presented.  Three separate spectral windows for each
frequency band in these 12m Array and ACA measurements
were employed.  For the Band 6 measurements each spectral window
spanned frequencies from 217.284692 to 219.159692\,GHz, from
218.971025 to 220.846025\,GHz, and from 233.7 to 235.7\,GHz.  For Band
7 the three separate spectral windows spanned frequencies from 350.831145 to
352.706145\,GHz, from 362.482149 to 364.357149\,GHz, and from 363.881785 to
365.756785\,GHz. For the Band 6 measurements of NGC\,253 the total
on-source integration time (t$_{on}$) for each type of measurement was
44.9\,minutes (12m Array) and 72.6\,min (ACA),
for a 12m:ACA on-source integration time ratio of 1:1.6.  At Band 7
the total on-source integration time toward 
NGC\,253 for each type of measurement was 34.3\,min (12m Array) and
65.6\,minutes (ACA),
for a 12m:ACA on-source integration time ratio of 1:1.9.  Both
on-source integration time ratios are consistent with the standard
integration time ratio for ALMA observations acquired in Cycle 2 of
12m Array:ACA:TP = 1:2:4 \citep{Mason2013}.

Amplitude, bandpass, phase, and pointing calibration
information for these measurements are listed in
Appendix~\ref{CalibrationDetails} (Table~\ref{tab:calibrators}).  For
Band 6 gain and bandpass calibrator 
fluxes ranged from 253 to 688\,mJy.  Uncertainties in the Band 6 gain
and bandpass calibrator measurements are in the
range 0.4--1.7\%.  For Band 7 gain and bandpass calibrator fluxes
ranged from 274 to 2244\,mJy.  Uncertainties in the Band 7 gain and
bandpass calibrator measurements are with one exception in the range
of 0.1--3.9\%.  The Band 7 12m Array measurements on 2014-05-19 have
higher gain calibration uncertainties that are in the range of 4.3--6.4\%.

At Band 6 flux calibration was effected using Uranus (28.9--34.8\,Jy),
Mars (78.5--89.5\,Jy), and Neptune (13.0--13.6\,Jy).  Band 6 flux
calibrator uncertainties are estimated to be 
5\% for Uranus \citep{Orton2014a}, 10\% for Mars
\citep{Weiland2011,Perley2013}, and 10\% for Neptune
\citep{Mueller2016}.  For Mars our estimate of the absolute flux
uncertainty derives from the 1--50\,GHz absolute flux
\citep{Perley2013} and 23--93\,GHz absolute brightness temperature
\citep{Weiland2011} measurements of Mars.  For our absolute flux
uncertainty estimate for Neptune we have increased the nominal 5\%
uncertainty 
quoted by
\cite{Mueller2016} owing to the presence of a CO absorption transition
at 230.538\,GHz present in the atmosphere of Neptune to 10\%.

At Band 7 flux calibration was effected using J2258$-$2758 (390\,mJy),
Neptune (25.8--27.8\,Jy), and Uranus (63.8--66.6\,Jy).
Band 7 flux calibrator uncertainties are estimated to be 10\% for
Neptune \citep{Mueller2016} and 5\% for Uranus \citep{Orton2014a}.
As was done for our Band 6 measurements, for our absolute flux
uncertainty estimate for Neptune we have increased the nominal 5\%
uncertainty 
quoted by
\cite{Mueller2016} owing to the presence of a CO absorption transition
at 345.796\,GHz present in the atmosphere of Neptune.  For unknown
reasons the 12m Array measurements of NGC\,253 at Band 7 employed a
nonstandard flux calibrator (J2258$-$2758).  The ALMA Calibration
Catalog lists the flux uncertainty for this quasar as 15\%, but no
further information regarding the time period over which this
uncertainty applies is provided.

All measurements were either manually or pipeline calibrated by
ALMA North American Regional Center staff using the CASA reduction
package.  Following delivery of each calibrated interferometric
measurement group (12m Array and ACA), self-calibration was attempted
to correct for residual phase errors.  For all measurement groups at
both Bands 6 and 7 one iteration of phase-only self-calibration
utilizing a 60 second averaging time was used. The bright
NGC\,253 continuum source at Bands 6 and 7 was used as a 
self-calibration source, which resulted in signal-to-noise improvement
by factors of 
2.9 and 1.9 in the Band 6 continuum of our 12m Array and ACA data,
respectively, and by factors of 3.0 and 3.5 in the Band 7 continuum of
our 12m Array and ACA data, respectively.

\section{Results}
\label{Results}

\subsection{Imaging and Spectral Baseline Fitting}
\label{Imaging}

Spectral cube imaging was performed using the CASA package.  Images of
each self-calibrated spectral window image cube for NGC\,253 were
created using the 
\texttt{clean} task within CASA.  An image may be characterized by the
spatial frequencies that are well represented in it; these
correspond to the regions in the Fourier transform (\textit{``uv''} space) of the
image that were well measured by the instrument. Interferometers
generally cover a range of spatial frequencies that are limited by the
longest and shortest baselines used in the interferometric measurement
(see Table~\ref{tab:obssum}). Single antennas, in principle, measure
all spatial frequencies down to those corresponding to the instrument
reflector diameter. Ideally, images with overlapping, well-sampled spatial
frequencies can be combined to derive an image that represents all
the sampled spatial frequencies. ``Feathering'' is a
technique by which images are combined in the \textit{uv}-plane to recover the
spatial frequencies in the input images \citep{Cotton2017}. Combination
of the 12m Array 
and ACA measurements for each spectral window was done
by employing the CASA implementation of the feathering technique, using the
task \texttt{feather}.  Table \ref{tab:images} lists the properties of
each 12m Array, ACA, and feathered image cube for each spectral window
from our measurements of NGC\,253.

\startlongtable
\begin{deluxetable*}{llll}
\tablewidth{0pt}
\tablecolumns{4}
\tablecaption{Spectral and Continuum Image Properties\label{tab:images}}
\tablehead{
\colhead{Rest Frequency} &
\colhead{$\Delta$v} &
\colhead{$\theta_{maj}\times\theta_{min}$@PA\tablenotemark{a}} &
\colhead{RMS per Channel} \\
\colhead{(GHz)} &
\colhead{(km/s)} &
\colhead{(arcsec,arcsec,deg)} &
\colhead{(mJy/beam)}
}
\startdata
\multicolumn{4}{l}{NGC\,253 Band 6 12m Array} \\
218.222192 & 5.5 & $1.52\times0.87$@$+82.15$ & 2.5 \\
219.908525 & 5.5 & $1.56\times0.89$@$+83.96$ & 2.0 \\
234.700 & 25.0 & $1.41\times0.81$@$+82.28$ & 0.9 \\
\tableline
\multicolumn{4}{l}{NGC\,253 Band 6 ACA} \\
218.222192 & 5.5 & $7.55\times4.70$@$+78.94$ & 12.5 \\
219.908525 & 5.5 & $7.61\times4.78$@$+79.08$ & 10.0 \\
234.700 & 25.0 & $7.19\times4.32$@$+76.34$ & 6.0 \\
\tableline
\multicolumn{4}{l}{NGC\,253 Band 6 12m Array and ACA Feather} \\
218.222192 & 5.5 & $1.52\times0.87$@$+82.15$ & 2.5 \\
\nodata & 55 (cont) & \nodata & 0.8 \\
219.908525 & 5.5 & $1.56\times0.89$@$+83.96$ & 1.8 \\
\nodata & 385 (cont) & \nodata & 0.3 \\
234.700 & 25.0 & $1.41\times0.81$@$+82.28$ & 0.9 \\
\nodata & 250 (cont) & \nodata & 0.4 \\
\tableline
\multicolumn{4}{l}{NGC\,253 Band 7 12m Array} \\
351.768645 & 3.5 & $0.45\times0.29$@$-85.64$ & 2.0 \\
363.419649 & 3.5 & $0.43\times0.28$@$-85.94$ & 2.8 \\
364.819285 & 3.5 & $0.43\times0.28$@$-86.00$ & 3.0 \\
\tableline
\multicolumn{4}{l}{NGC\,253 Band 7 ACA} \\
351.768645 & 3.5 & $4.99\times2.48$@$+79.58$ & 13.5 \\
363.419649 & 3.5 & $4.79\times2.39$@$+78.17$ & 13.5 \\
364.819285 & 3.5 & $4.78\times2.39$@$+78.96$ & 22.0 \\
\tableline
\multicolumn{4}{l}{NGC\,253 Band 7 12m Array and ACA Feather} \\
351.768645 & 3.5 & $0.45\times0.29$@$-85.64$ & 2.0 \\
\nodata & 35 (cont) & \nodata & 0.9 \\
363.419649 & 3.5 & $0.43\times0.28$@$-85.94$ & 3.0 \\
\nodata & 35 (cont) & \nodata & 0.8 \\
364.819285 & 3.5 & $0.43\times0.28$@$-86.00$ & 3.2 \\
\nodata & 35 (cont) & \nodata & 1.5 \\
\enddata
\tablenotetext{a}{~Continuum beam parameters are the same as their
  corresponding spectral line cubes, thus have been omitted.}
\end{deluxetable*}

\begin{figure}
\centering
\includegraphics[scale=0.50]{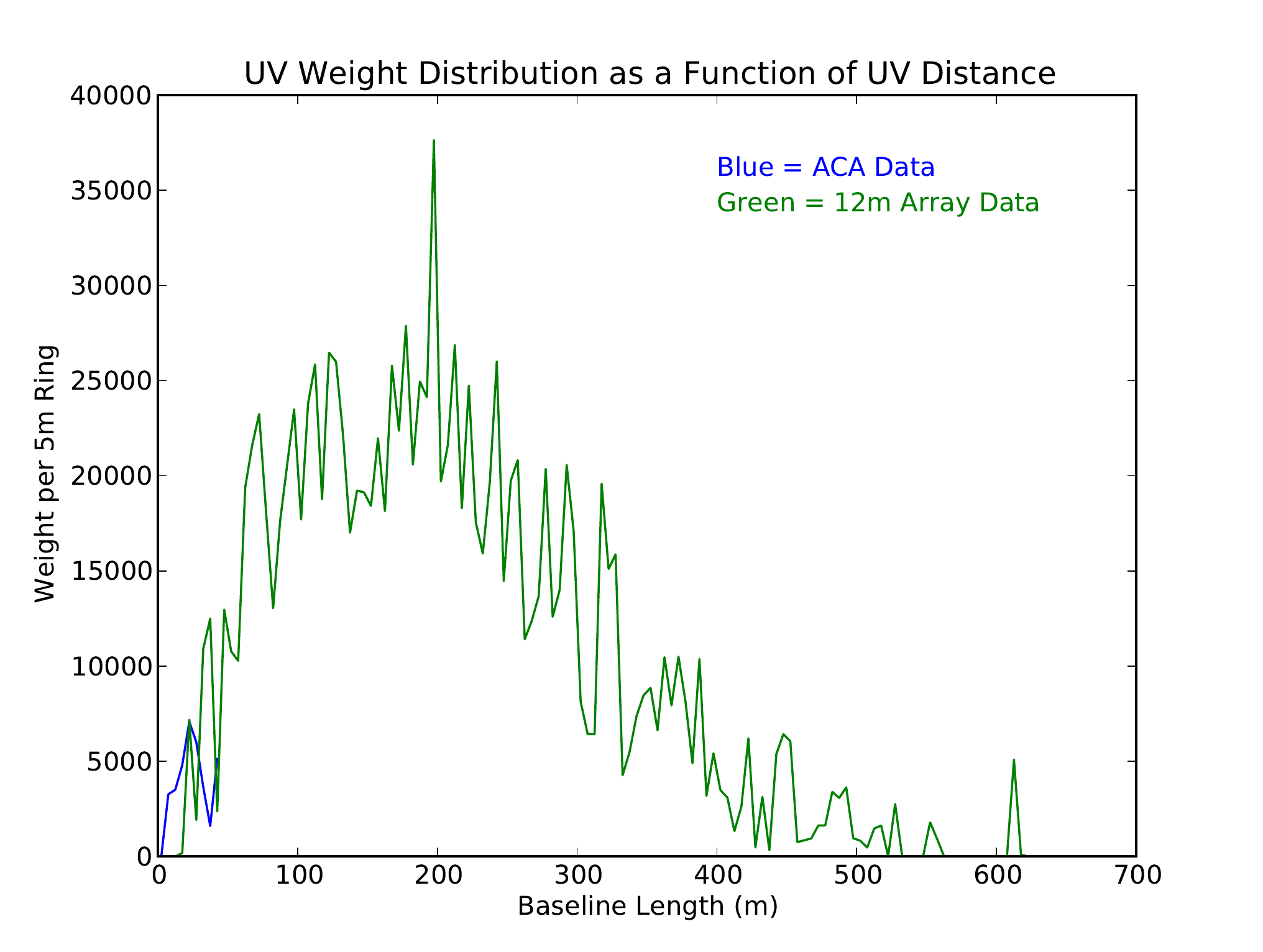}
\caption{Annularly averaged \textit{uv} data weights as function of baseline
  length for the 351\,GHz spectral window from our Band 7 12m Array
  and ACA measurements of NGC\,253.  The 12m Array and ACA uv data
  weights shown have been rescaled to place them on the same scale.
  This comparison suggests that the ACA data contribute minimally to
  the combined spatial frequency information in our imaging of
  NGC\,253.}
\label{fig:uvweights}
\end{figure}

To assess the relative import of each contributing image cube to the
combined (feathered) spectral image cube, we have plotted the \textit{uv} data
weights averaged over 5\,m annuli in the 
\textit{uv}-plane as a function of baseline length for corresponding 12m Array
and ACA data sets from our measurements.  Figure~\ref{fig:uvweights}
shows the annularly averaged \textit{uv} data weights for the 12m Array and ACA
measurements of our Band 7 spectral window near 351\,GHz.  This
comparison shows that the ACA measurements contribute minimally to the
overall spatial frequency sensitivity in our imaging measurements.

\begin{figure}
\centering
\includegraphics[scale=1.0]{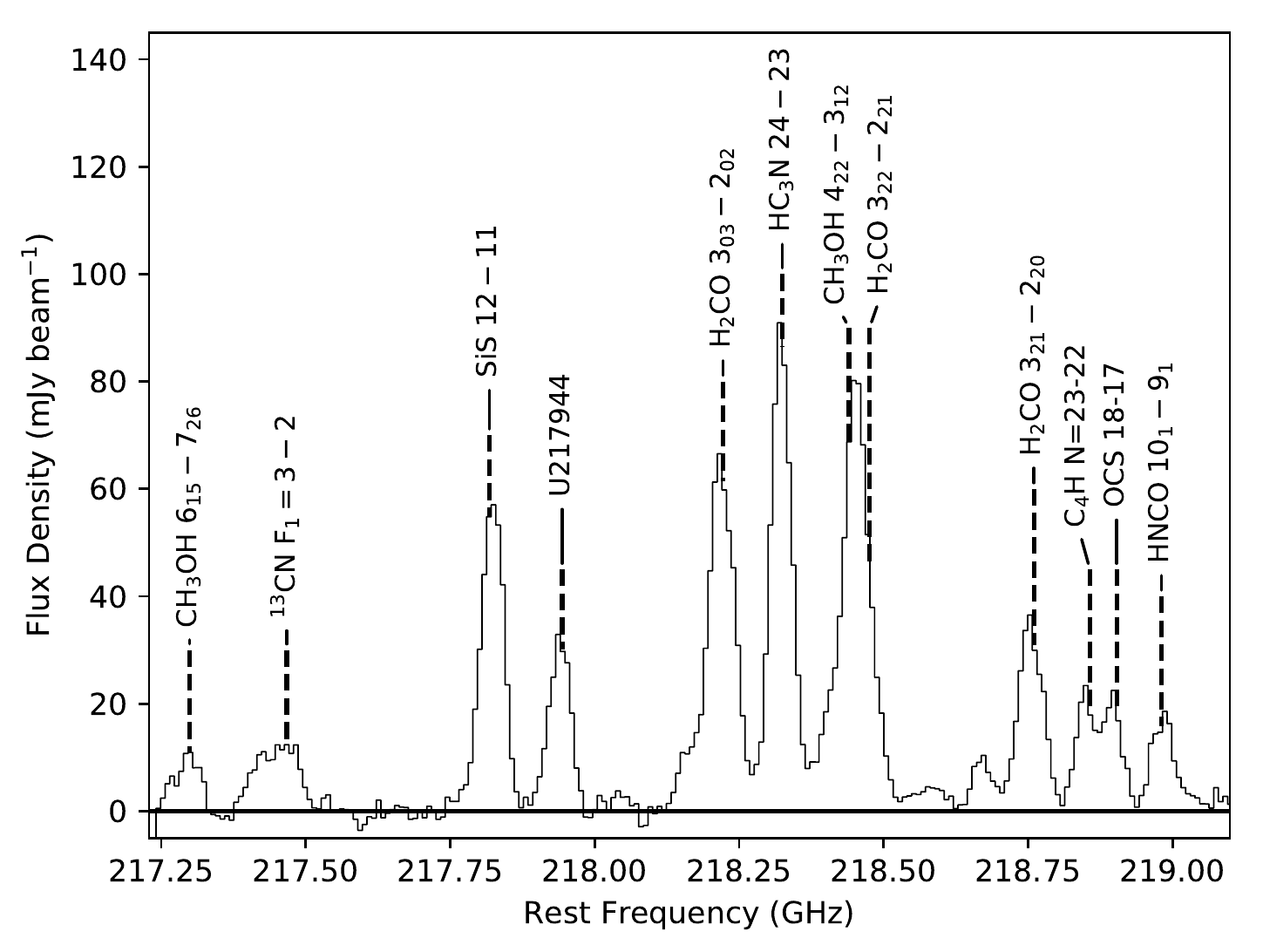}
\caption{Sample spectrum from our NGC\,253 Band 6 measurements
  centered at a rest frequency of 218.222192\,GHz.  Shown is the
  spectrum centered on Region 6 and averaged over a one arcsecond
  area and spectrally smoothed to 10\,km/s.  Identified molecular
  transitions are indicated.}
\label{fig:ExampleSpectrum}
\end{figure}

\begin{figure}
\centering
\includegraphics[trim=0mm 0mm 0mm 0mm,clip,scale=0.55]{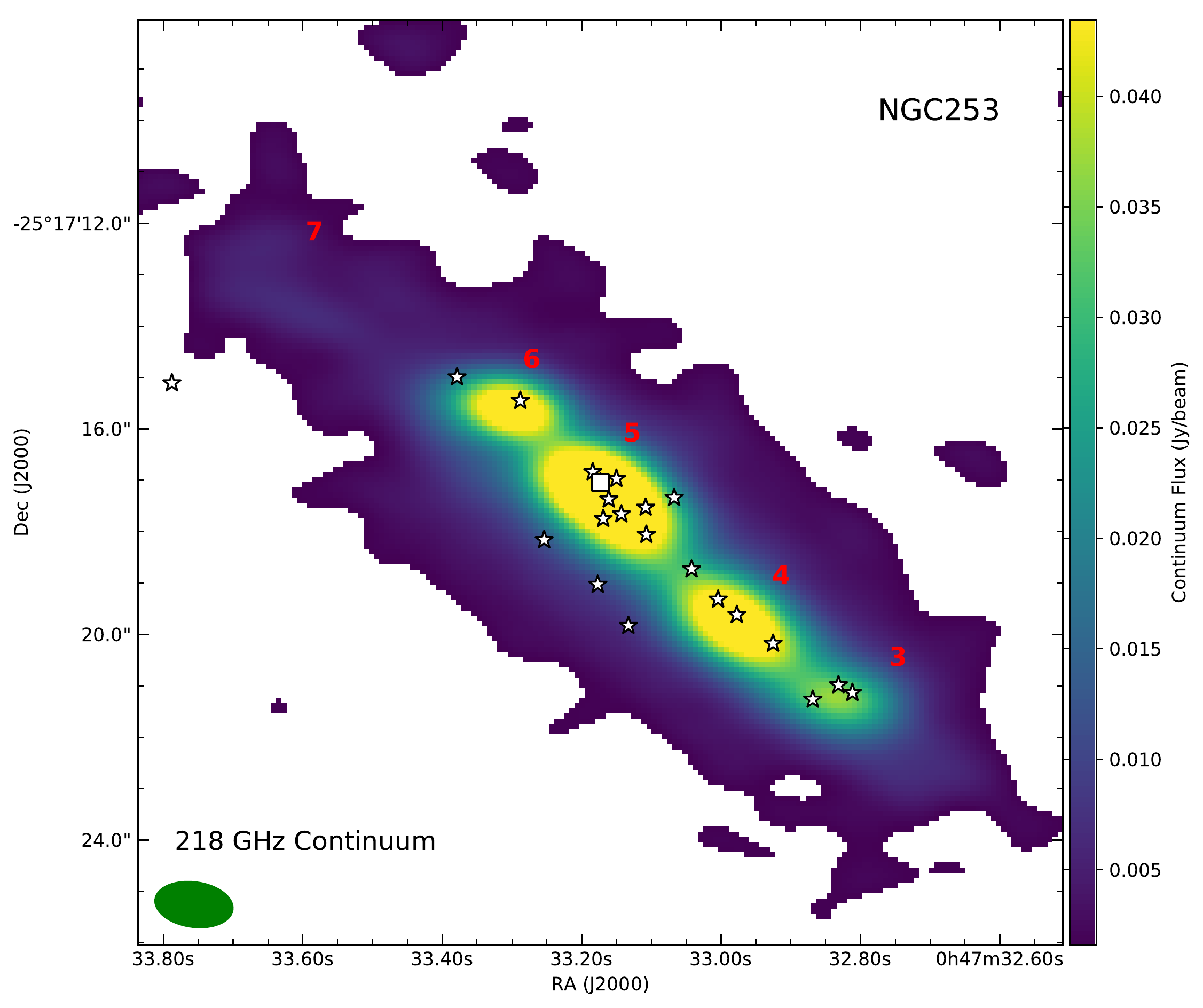}\\
\includegraphics[trim=0mm 0mm 0mm 0mm,clip,scale=0.55]{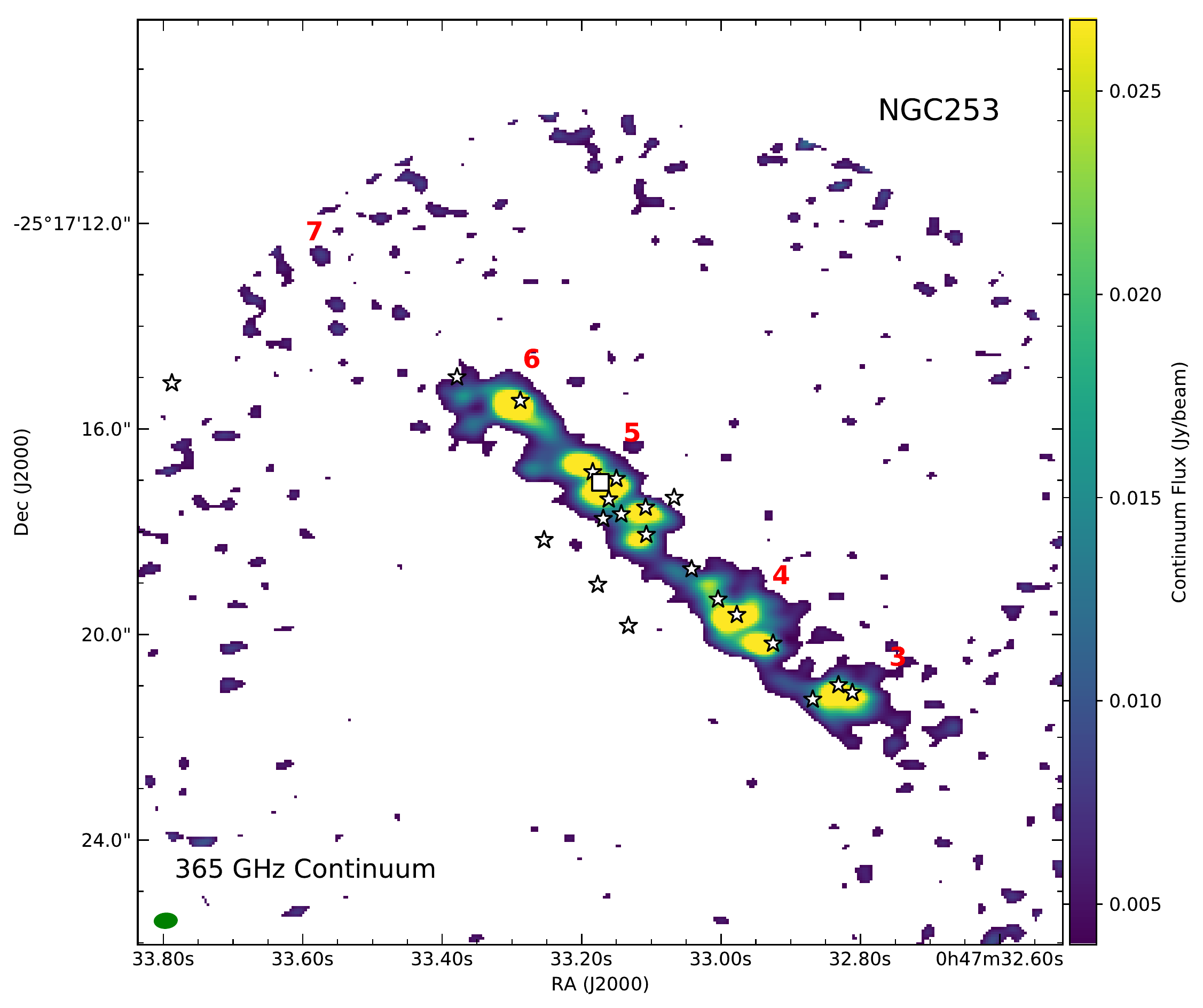}
\caption{Band 6 (218\,GHz with RMS = 0.8\,mJy/beam; top) and Band 7
  (365\,GHz with RMS = 1.5\,mJy/beam; bottom)
  continuum images of NGC\,253, shown on the same spatial scale and
  range. Red numbers indicate the locations of 
  the dense molecular emission regions identified by
  \citet[][Table~\ref{tab:ContPositions}]{Leroy2015}.  White
  black-bordered markers
  locate the positions of the 2\,cm radio continuum emission peaks
  \citep{Ulvestad1997}, with a square indicating the position of the
  strongest radio continuum peak identified by \cite{Turner1985} (TH2:
  RA(J2000) = 00$^h$ 47$^m$ 33$^s$.18, Dec(J2000) = $-25^\circ$
  17$^\prime$ 16$^{\prime\prime}$.93).}
\label{fig:NGC253Continuum}
\end{figure}

Due to the density of spectral lines in all of the spectral image
cubes for NGC\,253 (Figure~\ref{fig:ExampleSpectrum} presents an
example spectrum), spectral baseline fitting was an
iterative process using the CASA task \texttt{imcontsub}.  Almost all
baselines were fit with zeroth-order polynomials, with only a few
exceptions requiring the use of first-order polynomials.  Visual
inspections of the quality of the baseline removal for all spectral
windows indicate good quality overall.  For all but one spectral
window, only 10 spectral channels are believed to be line-free (the
Band 6 spectral window near 219\,GHz is estimated to have 70 line-free
channels). Following baseline subtraction, the line-free channels from
each spectral window were collapsed to form a pseudo-continuum image
of NGC\,253 in each spectral window.  The properties of these
pseudo-continuum images are listed in Table~\ref{tab:images} and
displayed in Figure~\ref{fig:NGC253Continuum}.  Note
that the RMS noise values listed for these continuum images are in
most cases slightly larger than a statistical averaging of the
chosen individual line-free channels would suggest, likely as a result
of low-level line contamination and imaging artifacts.

As it will be convenient to interchange between flux and brightness
temperature, we will use the general relation between the flux density
of a source (S$_\nu$) with solid angle $\Omega$ and its brightness
temperature (T$_B$): 
\begin{equation}
  S_\nu = {\frac{2k\nu^2}{c^2}}\int T_{B}(\Omega)d\Omega \\
\label{eq:svstb_gen}
\end{equation}
Note that Equation~\ref{eq:svstb_gen} assumes that the
Rayleigh-Jeans approximation ($h\nu \ll kT$) applies.  Assuming $\int
T_B d\Omega_s = T_B \Omega_s$ with $\Omega_s$ equal to the synthesized
gaussian beam solid angle $\Omega_B = \frac{\pi\theta_{maj}\theta_{min}}{4\ln{2}}$
in Equation~\ref{eq:svstb_gen} results in the following general
relation between flux density and measured brightness temperature for
a point source:
\begin{equation}
  T_B(K) \simeq 13.6\left(\frac{300}{\nu(GHz)}\right)^2
  \frac{S_\nu(Jy)}{\theta_{maj}(arcsec)\theta_{min}(arcsec)}
\label{eq:svstb_final}
\end{equation}

\subsection{Spatial Component Identification}
\label{CompID}

Using the continuum images described in Section~\ref{Imaging}, we have fit
single-component elliptical Gaussians to each of the continuum peaks
in these images.  Table~\ref{tab:ContPositions} lists the regions
identified in our measurements and also lists the corresponding
components noted by \cite{Leroy2015}, \cite{Meier2015},
\cite{Sakamoto2011}, \cite{Ando2017}, and \cite{Turner1985}.  Note
that components 4 and 
5 split into multiple components in our higher-resolution Band 7
measurements, as previously noted by \cite{Ando2017}.  In the
following we will use the regions noted in
Table~\ref{tab:ContPositions} as spatial reference positions for
further analysis of the spectral emission properties within the
nuclear disk of NGC\,253.

\begin{deluxetable}{lcclccc}
\tablewidth{0pt}
\tablecolumns{7}
\tablecaption{NGC\,253 Continuum Source Positions and Component Correspondence\label{tab:ContPositions}}
\tablehead{
\colhead{Region\tablenotemark{a}} & 
\colhead{RA(J2000)\tablenotemark{b}} & 
\colhead{Dec(J2000)\tablenotemark{b}} &
\colhead{M2015\tablenotemark{c}} &
\colhead{S2011\tablenotemark{c}} &
\colhead{A2017\tablenotemark{c}} &
\colhead{TH1985\tablenotemark{c}} \\
\colhead{} &
\colhead{(00$^h$ 47$^m$)} &
\colhead{($-25^\circ$ 17$^\prime$)} &
\colhead{} &
\colhead{} &
\colhead{} &
\colhead{}
}
\startdata
3 & 32.$^s$848 & 21.$^{\prime\prime}$05 & 4 & S1 & 8 & 9 \\
4 & 32.$^s$976 & 19.$^{\prime\prime}$79 & 5 & S2 & \nodata & 8 \\
4a & 32.$^s$982 & 19.$^{\prime\prime}$70 & 5 (subcomponent) & \nodata
& 6 & 8 \\
4b & 32.$^s$950 & 20.$^{\prime\prime}$00 & 5 (subcomponent) & \nodata
& 7 & \nodata \\
5 & 33.$^s$166 & 17.$^{\prime\prime}$29 & 6 & \nodata & \nodata & 2--6\\
5a & 33.$^s$118 & 17.$^{\prime\prime}$63 & 6 (subcomponent) & \nodata
& 4 & 6 \\
5b & 33.$^s$129 & 17.$^{\prime\prime}$89 & 6 (subcomponent) & \nodata
& 5 & 6 \\
5c & 33.$^s$165 & 17.$^{\prime\prime}$18 & 6 (subcomponent) & \nodata
& 2 & 2--5 \\
5d & 33.$^s$196 & 16.$^{\prime\prime}$80 & 6 (subcomponent) & \nodata
& 3 & 2 \\
6 & 33.$^s$297 & 15.$^{\prime\prime}$56 & 7 & S4 & 1 & 1 \\
7 & 33.$^s$637 & 13.$^{\prime\prime}$01 & 8 & S5 & \nodata & \nodata \\
\enddata
\tablenotetext{a}{~Nomenclature adopts \cite{Leroy2015} component numbering.}
\tablenotetext{b}{~Position errors are all $< 0.25^{\prime\prime}$.}
\tablenotetext{c}{~M2015:\citet{Meier2015}, S2011:\citet{Sakamoto2011},
  A2017:\citet{Ando2017}, TH1985:\citet{Turner1985}.}
\end{deluxetable}

\subsection{Spatial Component Hydrogen Column Density and Mass}
\label{DustColumnMass}

Using the peak continuum flux measurements derived from our spatial
gaussian fits to our continuum images (Section~\ref{CompID}), and
assuming that the continuum emission is dominated by thermal dust
emission, we have
calculated the hydrogen column densities and masses using the
well-worn dust emission assumptions elucidated by
\cite{Hildebrand1983}.  Assuming that the hydrogen column densities
and masses are dominated by molecular hydrogen, the total column
density and mass of hydrogen are given by
\begin{flalign}
  N(H_2) &\simeq 7.0\times10^{22} R_{gd} \left(\frac{\lambda
    (mm)}{0.4}\right)^\beta \frac{T_R (K)}{T_d (K)}\,cm^{-2} \\
  M(H_2) &\simeq 2\times10^{-4} R_{gd} \left(\frac{\lambda
    (mm)}{0.25}\right)^{\beta+3} D^2 (kpc) S_\nu (Jy)
  \exp\left(\frac{14.4}{\lambda (mm) T_d (K)} - 1\right)\,M_\odot
\label{eq:DustColumnMass}
\end{flalign}
where we have assumed optically thin dust
emission ($T_R = T_d\left[1-\exp(-\tau_\lambda)\right] \simeq T_d \tau_\lambda$) and
parameterized the gas-to-dust mass ratio as $R_{gd}$.  The assumption
of optically thin dust emission appears to be justified given the
moderate continuum brightness temperatures of $\sim 0.5$\,K that we
measure.  This optically thin assumption is also consistent with the
submillimeter through infrared dust continuum measurements presented
in \cite{PerezBeaupuits2018}.  The other variables
are the wavelength of observation ($\lambda$) in millimeters, the dust
emissivity power law ($\beta$), the radiation temperature
corresponding to the measured 
continuum flux in kelvin ($T_R$), and the dust temperature ($T_d$),
also in Kelvin.  We also use
Equation~\ref{eq:svstb_final} to convert our measured continuum fluxes
($S_\nu$) to radiation temperatures ($T_R$) assuming the spatial
resolutions associated with each continuum image.  Assuming then that
$R_{gd} = 150$, $\beta = 1.5$, and $T_d = 35$\,K 
\citep[see discussion in][Section 3.1.1]{Leroy2015}, with the
associated region dust continuum fluxes, we derive the hydrogen column
densities and masses listed in Table~\ref{tab:DustColumnMass}.  The
GMC hydrogen masses we derive are consistent with those measured by
\cite{Leroy2015}, and the sum of our GMC hydrogen masses is consistent
with the total CMZ hydrogen mass of $4.5\pm1.3\times10^8$\,M$_\sun$
measured by \cite{PerezBeaupuits2018}.

Our assumption of $R_{gd} = 150$ follows that of \cite{Leroy2015} and
\cite{Weiss2008}, which leverages the elemental (carbon through
nickel) depletion analysis 
presented in \cite{Draine2011}, which itself uses the Milky Way
elemental depletion analysis presented in \cite{Jenkins2009}.  The average
depletion of all sight lines analyzed by \cite{Jenkins2009} is $F_* =
0.36$, which implies a value for the gas-to-dust mass ratio of the
diffuse Milky Way of $R_{gd} = 150$.  Even though it is not clear
whether a diffuse gas Milky Way value for $R_{gd}$ is appropriate to the
dense CMZ of NGC\,253, it is the only properly calibrated (through
line-of-sight UV absorption measurements) value for this quantity.

\begin{deluxetable}{llll}
\tablewidth{0pt}
\tablecolumns{4}
\tablecaption{NGC\,253 GMC Hydrogen Column Densities and Masses\label{tab:DustColumnMass}}
\tablehead{
\colhead{Region\tablenotemark{a}} & 
\colhead{S$_\nu$ (mJy)} & 
\colhead{N(H$_2$) ($\times10^{23}$~cm$^{-2}$)\tablenotemark{b}} &
\colhead{M(H$_2$) ($\times10^6$\,M$_\sun$)\tablenotemark{b}}
}
\startdata
\multicolumn{4}{l}{218\,GHz Continuum; $1.52\times0.87$\,arcsec} \\
3 & $31.07\pm3.82$ & $11.53\pm1.42$ & $12.11\pm1.49$ \\
4 & $45.77\pm4.48$ & $16.99\pm1.66$ & $17.84\pm1.75$ \\
5 & $68.24\pm6.00$ & $25.33\pm2.23$ & $26.60\pm2.34$ \\
6 & $43.54\pm5.52$ & $16.16\pm2.05$ & $16.98\pm2.15$ \\
7 & $4.64\pm0.70$ & $1.73\pm0.26$ & $1.82\pm0.27$ \\
\hline
\multicolumn{4}{l}{220\,GHz Continuum; $1.56\times0.89$\,arcsec} \\
3 & $25.96\pm3.10$ & $8.93\pm1.07$ & $9.80\pm1.17$ \\
4 & $37.79\pm3.54$ & $13.00\pm1.22$ & $14.27\pm1.34$ \\
5 & $54.71\pm4.39$ & $18.26\pm1.51$ & $20.65\pm1.66$ \\
6 & $38.25\pm4.18$ & $13.16\pm1.44$ & $14.44\pm1.58$ \\
7 & $2.48\pm0.36$ & $0.85\pm0.12$ & $0.93\pm0.13$ \\
\hline
\multicolumn{4}{l}{351\,GHz Continuum; $0.45\times0.29$\,arcsec} \\
3 & $38.66\pm2.98$ & $27.34\pm2.10$ & $2.11\pm0.16$ \\
4a & $42.41\pm5.83$ & $29.99\pm4.12$ & $2.32\pm0.32$ \\
4b & $60.98\pm4.76$ & $43.12\pm3.37$ & $3.33\pm0.26$ \\
5a & $25.99\pm3.32$ & $18.38\pm2.35$ & $1.42\pm0.18$ \\
5b & $41.54\pm6.37$ & $29.37\pm4.51$ & $2.27\pm0.35$ \\
5c & $48.15\pm4.89$ & $34.05\pm3.46$ & $2.63\pm0.27$ \\
5d & $52.51\pm8.08$ & $37.13\pm5.71$ & $2.87\pm0.44$ \\
6 & $95.62\pm7.31$ & $67.62\pm5.17$ & $5.22\pm0.40$ \\
7 & $2.74\pm1.11$ & $1.94\pm0.79$ & $0.15\pm0.06$ \\
\hline
\multicolumn{4}{l}{363\,GHz Continuum; $0.43\times0.28$\,arcsec} \\
3 & $40.21\pm3.32$ & $27.50\pm2.27$ & $1.93\pm0.16$ \\
4a & $45.27\pm5.70$ & $30.96\pm3.90$ & $2.17\pm0.27$ \\
4b & $65.97\pm5.44$ & $45.12\pm3.72$ & $3.16\pm0.26$ \\
5a & $28.32\pm5.52$ & $19.37\pm3.78$ & $1.36\pm0.26$ \\
5b & $46.02\pm7.01$ & $31.47\pm4.79$ & $2.21\pm0.34$ \\
5c & $51.79\pm5.52$ & $35.42\pm3.78$ & $2.48\pm0.26$ \\
5d & $56.16\pm8.67$ & $38.41\pm5.93$ & $2.69\pm0.42$ \\
6 & $99.66\pm7.85$ & $68.16\pm5.37$ & $4.78\pm0.38$ \\
7 & $4.57\pm0.93$ & $3.13\pm0.64$ & $0.22\pm0.04$ \\
\hline
\multicolumn{4}{l}{365\,GHz Continuum; $0.43\times0.28$\,arcsec} \\
3 & $40.41\pm3.52$ & $27.27\pm2.38$ & $1.91\pm0.17$ \\
4a & $45.54\pm6.43$ & $30.73\pm4.34$ & $2.15\pm0.30$ \\
4b & $66.44\pm5.49$ & $44.83\pm3.70$ & $3.14\pm0.26$ \\
5a & $29.94\pm5.60$ & $20.20\pm3.78$ & $1.41\pm0.26$ \\
5b & $43.96\pm7.71$ & $29.66\pm5.20$ & $2.08\pm0.36$ \\
5c & $53.62\pm5.72$ & $36.18\pm3.86$ & $2.53\pm0.27$ \\
5d & $59.93\pm9.74$ & $40.44\pm6.57$ & $2.83\pm0.46$ \\
6 & $101.48\pm7.99$ & $68.48\pm5.39$ & $4.79\pm0.38$ \\
7 & $4.15\pm3.79$ & $2.80\pm2.56$ & $0.20\pm0.18$ \\
\enddata
\tablenotetext{a}{~Nomenclature adopts \cite{Leroy2015} component numbering.}
\tablenotetext{b}{~Assuming R$_{gd} = 150$, T$_d = 35$\,K, $\tau_d
  \ll 1$, $\beta = 1.5$, and D = 3.5\,Mpc.}
\end{deluxetable}

In order to calculate molecular abundances using our total molecular
column densities (Section~\ref{Colden}), we average our GMC-specific
hydrogen column densities per receiver band over any subcomponents
that compose a main GMC in Table~\ref{tab:DustColumnMass}.  These
averaged hydrogen column densities (and masses) are listed in
Table~\ref{tab:DustColumnMassAverage}.  As will be done for our
calculations of the total molecular column density
(Section~\ref{Colden}), the uncertainties associated with each
hydrogen column density and mass represent the larger of the 
statistical uncertainty and the standard deviation of the individual
column densities or masses derived.  These averaged hydrogen column
densities will be used to derive molecular abundances within the GMCs
of NGC\,253.

\begin{deluxetable}{lll}
\tablewidth{0pt}
\tablecolumns{3}
\tablecaption{NGC\,253 Average GMC Hydrogen Column Densities and Masses\label{tab:DustColumnMassAverage}}
\tablehead{
\colhead{Region\tablenotemark{a}} & 
\colhead{$\langle$N(H$_2$)$\rangle$ ($\times10^{23}$~cm$^{-2}$)\tablenotemark{b}} &
\colhead{$\langle$M(H$_2$)$\rangle$ ($\times10^6$\,M$_\sun$)\tablenotemark{b}}
}
\startdata
\multicolumn{3}{l}{220\,GHz Continuum; $1.5\times0.9$\,arcsec} \\
3 & $10.23\pm1.30$ & $10.96\pm1.15$ \\
4 & $15.00\pm2.00$ & $16.05\pm1.79$ \\
5 & $21.80\pm3.53$ & $23.63\pm2.98$ \\
6 & $14.66\pm1.50$ & $15.71\pm1.33$ \\
7 & $1.29\pm0.44$ & $1.38\pm0.44$ \\
\hline
\multicolumn{3}{l}{360\,GHz Continuum; $0.4\times0.3$\,arcsec} \\
3 & $27.38\pm1.31$ & $1.98\pm0.09$ \\
4 & $37.46\pm6.93$ & $2.71\pm0.50$ \\
5 & $30.84\pm7.38$ & $2.23\pm0.53$ \\
6 & $68.09\pm3.06$ & $4.93\pm0.22$ \\
7 & $2.63\pm0.92$ & $0.19\pm0.06$ \\
\enddata
\tablenotetext{a}{~Nomenclature adopts \cite{Leroy2015} component numbering.}
\tablenotetext{b}{~Assuming R$_{gd} = 150$, T$_d = 35$\,K, $\tau_d
  \ll 1$, $\beta = 1.5$, and D = 3.5\,Mpc.}
\end{deluxetable}

\section{Analysis}
\label{Analysis}

\subsection{Spectral Line Identification}
\label{LineID}

Molecular spectral line identification within each spectral window of
our NGC\,253 measurements was done in two steps.  In a
first step the ALMA Data-Mining Toolkit \citep[ADMIT;][]{Teuben2015} was
used to search for the most appropriate identification of spectral
features based on the known velocity structure within the nucleus of
NGC\,253.  Once an initial set of molecular species was
identified, residual species within each spectral window were
identified by eye using lists of line rest frequencies
\citep{Lovas1992,Muller2001} and anticipated
general abundances for potential species. 
Table~\ref{tab:SpectralProperties} in Appendix~\ref{Spectral} lists the
molecular transitions and frequencies measured toward NGC\,253.

The Band 6 low spectral resolution spectral window centered at
a rest frequency of 234.7\,GHz was anticipated to be line free, and
thus would have served as a sensitive continuum measurement.  It was determined,
though, that spectral lines existed in this spectral window with rest
frequencies near 234.69 and 235.15\,GHz.  These spectral line
frequencies are consistent with CH$_3$OH $4_{2}-5_{1}$~A at
234683.39\,MHz and/or CH$_3$OH $5_{-4}-6_{-3}$~E at 234698.45\,MHz.

\subsection{Spectral Line Signal Extraction and Spatial Component Fitting}
\label{Extraction}

In order to extract integrated spectral line intensities from our
measurements, we have developed a python script, called
\texttt{CubeLineMoment}\footnote{\url{https://github.com/keflavich/mangum_galaxies/blob/master/CubeLineMoment.py}},
which uses a series of spectral and spatial 
masks to extract integrated intensities for a defined list of target
spectral frequencies.  \texttt{CubeLineMoment} makes extensive use of
\texttt{spectral-cube}\footnote{\url{https://zenodo.org/record/1213217}}.

The masking process begins by selecting a bright spectral line whose velocity
structure is representative of the emission across the galaxy.  Preferably, it
should be maximally inclusive, such that all other lines emit over a smaller
area in position-position-velocity (PPV) space, which here has right
ascension, declination and heliocentric velocity as axes.  Various
images are computed 
based on this line, including noise, peak intensity, position of peak
intensity, and second moment (velocity dispersion).  The C$^{18}$O $2-1$ and
H$_2$CO $5_{15}-4_{14}$ transitions were found to be appropriate
choices for the bright ``tracer'' transitions in our Band 6 and 7
spectral windows, respectively.

These maps are then converted into a PPV mask cube by producing Gaussian
profiles at each spatial pixel with peak intensity, centroid, and width defined
by the appropriate masks.  The Gaussians are sampled onto the PPV grid defined
by the target emission line.  
For each spatial pixel, spectral pixels are masked out below the
1-$\sigma$ level evaluated on the model Gaussian.  By evaluating only
on the model Gaussian, we exclude pixels above 1-$\sigma$ at other
parts of the spectrum, which otherwise would contribute significantly
to the included region.  The mask is then applied to the target
emission-line data cube, and moment 0, 1, and 2 maps are produced.

Integrated spectral line intensity images derived from our
\texttt{CubeLineMoment} analysis are listed in Appendix~\ref{IntegInt}.  In
order to associate spectral line integrated intensities with each of the
spatial components noted in both previous and the current measurements, we have
used the
\texttt{gaussfit\_catalog}\footnote{\url{https://github.com/radio-astro-tools/gaussfit_catalog}} application,
which uses
\texttt{pyspeckit}\footnote{\url{https://pyspeckit.readthedocs.io/}} to perform gaussian fits of the spatial
molecular spectral line components associated with the regions listed
in Table~\ref{tab:ContPositions}.  As a characterization of the
high-density component structure within the NGC\,253 nucleus, and to
provide an example of the spatial gaussian fits performed,
Table~\ref{tab:H2COGaussFit} lists the derived averaged peak position
and size for the nuclear regions derived from our H$_2$CO
$3_{03}-2_{02}$, $3_{21}-2_{20}$, $5_{15}-4_{14}$, $5_{05}-4_{04}$,
$5_{24}-4_{23}$, $5_{23}-4_{22}$, $5_{3}-4_{3}$ and
$5_{4}-4_{4}$ integrated\footnote{Since the $5_{33}-4_{32}$ and $5_{32}-4_{31}$
  transition pair, and the $5_{42}-4_{41}$ and $5_{41}-4_{40}$
  transition pair are both spectrally blended, we use the shorthand
  notation which drops the K$_{+1}$ quantum number.}
intensity images.  We have compared the derived peak position for
each H$_2$CO component to those that correspond to positions
derived from our Band 6 and 7 dust continuum and the spectral line
measurements of \cite{Leroy2015}, \cite{Meier2015},
\cite{Sakamoto2011}, \cite{Ando2017}, and \cite{Turner1985}
(Table~\ref{tab:ContPositions}).  We find that with but one  
exception, all peak H$_2$CO positions are consistent within respective
measurement uncertainties with their
corresponding ALMA Band 6 and 7 continuum positions and with
component positions derived from the earlier works listed.  For the
one exception, Region 5, the H$_2$CO position differs by
($\Delta$RA,$\Delta$Dec) = $(+0.64,+0.69)$\,arcsec ((11,12)\,pc) from its reference
position.  The position for Region 5 is derived from our
Band 6 measurements, whose spatial resolution is
$\theta_{maj}\times\theta_{min} \simeq 1.5\times0.9$\,arcsec).  Region
5 is also known to have substructure in higher-resolution measurements
(including our Band 7 imaging) and has been suggested as a component
that suffers from self-absorption in lower-excitation molecular
spectral line measurements \citep{Meier2015}.  Even though real
molecular abundance gradients within Region 5 cannot be excluded, the
currently most plausible explanation for this position shift between
our H$_2$CO and dust continuum plus previous low-excitation molecular
emission measurements is a complex emission structure below the
spatial resolution and sensitivity of our measurements.

\startlongtable
\begin{deluxetable}{lccc}
\tablewidth{0pt}
\tablecolumns{4}
\tablecaption{Averaged Gaussian Fits to NGC\,253 Formaldehyde Components\label{tab:H2COGaussFit}}
\tablehead{
\colhead{Region} & 
\colhead{RA(J2000)} & 
\colhead{Dec(J2000)} &
\colhead{FWHM (arcsec,deg)} \\
\colhead{} &
\colhead{(00$^h$ 47$^m$)} &
\colhead{($-25^\circ$ 17$^\prime$)} &
\colhead{($\theta_{maj}\times\theta_{min}$ @ PA)}
}
\startdata
3 & 32.8258$\pm$0.0077 & 21.1676$\pm$0.0671 &
$1.19\pm0.78 \times 0.52\pm0.29$ @$-59\pm169$ \\
4 & 32.9644$\pm$0.0004 & 19.7354$\pm$0.0231 &
$1.90\pm0.14 \times 1.01\pm0.05$ @ $160\pm11$ \\
4a & 32.9833$\pm$0.0048 & 19.7554$\pm$0.1313 &
$0.61\pm0.30 \times 0.36\pm0.13$ @$20\pm38$ \\
4b & 32.9537$\pm$0.0059 & 20.0595$\pm$0.1917 &
$0.85\pm0.37 \times 0.46\pm0.25$ @$-38\pm49$ \\
5 & 33.2129$\pm$0.0094 & 16.5954$\pm$0.0619 &
$2.89\pm0.20 \times 1.09\pm0.01$ @$144\pm0$ \\
5a & 33.113$\pm$0.0014 & 17.6237$\pm$0.0173 &
$0.54\pm0.22 \times 0.26\pm0.05$ @$-81\pm84$ \\
5b & 33.1298$\pm$0.0004 & 17.8931$\pm$0.0624 &
$0.51\pm0.15 \times 0.27\pm0.06$ @$-49\pm24$ \\
5c & 33.1666$\pm$0.0044 & 17.2551$\pm$0.0819 &
$0.52\pm0.10 \times 0.38\pm0.07$ @$-96\pm151$ \\
5d & 33.1979$\pm$0.0043 & 16.7585$\pm$0.0709 &
$0.68\pm0.34 \times 0.32\pm0.10$ @$15\pm70$ \\
6 & 33.2994$\pm$0.0063 & 15.5952$\pm$0.0869 &
$0.89\pm0.70 \times 0.54\pm0.32$ @$39\pm70$ \\
7 & 33.6378$\pm$0.0063 & 13.0969$\pm$0.1072 &
$0.88\pm0.46 \times 0.50\pm0.32$ @$-78\pm144$ \\
\enddata
\end{deluxetable}

Table~\ref{tab:IntegInt} lists the
spectral line integrated intensities derived from the
\texttt{CubeLineMoment} output and \texttt{gaussfit\_catalog} gaussian
fit analysis from all detected transitions
that are not significantly blended and whose derived integrated
intensity is larger than $2\sigma$.  Full-width half-maximum
(FWHM) line widths derived from this analysis range from $\sim50$ to
$\sim80$\,km/s for most of our unblended spectral line
measurements.  Furthermore, these line widths show little variation
over the Regions that compose the NGC\,253 CMZ.
In the following sections we will
use these integrated intensities to derive molecular column densities
that we will subsequently use to study the kinetic temperature and
molecular abundance ratios within the starburst nuclear components of
NGC\,253. 

\begin{longrotatetable}
\begin{deluxetable}{lcp{2.5cm}p{2.5cm}cc}
\tablewidth{0pt}
\tablecolumns{6}
\tablecaption{NGC\,253 Integrated Spectral Line Intensities (Jy km/s)\tablenotemark{a}\label{tab:IntegInt}}
\tablehead{
\colhead{Transition} & 
\colhead{Region 3} & 
\colhead{Region 4} & 
\colhead{Region 5} &
\colhead{Region 6} &
\colhead{Region 7}
}
\startdata
$^{13}$CO $2-1$ & 16.15$\pm$4.11 & 19.51$\pm$4.00 & (3.30) &
26.53$\pm$4.55 & 11.92$\pm$3.20 \\
C$^{18}$O $2-1$ & 6.80$\pm$1.60 & 7.17$\pm$1.34 & (1.07) &
11.91$\pm$1.67 & 4.48$\pm$1.09 \\
$^{13}$CN, F$_1$=3-2 & (0.12) & 1.42$\pm$0.17 & (0.15) &
1.29$\pm$0.13 & (0.04) \\
SiS $12-11$ & 2.22$\pm$0.32 & 2.37$\pm$0.30 & (0.29) &
3.27$\pm$0.37 & 1.42$\pm$0.11 \\
SO 6$_5$-5$_4$ & 1.46$\pm$0.19 & 2.49$\pm$0.20 &
1.70$\pm$0.22 & 4.68$\pm$0.32 & 0.94$\pm$0.05 \\
SO$_2$ $4_{22}-3_{13}$ & 1.05$\pm$0.11 & 0.94$\pm$0.10 & 0.54$\pm$0.09
& 1.32$\pm$0.15 & 0.20$\pm$0.07 \\
SO$_2$ $5_{33}-4_{22}$ & 0.31$\pm$0.05 &
4a:0.25$\pm$0.06, 4b:0.66$\pm$0.12 &
5a:0.15$\pm$0.03, 5b:0.25$\pm$0.04, 5c:0.28$\pm$0.08, 5d:0.38$\pm$0.13 &
0.90$\pm$0.04 & (0.04) \\
SO$_2$ $14_{4,10}-14_{3,11}$ & 0.21$\pm$0.07 &
4a:0.30$\pm$0.07, 4b:0.80$\pm$0.24 & 
5a:0.37$\pm$0.10, 5b:(0.09), 5c:0.99$\pm$0.12, 5d:0.44$\pm$0.08 &
1.34$\pm$0.08 & (0.03) \\
HNC $4-3$ & 3.90$\pm$0.81 & 4a:17.99$\pm$1.27, 4b:(1.61) & 
5a:6.03$\pm$2.22, 5b:3.62$\pm$2.16, 5c:6.05$\pm$1.47, 5d:(3.45) &
28.77$\pm$1.38 & 1.54$\pm$0.22 \\
HNC $4-3~\nu_2 = 1f$ & 0.68$\pm$0.07 & 4a:1.81$\pm$0.07, 4b:(0.07) &
5a:1.21$\pm$0.13, 5b:(0.17), 5c:(0.09), 5d:(0.56) & 
0.54$\pm$0.13 & 0.09$\pm$0.04 \\
OCS $30-29$ & 1.30$\pm$0.25 & 4a:2.55$\pm$0.31, 4b:(0.38) &
5a:1.08$\pm$0.10, 5b:(0.78), 5c:2.97$\pm$0.35, 5d:(0.46) &
5.74$\pm$0.68 & 0.71$\pm$0.17 \\
HNCO $10_0-9_0$\tablenotemark{b} & 4.58$\pm$0.52 & 2.45$\pm$0.32 & (0.27) &
4.70$\pm$0.38 & 3.43$\pm$0.20 \\
HNCO $10_1-9_1$\tablenotemark{b} & 1.29$\pm$0.12 & 0.82$\pm$0.10 & (0.14) &
1.14$\pm$0.09 & 0.18$\pm$0.05 \\
HNCO $10_2-9_2$\tablenotemark{b} & \nodata & 0.39$\pm$0.04 & (0.06) &
0.55$\pm$0.07 & (0.03) \\
H$_2$CO $3_{03}-2_{02}$ & 2.52$\pm$0.37 & 2.99$\pm$0.39 &
(0.35) & 4.22$\pm$0.49 & 2.96$\pm$0.27 \\
H$_2$CO $3_{21}-2_{20}$ & 1.18$\pm$0.19 & 2.01$\pm$0.20 &
(0.19) & 2.74$\pm$0.20 & 1.20$\pm$0.08 \\
H$_2$CO $5_{05}-4_{04}$\tablenotemark{c} & \nodata & \nodata &
\nodata & \nodata & \nodata \\
H$_2$CO $5_{24}-4_{23}$ & 0.85$\pm$0.08 &
4a:1.73$\pm$0.10, 4b:0.50$\pm$0.12 &
5a:0.91$\pm$0.15, 5b:(0.16), 5c:0.45$\pm$0.10, 5d:0.91$\pm$0.34 &
3.83$\pm$0.10 & 0.21$\pm$0.07 \\
H$_2$CO $5_{23}-4_{22}$ & 0.46$\pm$0.06 &
4a:1.41$\pm$0.07, 4b:0.28$\pm$0.09 &
5a:0.92$\pm$0.10, 5b:(0.13), 5c:0.33$\pm$0.09, 5d:0.60$\pm$0.28 &
2.52$\pm$0.11 & 0.64$\pm$0.12 \\
H$_2$CO $5_4-4_4$\tablenotemark{d} & 0.20$\pm$0.04 & 4a:0.54$\pm$0.06, 4b:0.34$\pm$0.17 &
5a:0.35$\pm$0.08, 5b:(0.12), 5c:0.56$\pm$0.05, 5d:0.20$\pm$0.10 &
(0.04) & (0.07) \\
H$_2$CO $5_{15}-4_{14}$ & 1.73$\pm$0.14 &
4a:3.74$\pm$0.24, 4b:1.23$\pm$0.28 &
5a:1.62$\pm$0.29, 5b:1.26$\pm$0.42, 5c:2.18$\pm$0.30, 5d:1.80$\pm$0.38 & 
4.71$\pm$0.42 & 0.63$\pm$0.10 \\
H$_2$CO $5_3-4_3$\tablenotemark{d} & 1.29$\pm$0.11 & 4a:3.30$\pm$0.13, 4b:0.37$\pm$0.15 &
5a:1.75$\pm$0.32, 5b:(0.32), 5c:1.25$\pm$0.24, 5d:1.45$\pm$0.59 & 
5.18$\pm$0.23 & 0.93$\pm$0.10 \\
H$_3$O$^+$ $3_2-2_2$ & 1.08$\pm$0.23 &
4a:2.35$\pm$0.30, 4b:(0.37) &
5a:\nodata, 5b:0.64$\pm$0.58, 5c:2.63$\pm$0.33, 5d:1.09$\pm$0.41 &
5.46$\pm$0.68 & 0.65$\pm$0.15 \\
HC$_3$N $24-23$ & 2.38$\pm$0.27 & 2.85$\pm$0.27 & 2.66$\pm$0.35 &
6.63$\pm$0.54 & 1.29$\pm$0.10 \\
HC$_3$N $24-23 \nu_7 = 2$ & (0.01) & 0.06$\pm$0.02 & 2.44$\pm$0.39 &
0.55$\pm$0.14 & 0.09$\pm$0.04 \\
HC$_3$N $40-39$ & 0.77$\pm$0.07 &
4a:0.90$\pm$0.09, 4b:0.46$\pm$0.10 & 
5a:0.52$\pm$0.14, 5b:(0.14), 5c:0.31$\pm$0.11, 5d:0.78$\pm$0.30 &
4.29$\pm$0.13 & (0.13) \\
C$_4$H N=23-22\tablenotemark{e} & 0.36$\pm$0.06 & 0.51$\pm$0.05 &
0.56$\pm$0.09 & 1.71$\pm$0.10 & 0.045$\pm$0.03 \\
CH$_3$CN $12-11$ & 1.58$\pm$0.12 & 0.88$\pm$0.09 & \nodata &
2.39$\pm$0.11 & 0.65$\pm$0.04 \\
CH$_3$OH $9_{5}$-$10_{4}$E & 0.28$\pm$0.08 &
4a:0.25$\pm$0.14, 4b:1.09$\pm$0.03 & 
5a:0.17$\pm$0.04, 5b:0.32$\pm$0.06, 5c:0.58$\pm$0.09, 5d:0.38$\pm$0.12 &
0.89$\pm$0.06 & (0.03) \\
U217944\tablenotemark{f} & 1.17$\pm$0.15 & \nodata & \nodata & 
2.10$\pm$0.20 & 0.47$\pm$0.06 \\
U351047 & 0.96$\pm$0.13 & 4a:1.61$\pm$0.12, 4b:\nodata &
5a:0.77$\pm$0.12, 5b:0.55$\pm$0.23, 5c:1.07$\pm$0.16, 5d:1.06$\pm$0.20 &
2.37$\pm$0.20 & \nodata \\
U352199\tablenotemark{f} & 0.19$\pm$0.04 & 4a:0.36$\pm$0.04,
4b:(0.04) &
5a:0.16$\pm$0.05, 5b:(0.04), 5c:0.36$\pm$0.05, 5d:0.24$\pm$0.08 &
0.87$\pm$0.04 & \nodata \\
U365185\tablenotemark{f} & 0.66$\pm$0.07 & 4a:1.84$\pm$0.07, 4b:(0.07) &
5a:1.22$\pm$0.12, 5b:(0.17), 5c:0.20$\pm$0.09, 5d:0.91$\pm$0.59 &
7.41$\pm$0.12 & \nodata \\
\enddata
\tablenotetext{a}{~For values less than $2\sigma$, RMS
  uncertainty listed in parentheses.}
\tablenotetext{b}{~All three HNCO transitions are blended with nearby
  species, making these HNCO integrated intensities uncertain.}
\tablenotetext{c}{~Blend with HNC $4-3$, making H$_2$CO integrated
  intensities unrecoverable.}
\tablenotetext{d}{~Since this is a blend of two transitions whose
  intensities should be equal, the measured intensity from
  Table~\ref{tab:IntegInt} has been divided by 2 to calculate the
  column density.}
\tablenotetext{e}{~Composed of the blended (J=47/2-45/2,F=23-22),
  (J=47/2-45/2,F=24-23), (J=45/2-43/2,F=22-21), and
  (J=45/2-43/2,F=23-22) multiplet at 218837.00950\,MHz.}
\tablenotetext{f}{~May be assigned as CH$_3$OCHO J=34-33 (U365185),
  CH$_3$OCHO $45_{14,32}-45_{13,33}$ (U352199), and CH$_3$OCHO
  $45_{29,16}-46_{28,18}$ (U217944)}
\end{deluxetable}
\end{longrotatetable}

\section{Kinetic Temperature Derivation Using Formaldehyde}
\label{H2COTk}
As described by \cite{Mangum1993}, the formaldehyde molecule possesses
structural properties that allow for its rotational transitions to be
used as probes of the kinetic temperature in dense molecular gas
environs.  H$_2$CO is a slightly asymmetric rotor molecule, so that
its energy levels are defined by three quantum numbers: total
angular momentum J, the projection of J along the symmetry axis for a
limiting prolate symmetric top, K$_{-1}$, and the projection of J
along the symmetry axis for a limiting oblate symmetric top,
K$_{+1}$. The H$_2$CO energy level diagram that shows all energy
levels below 300\,K is shown in Figure~12 of \cite{Mangum1993}.

For radiative excitation in a symmetric rotor molecule, dipole
selection rules dictate that $\Delta$K = 0. Transitions between energy levels 
when $\Delta$K$\neq$0 can only occur via collisional excitation. This,
then, is the fundamental reason why symmetric rotor molecules are
tracers of kinetic temperature in dense molecular clouds. A comparison
between the energy level populations from different K-levels within
the same molecular symmetry species (ortho or para) should allow a
direct measure of the 
kinetic temperature in the gas. The asymmetry in H$_2$CO ($\kappa =
-0.96$) makes it structurally similar to a prolate symmetric rotor molecule
($\kappa = -1.0$). Therefore, measurements of 
the relative intensities of two transitions whose K-levels originate
from the same $\Delta$J = 1 transition provide a direct measure of the
kinetic temperature.

To connect the kinetic temperature in the dense nuclear gas in
starburst galaxies to the intensity of molecular transitions which
originate from these nuclei, one needs to solve for the coupled
statistical equilibrium and radiative transfer equations.  A simple
solution to these coupled equations is afforded by the large velocity
gradient (LVG) approximation \citep{Sobolev1960}.  The detailed
properties of our implementation of the LVG approximation 
are described in \cite{Mangum1993}.  From the \cite{Mangum1993}
summary of the uncertainties associated with LVG model results, we note
that uncertainties in the collisional excitation rates \citep{Green1991}, which can be
as high as 50\% for state-to-state rates, are not included in our
analysis uncertainties.  This contribution to the uncertainties
of our derived physical conditions is traditionally
ignored, and we only mention it here to provide context to our
analysis.  The simplified solution to the radiative transfer equation
that the LVG approximation provides allows for a calculation of the
global dense gas properties in a range of environments.

We have applied our LVG model formalism to the unblended H$_2$CO
transitions measured toward NGC\,253.  With the nine
transitions for which we have measured integrated intensities
(Table~\ref{tab:IntegInt}) we can form five unique H$_2$CO transition ratios that
can be used to derive the kinetic temperature in the dense nuclear
regions of NGC\,253.
The analysis of the limits in kinetic temperature, volume density, and
H$_2$CO column density to our measured H$_2$CO kinetic temperature
sensitive ratios derived by \cite{Mangum1993} are directly applicable
to our NGC\,253 measurements.  That analysis concluded
that, over a volume density range of n(H$_2$) =
  $10^{4.5}-10^{8.5}$\,cm$^{-3}$ in a molecular cloud
core, the following integrated intensity ratios measure kinetic
temperature to an uncertainty of $\lesssim 25$\,\%:
\begin{enumerate}
  \item $\frac{Sdv(3_{03}-2_{02})}{Sdv(3_{21}-2_{20})}$ when T$_K
    \lesssim 50$\,K and N(para-H$_2$CO)/$\Delta$v $\lesssim
    10^{13.5}$\,cm$^{-2}$/(km\,s$^{-1}$).  This rule also applies to
    the same ratio involving the $3_{22}-2_{21}$ transition;
  \item $\frac{Sdv(5_{05}-4_{04})}{Sdv(5_{23}-4_{22})}$ when T$_K
    \lesssim 75$\,K and N(para-H$_2$CO)/$\Delta$v $\lesssim
    10^{14.0}$\,cm$^{-2}$/(km\,s$^{-1}$).  This rule also applies to
    the same ratio involving the $5_{24}-4_{23}$ transition;
  \item $\frac{Sdv(5_{24}-4_{23})}{Sdv(5_{4}-4_{4})}$ when T$_K
    \lesssim 150$\,K and N(para-H$_2$CO)/$\Delta$v $\lesssim
    10^{14.5}$\,cm$^{-2}$/(km\,s$^{-1}$);
  \item $\frac{Sdv(5_{15}-4_{14})}{Sdv(5_{3}-4_{3})}$ when T$_K
    \lesssim 100$\,K and N(ortho-H$_2$CO)/$\Delta$v $\lesssim
    10^{14.0}$\,cm$^{-2}$/(km\,s$^{-1}$);
\end{enumerate}
and to an upper limit, defined as the point at which the uncertainty
in T$_K$ becomes 50\%, of the following:
\begin{enumerate}
  \item T$_K \lesssim 150$\,K for $\frac{Sdv(3_{03}-2_{02})}{Sdv(3_{21}-2_{20})}$
    (this rule also applies to the same ratio involving the
    $3_{22}-2_{21}$ transition);
  \item T$_K \lesssim 200$\,K $\frac{Sdv(5_{05}-4_{04})}{Sdv(5_{23}-4_{22})}$
    (this rule also applies to the same ratio involving the
    $5_{24}-4_{23}$ transition);
  \item T$_K \lesssim 300$\,K
    $\frac{Sdv(5_{24}-4_{23})}{Sdv(5_{4}-4_{4})}$ (this rule also applies to
    the same ratio involving the $5_{23}-4_{22}$ transition);
  \item T$_K \lesssim 250$\,K  $\frac{Sdv(5_{15}-4_{14})}{Sdv(5_{3}-4_{3})}$;
\end{enumerate}
for the limits to the column densities per line width (FWZI) listed.

Due to velocity blending with the HNC $4-3$ transition, the H$_2$CO
$5_{05}-4_{04}$ transition cannot be used as a reliable kinetic
temperature diagnostic in NGC\,253.  We rely then on the 
$\frac{Sdv(3_{03}-2_{02})}{Sdv(3_{21}-2_{20})}$,
$\frac{Sdv(5_{15}-4_{14})}{Sdv(5_{3}-4_{3})}$,
$\frac{Sdv(5_{24}-4_{23})}{Sdv(5_{4}-4_{4})}$, and
$\frac{Sdv(5_{23}-4_{22})}{Sdv(5_{4}-4_{4})}$ ratios to derive kinetic
temperature images of the NGC\,253 CMZ.

\begin{figure}
\centering
\includegraphics[trim=0mm 8mm 0mm 10mm,clip,scale=0.40]{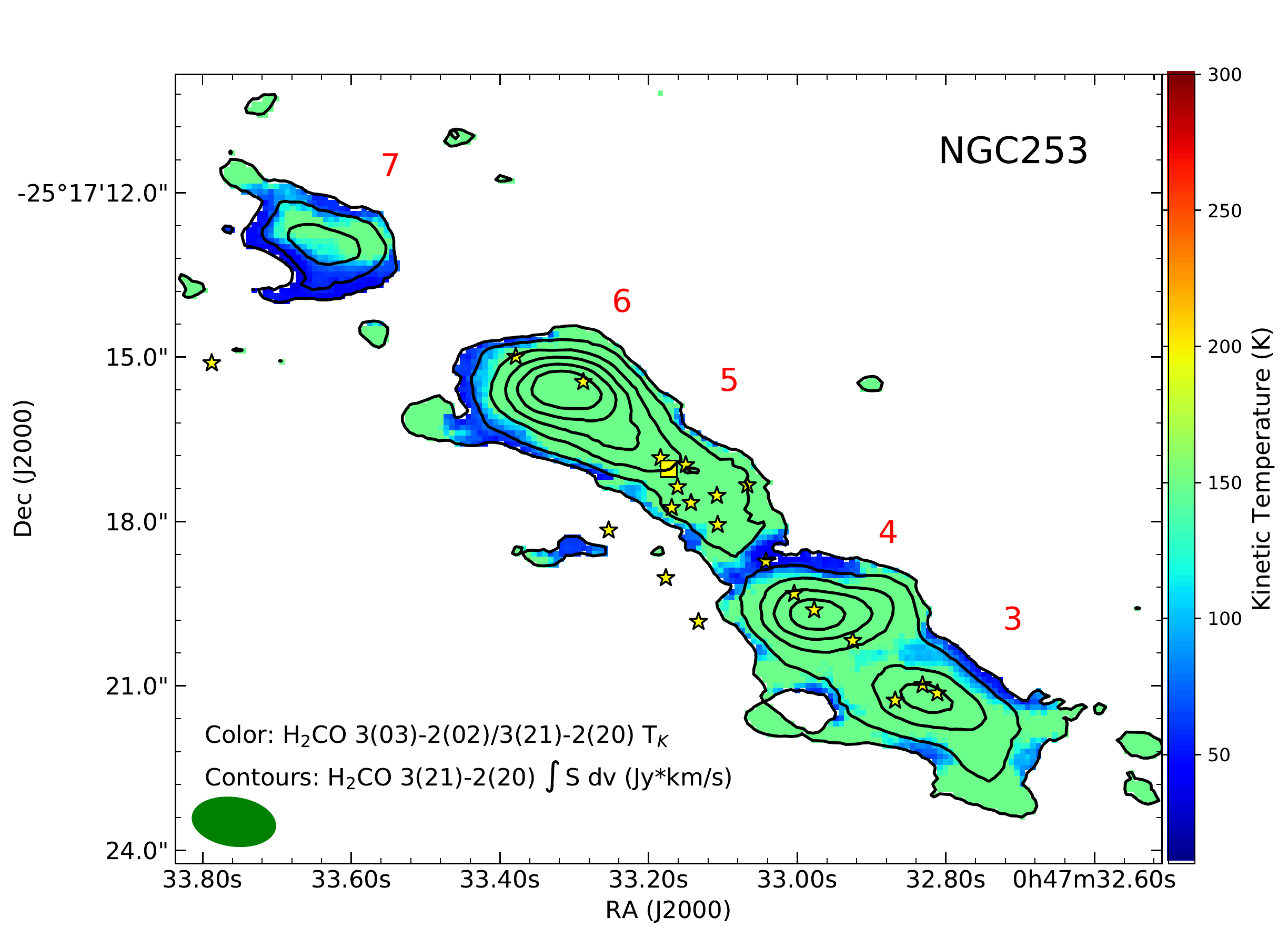}
\includegraphics[trim=0mm 12mm 0mm 10mm,clip,scale=0.40]{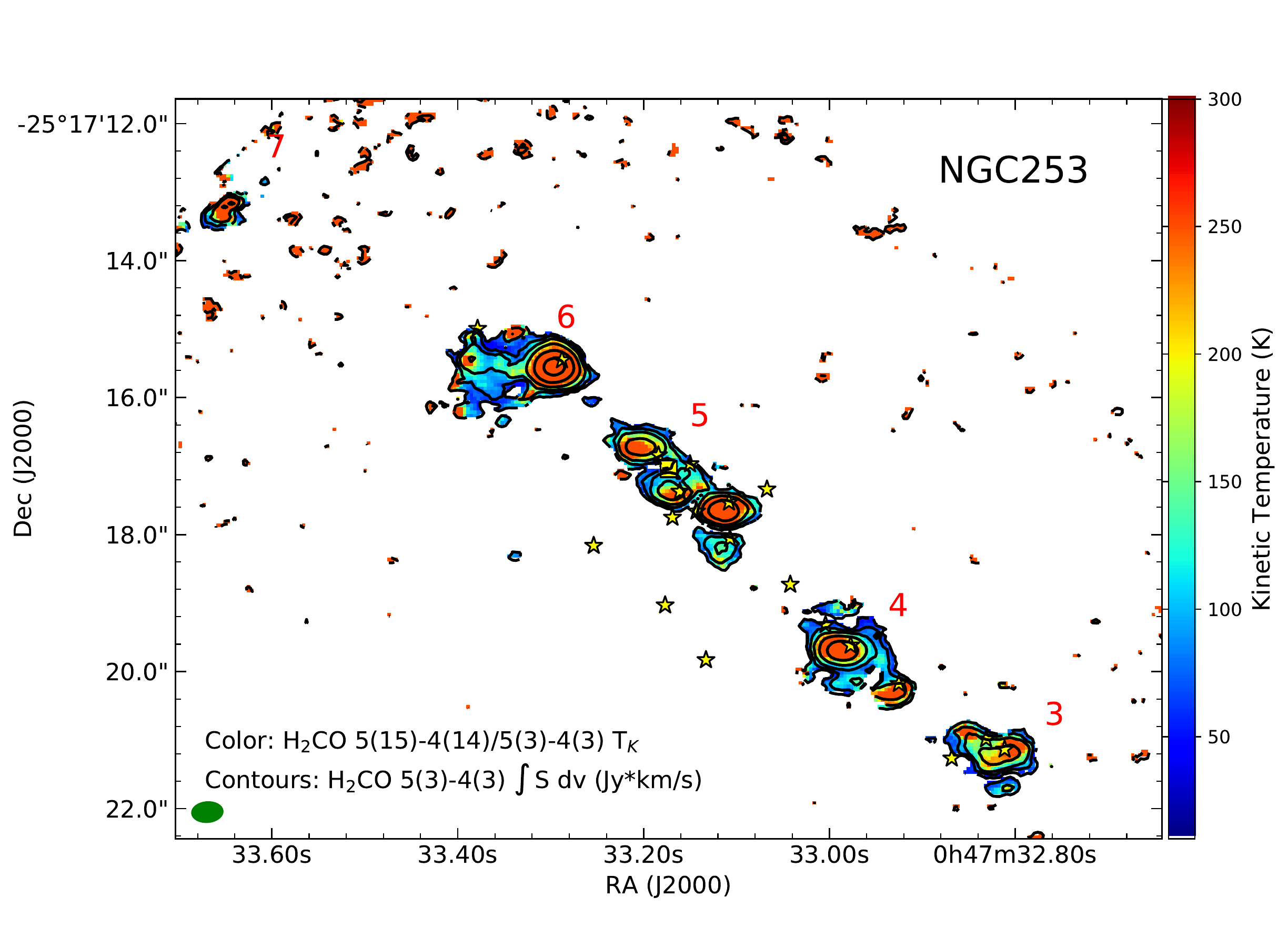} 
\includegraphics[trim=0mm 6mm 0mm 10mm,clip,scale=0.40]{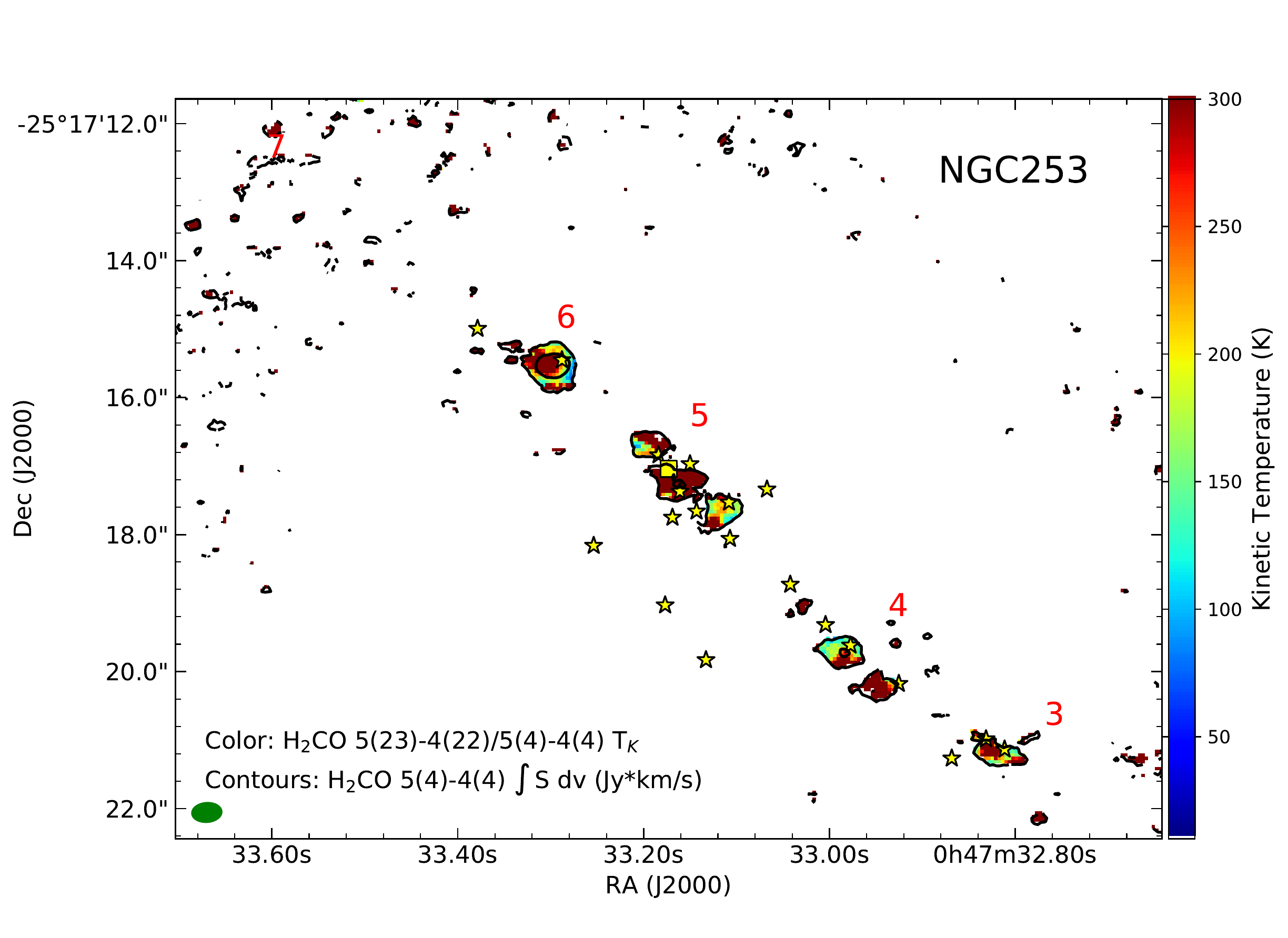}
\caption{Kinetic temperature images derived from H$_2$CO integrated
  intensity ratios: $\frac{Sdv(3_{03}-2_{02})}{Sdv(3_{21}-2_{20})}$
  (top), $\frac{Sdv(5_{15}-4_{14})}{Sdv(5_{3}-4_{3})}$ (middle),
  $\frac{Sdv(5_{23}-4_{22})}{Sdv(5_{4}-4_{4})}$ (equivalent to
  $\frac{Sdv(5_{24}-4_{23})}{Sdv(5_{4}-4_{4})}$; bottom).  Note that
  the appropriate upper sensitivity limit to the kinetic temperature
  scale has been applied based on the discussion in
  Section~\ref{H2COTk}: 150\,K (top), 250\,K (middle), and 300\,K
  (bottom).  The synthesized beam is displayed in the lower left of
  each panel.  Red numbers indicate the locations of 
  the dense molecular transition regions identified by
  \citet[][Table~\ref{tab:ContPositions}]{Leroy2015}.  Yellow stars
  locate the positions of the 2\,cm radio continuum emission peaks
  \citep{Ulvestad1997}, with a square indicating the position of the
  strongest radio continuum peak identified by \cite{Turner1985} (TH2:
  RA(J2000) = 00$^h$ 47$^m$ 33$^s$.18, Dec(J2000) = $-25^\circ$
  17$^\prime$ 16$^{\prime\prime}$.93).  Contour levels are 0.25, 0.50,
  1.0, 1.5, 2.0, and 2.5\,Jy*km/s (H$_2$CO $3_{21}-2_{20}$); 0.10, 0.25,
  0.50, 1.0, 2.0, and 4.0\,Jy*km/s (H$_2$CO $5_{3}-4_{3}$); and 0.10,
  0.50\,Jy*km/s (H$_2$CO $5_{4}-4_{4}$).}
\label{fig:TkRat}
\end{figure}

Over the range of H$_2$CO volume densities and kinetic temperatures
appropriate to our NGC\,253 measurements our measured
kinetic-temperature-sensitive integrated intensity ratios are
relatively 
insensitive to changes in H$_2$CO column density \citep{Mangum1993}.
Therefore, we have interpolated our measured integrated intensity
ratios onto our LVG model grid assuming $\log(N(species-H_2CO)/\Delta
v) = 12.5$\,cm$^{-2}$/(km s$^{-1}$) and $\log(n(H_2)) = 5.0$
\citep{Mangum2013}, where ``species'' is ortho or para.
We have 
also applied the sensitivity limits listed above to properly identify
the upper limit to the kinetic temperature sensitivity for each
H$_2$CO transition ratio.  This
then allows us to convert our measured H$_2$CO integrated intensity
ratios to kinetic temperatures.  Figure~\ref{fig:TkRat} shows the
results deduced from the above-mentioned ratios and the resulting LVG
model interpolation. 

From our NGC\,253 kinetic temperature images shown in
Figure~\ref{fig:TkRat} we can conclude the following:
\begin{enumerate}
  \item Based on the $\frac{Sdv(3_{03}-2_{02})}{Sdv(3_{21}-2_{20})}$ and
    $\frac{Sdv(5_{15}-4_{14})}{Sdv(5_{3}-4_{3})}$ ratios,
    the kinetic temperature ranges from 50 to $\gtrsim 150$\,K over
    dense gas regions as large as $\sim 5$\,arcsec ($\sim 80$\,pc).
  \item On smaller physical scales, $\lesssim
    1$\,arcsec ($\lesssim 16$\,pc), our
    $\frac{Sdv(5_{15}-4_{14})}{Sdv(5_{3}-4_{3})}$ and
    $\frac{Sdv(5_{24}-4_{23})}{Sdv(5_{4}-4_{4})}$ or
    $\frac{Sdv(5_{23}-4_{22})}{Sdv(5_{4}-4_{4})}$ ratios indicate that 
    the kinetic temperature is greater than 300\,K.
\end{enumerate}
An analysis of the H$_2$CO $1_{10}-1_{11}$ and $2_{11}-2_{12}$
emission toward NGC\,253 \citep{Mangum2013} measured volume
densities n(H$_2$) $\geq 10^{4-5}$\,cm$^{-3}$.  Furthermore, the
effective critical density \citep{Shirley2015} for the H$_2$CO
transitions considered in the current analysis is n(H$_2$) $\gtrsim
10^4$\,cm$^{-3}$.  Our dense gas imaging of NGC\,253 has therefore
revealed volume densities that are greater than
$10^{4-5}$\,cm$^{-3}$, as well as the existence of very high kinetic
temperatures within the dense star-forming gas that encompasses a large area
within the starburst nucleus of NGC\,253.  In Section~\ref{HighTK} we
will investigate the potential sources for these high dense gas
kinetic temperatures.

\subsection{Comparison to Previous Kinetic Temperature Measurements}
\label{Comparison}

As was summarized in \citet{Mangum2008}, \citet{Mangum2013}, and
\citet{Mangum2013b} numerous measurements of dense gas molecular
tracers have pointed to the existence of multiple temperature
components in NGC\,253.  Three of the more recent studies of the nuclear
kinetic temperature structure within NGC\,253
\citep{Mangum2013b,Gorski2017,PerezBeaupuits2018} suggest the
existence of at least two kinetic temperature components: 
\begin{enumerate}
\item A warm component with T$_K \simeq 75$\,K
\item A hot component with T$_K \gtrsim 150$\,K
\end{enumerate}
A third cooler component with T$_K \simeq 40-50$\,K is also required
by the dust and gas spectral energy distribution (SED) fits of
\cite{PerezBeaupuits2018}, though it may be difficult to distinguish
this component from the cooler wings of a 75\,K component in many of
the previous high-excitation molecular spectral line measurements.
These previous dense molecular gas 
kinetic temperature measurements are consistent with our
H$_2$CO-derived kinetic temperatures in NGC\,253.  The warm component
measured 
in previous low spatial resolution studies is associated with dense
gas on $\sim 80$\,pc scales, while the hot component appears to
originate in dense gas on $\lesssim 16$\,pc scales.

\section{Molecular Spectral Line Column Density}
\label{Colden}

In order to measure the relative abundances of the molecular species
sampled by our ALMA imaging of NGC\,253 we
need to calculate the molecular column density for each species.  In
all but a handful of species we have sampled only one transition,
which dictates that we perform a rather simplistic analysis of the
molecular column density.  With little information beyond the
intensity and spatial distribution of each measured transition, to
calculate molecular column densities we assume a kinetic temperature
derived from our H$_2$CO measurements (see Section~\ref{H2COTk}) in
the optically thin limit. 
With the additional assumptions that the excitation temperature of the
measured position is much
larger than the background temperature and equal to the kinetic
temperature and that the source filling
factor is unity, from \cite{Mangum2015} we adopt the
total molecular column density in the optically thin limit with
$T_{ex} \gg T_{bg}$ given by
\begin{equation}
N^{thin}_{tot} = \left(\frac{3k}{8 \pi^3 S \mu^2 \nu
  R_i}\right)\left(\frac{Q_{rot}}{g_J g_K g_I}\right)
\exp\left(\frac{E_u}{T_K}\right)\int T_B dv~cm^{-2},
\label{eq:ntotthin0}
\end{equation}
where $S$ is the transition line strength; $\mu$ is the molecular
dipole moment in Debye; $\nu$ is the transition frequency in GHz; $R_i$ is the
relative transition intensity (for hyperfine transitions); $g_J$,
$g_I$, and $g_K$ are the rotational, nuclear spin, and K degeneracies;
$E_u$ is the transition upper energy level in kelvin; $T_K$ is the kinetic
temperature in kelvin, T$_B$ is the measured transition brightness
temperature; and $Q_{rot}$ is the rotational partition function,
\begin{equation}
Q_{rot} = \sum^\infty_{J=0}\sum^J_{K=-J} g_K g_I (2J + 1)\exp\left(-\frac{E_{JK}}{T_{rot}}\right)
\label{eq:qrotsymmetric}
\end{equation}
for linear (where the summation over K is removed), symmetric, and
slightly asymmetric rotor molecules.  $E_{JK}$ represents the
energy above the ground state in kelvin for a transition with quantum numbers (J,K).
We assume that the rotational temperature, $T_{rot}$, is equal to the
kinetic temperature. 
Inserting Equation~\ref{eq:svstb_final} for $T_B$ into
Equation~\ref{eq:ntotthin0} results in the following:
\begin{equation}
N^{thin}_{tot} \simeq \frac{2.04\times10^{20}
  Q_{rot}\exp\left(\frac{E_u}{T_K}\right)\int S_\nu(Jy) dv(km/s)}{S
  \mu^2(Debye) \nu^3(GHz) R_i g_J g_K g_I \theta_{maj}(arcsec)\theta_{min}(arcsec)}~cm^{-2}
\label{eq:ntotthin1}
\end{equation}

To check whether our assumption of optically thin emission is reasonable,
we compared our measured spectral line peak intensities to those
derived from an LVG model prediction of those intensities.  For
example, our measured $^{13}$CO and C$^{18}$O $3-2$ integrated
intensities are $\sim 10-20$\,Jy km/s (Table~\ref{tab:IntegInt}),
with FWHM of $\sim 50$\,km/s.  For an LVG model that assumes T$_K =
150$\,K (Section~\ref{H2COTk}), n(H$_2$) = $10^4$\,$cm^{-3}$, and N($^{13}$CO or C$^{18}$O) =
$10^{17}$\,cm$^{-2}$ (Table~\ref{tab:ColumnDensity}), we find that
$\tau\simeq 0.1$ for both $^{13}$CO and C$^{18}$O $3-2$, with
predicted brightness temperatures similar to those that we measure.
Furthermore, if one assumes a lower kinetic temperature of 50\,K in
these LVG calculations, $\tau$ increases modestly to $\simeq 0.3$.
Similar estimates using LVG model calculations that use our measured
peak brightness temperatures and column densities to estimate
transition optical depths indicate that most of our measurements are
well within the optically thin regime, and only reach moderate optical
depths in a few cases.  Two such moderate optical depth cases are
the H$_2$CO $3_{03}-2_{02}$ and $3_{22}-2_{21}$ transitions, for which
$\tau \simeq 0.8$ and 0.4, respectively.  As the H$_2$CO column
density is based on a sample of seven transitions
(Table~\ref{tab:ColumnDensity}), the moderate optical depths within
these two transitions are unlikely to significantly affect the total
H$_2$CO column density derived.

Using $\theta_{maj}$, $\theta_{min}$, and $\int S_\nu(Jy)dv(km/s)$
from Table~\ref{tab:IntegInt} and the spectral line frequencies
($\nu$), upper-state energies above ground ($E_u$), dipole moments
($\mu$), line strengths ($S$), transition degeneracies ($g_J$, $g_K$,
$g_I$), and partition function values at representative kinetic
temperatures ($Q_{rot}(T)$; Table~\ref{tab:SpectralProperties}), we
calculate $N^{thin}_{tot}$ in 
Table~\ref{tab:ColumnDensity} using the indicated variable values and
assuming T$_K = 150$\,K.  We have chosen T$_K = 150$\,K as the
representative kinetic temperature for our molecular column density
calculations, as it accounts for both the warm gas measured on GMC
($\sim 80$\,pc) spatial scales and the hot gas measured on smaller
($\lesssim 16$\,pc) scales (see Section~\ref{H2COTk}). 

Note that to scale these total
optically thin molecular column densities to an assumed kinetic
temperature other than 150\,K, one simply needs to apply
Equation~\ref{eq:ntotthinscaled} to the total molecular column
densities listed in Table~\ref{tab:ColumnDensity}:
\begin{eqnarray}
\frac{N^{thin}_{tot}(T_{K1})}{N^{thin}_{tot}(T_{K2})} & = &
\frac{Q_{rot}(T_{K1})}{Q_{rot}(T_{K2})}\exp\left(\frac{E_u}{T_{K1}} -
\frac{E_u}{T_{K2}}\right) \nonumber \\
& \simeq & \left(\frac{T_{K1}}{T_{K2}}\right)^{\frac{n}{2}}\exp\left(\frac{E_u}{T_{K1}} -
\frac{E_u}{T_{K2}}\right)
\label{eq:ntotthinscaled}
\end{eqnarray}
where we have used the fact that, to a very good approximation,
$Q_{rot}(T_K) \propto T^{\frac{n}{2}}_K$, where $n = 2$ for linear
molecules and $n = 3$ for symmetric and slightly asymmetric rotor
molecules \citep[see][Section~7]{Mangum2015}.

When multiple transitions are available
(indicated by the N$_{trans}$ column in
Table~\ref{tab:ColumnDensity}), a spatially averaged molecular column
density is listed whose uncertainty is the larger of the statistical
uncertainty and the standard deviation of the individual transition
column densities derived.  Spatial averaging is done in order to
include measurements from both Bands 6 and 7, in that we have averaged
over the Band 7 subcomponents within Regions 4 and 5.  
In Section~\ref{Chemical} we will
compare the relative abundances of the molecular species identified in
our column density analysis with a goal of using these abundance
ratios as a diagnostic of the heating processes in the NGC\,253
starburst nuclear subcomponents.

\begin{longrotatetable}
\begin{deluxetable}{lcccccc}
\tablewidth{0pt}
\tablecolumns{7}
\tablecaption{NGC\,253 Component Total Molecular Column Densities\label{tab:ColumnDensity}\tablenotemark{a}}
\tablehead{
&& \multicolumn{5}{c}{$N^{thin}_{tot}\times 10^{14}$\,cm$^{-2}$ at
    T$_K$ = 150\,K} \\
\cline{3-7}
\colhead{Molecule} & \colhead{N$_{trans}$} & \colhead{Region 3} & \colhead{Region 4}
& \colhead{Region 5} & \colhead{Region 6} & \colhead{Region 7}}
\startdata
$^{13}$CO & 1 & $2.58\pm0.97(3)$ & $2.70\pm1.07(3)$ &
(840) & $4.50\pm2.33(3)$ & $2.21\pm0.80(3)$ \\
C$^{18}$O & 1 & $1.15\pm0.48(3)$ & $1.05\pm0.37(3)$ &
(288) & $2.37\pm0.68(3)$ & $0.90\pm0.31(3)$ \\
$^{13}$CN & 1 & (0.22) & $2.29\pm0.46$ & (0.28) &
$1.50\pm0.23$ & (0.41) \\
SiS & 1 & $2.61\pm0.66$ & $2.99\pm0.67$ & (0.53) &
$3.71\pm0.79$ & $2.43\pm0.31$ \\
SO & 1 & $2.60\pm0.64$ & $6.68\pm0.91$ & $2.98\pm0.70$ &
$12.71\pm1.51$ & $2.99\pm0.28$ \\
SO$_2$ & 3 & $39.84\pm6.12$ & $27.47\pm10.72$ & $26.72\pm9.68$ &
$112.23\pm22.20$ & (4.50) \\
HNC & 1 & $0.88\pm0.41$ & $5.84\pm0.96$ & (4.27) & $11.66\pm0.97$ &
$0.34\pm0.15$ \\
OCS & 1 & $24.82\pm8.57$ & (27.94) &
$104.62\pm28.08$ & $237.69\pm50.83$ & (969.61) \\
HNCO & 3 & $18.06\pm6.54$ & $7.04\pm1.75$ & (1.25) & $13.71\pm4.55$ &
$10.94\pm7.18$ \\
H$_2$CO & 7 & $11.12\pm5.11$ & $38.04\pm19.60$ & $41.34\pm10.28$ &
$47.11\pm29.56$ & $22.68\pm5.14$ \\
H$_3$O$^+$ & 1 & $2.73\pm1.02$ & (4.09) & $13.26\pm4.04$ &
$35.09\pm8.30$ & (3.47) \\
HC$_3$N & 2 & $1.26\pm0.31$ & $2.72\pm1.26$ &
$2.23\pm1.24$ & $11.22\pm7.38$ & $0.89\pm0.12$ \\
C$_4$H & 1 & $0.07\pm0.02$ & $0.10\pm0.02$ & $0.05\pm0.01$ &
$0.30\pm0.03$ & (0.02) \\
CH$_3$CN\tablenotemark{b} & 1 & (0.35) & (0.20) & 
\nodata & $0.92\pm0.36$ & $0.41\pm0.20$ \\
CH$_3$OH & 1 & (13.98) & (27.20) & (31.84) & 
$87.08\pm9.94$ & (73.57) \\
\enddata
\tablenotetext{a}{Column density calculation includes both integrated
  intensity and gaussian fit uncertainties.}
\tablenotetext{b}{Since the CH$_3$CN $12-11$ transition is a
  combination of the K=0 through 3 levels, the column density has been
  scaled by a factor of $\frac{72}{349}$ (using Sections 5 and 9 in
  \cite{Mangum2015}) to account for spin degeneracy and line strength
  differences.}
\end{deluxetable}
\end{longrotatetable}

\section{What Drives the High Kinetic Temperatures in NGC\,253?}
\label{HighTK}

As was shown in Section~\ref{H2COTk} the kinetic temperature within
the starburst nucleus of NGC\,253 is $\sim 50$ to $\gtrsim 150$\,K on
$\sim 80$\,pc scales and rises to kinetic temperatures greater than
300\,K on $\lesssim 16$\,pc scales.  NH$_3$ measurements yield similar
values \citep{Mangum2013b}.  Furthermore, note that the
Galactic CMZ possesses similarly high kinetic temperatures over GMC
($\sim 80$\,pc) size scales \citep{Ao2013,Ginsburg2016}.  \citet{Mills2013},
from observations of high energy level NH$_3$ absorption, find
evidence for an even hotter gas component ($T_K > 350$\,K) that is
widespread in the Galactic CMZ. This component most likely originates
in a lower-density gas component that is not sampled by our H$_2$CO
data, which exclusively trace dense gas. Our $T_K > 300$\,K gas thus
\textit{is not} a counterpart of the Galactic CMZ dilute gas 
component.  What physical processes can maintain such high kinetic
temperatures?

As we discussed previously in the context of the high dense gas
kinetic temperatures measured using NH$_3$ emission within a sample of
starburst galaxies \citep{Mangum2013b}, high kinetic temperatures can
be generated by cosmic-ray (CR) and/or mechanical heating.  CR heating
can be effective at high column densities owing to the small ($\sim
3\times10^{-26}$\,cm$^{-2}$) H$_2$ CR dissociation cross section
\citep{PerezBeaupuits2018}.  As noted
in \citet{Mangum2013b}, adapted chemical PDR
models have been used by a number of groups
\citep[\eg][]{Bayet2011,Meijerink2011} to study the effects of CR and
mechanical heating on the chemical abundances within starburst
galaxies.  In these models, kinetic temperatures ranging up to 150\,K
can be generated by injecting varying amounts of CR and/or mechanical
energy.  For example, in the CR plus mechanical heating models of
\cite{Meijerink2011}, T$_K \simeq 150$\,K is attained for a mechanical heating
rate of $3\times 10^{-18}$\,erg\,cm$^{-3}$\,s$^{-1}$ in a high-density
(n(H$_2$) = $10^{5.5}$\,cm$^{-3}$) high column density (N(H$_2$) 
$>10^{22}$\,cm$^{-2}$) environment with CR rates ranging from $5\times 10^{-17} -
5\times 10^{-14}$\,s$^{-1}$.  In order to distinguish between
different physical processes that can produce the signatures of CR
and/or mechanical energy, in the following we evaluate the radiative
and chemical diagnostics of energy input within starburst galaxies.

\subsection{Radio, Near-Infrared, and X-Ray Diagnostics of Dense Gas Heating}
\label{RadioNearIRXray}

\begin{figure}
\centering
\includegraphics[scale=0.60]{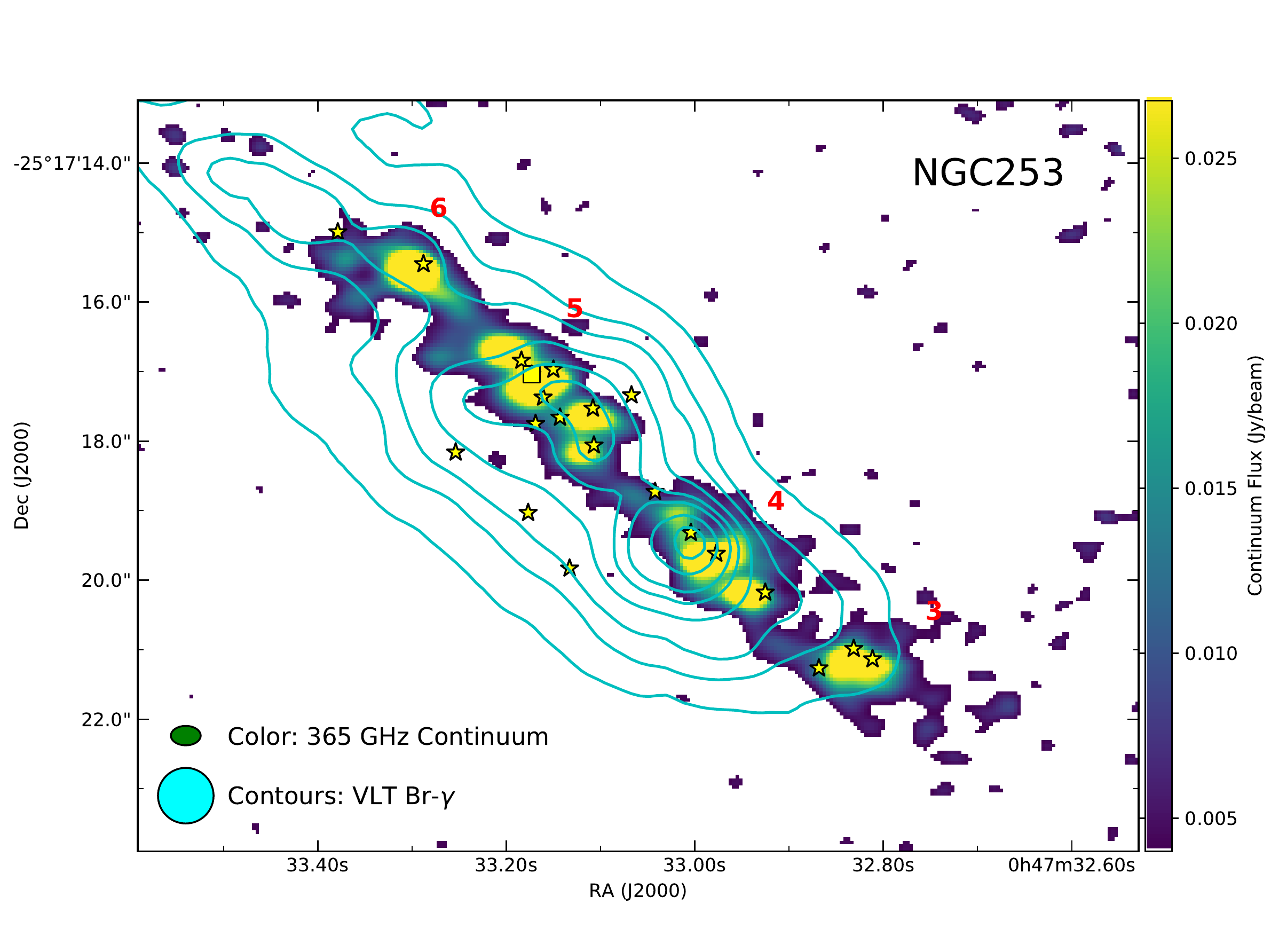}
\caption{Comparison of the 365\,GHz continuum (color) and VLT
  Br$\gamma$ \citep[contours;][]{Rosenberg2013} emission distributions
  toward the CMZ of NGC\,253.  Contour levels are 0.2, 0.4, 0.6, 1.0,
  1.5, 2.0, 3.0, and 4.0 $\times
  10^{-13}$\,erg\,cm$^{-2}$\,s$^{-1}$\,arcsec$^{-2}$.  The spatial
  resolution for both images is shown in the lower left corner.  Red
  numbers indicate the locations of 
  the dense molecular transition regions identified by
  \citet[][Table~\ref{tab:ContPositions}]{Leroy2015}.  Note that we
  have adopted the astronometry from \cite{FernandezOntiveros2009} to
  define the absolute positions in the Br$\gamma$ image.}
\label{fig:Band7ContBrgamma}
\end{figure}

Radio wavelength measurements
\citep{Turner1985,Ulvestad1997,Brunthaler2009} ranging from $\lambda
=$ 1.3 to 20\,cm have identified over 60 individual compact continuum
sources within the NGC\,253 CMZ.
Figures~\ref{fig:NGC253Continuum} through \ref{fig:IntegIntUnid} show
the locations of the compact 2\,cm continuum sources identified by
\cite{Ulvestad1997}.  The brightest of these radio continuum sources
(S(2\,cm) $\simeq 30$\,mJy/beam within $\theta \simeq 0.05$\,arcsec)
is located at the dynamical center of NGC\,253 and has
been suspected to be either a low-luminosity active galactic nucleus
(LLAGN) or a compact supernova remnant
\citep{Turner1985,Ulvestad1997,Brunthaler2009}.  This nuclear
continuum source (designated ``TH2'' in this article and indicated by
a filled square in the figures presented) has a $\lambda = 2$\,cm
brightness temperature of $\gtrsim 4\times10^4$\,K and a size
$\lesssim 1$\,pc 
\citep{Turner1985,Ulvestad1997}.  This brightness temperature over
such a compact region would suggest that the emission is due to
synchrotron radiation from a LLAGN \citep{Condon1992}.  There are
several additional observations, though, which argue against the LLAGN
explanation for TH2:
\begin{enumerate}
  \item Over the 2 to 6\,cm wavelength range the spectral index is
    $\alpha \simeq -0.2$ to $-0.3$ \citep[S$_\nu \propto
    \nu^\alpha$;][]{Turner1985,Ulvestad1997}, which is more consistent
    with bremsstrahlung emission from HII regions than that from
    optically thin synchrotron emission ($\alpha \simeq -0.75$).
  \item \cite{Brunthaler2009} failed to detect subparsec-scale
    structure at 22\,GHz toward TH2, suggesting that TH2 is a radio
    supernova or young supernova remnant.
  \item Even though \textit{Chandra} X-ray observations of the nuclear region within
    NGC\,253 \citep{Weaver2002} suggest the presence of a LLAGN, the
    spatial resolution is not sufficient to distinguish
    between TH2 and TH4.  The X-ray source detected by Chandra is in
    fact consistent with emission from ultraluminous X-ray sources
    (\ie, X-ray binaries) in
    other galaxies \citep{Brunthaler2009}.
  \item \cite{FernandezOntiveros2009} found no IR or optical
    counterpart to TH2, suggesting that there is no AGN associated
    with TH2.
  \item Over an 8 year period \cite{Ulvestad1997} monitored the
    stability of the fluxes of the compact continuum sources within
    NGC\,253 at wavelengths from 20 to 1.3\,cm.  They found no
    compelling evidence for variability in any of the compact source
    fluxes, variability that one might expect to see from an LLAGN.
    This flux constancy timescale of 8 years also helps to constrain
    the radio supernova rate to $\lesssim 0.3$\,yr$^{-1}$,
    consistent with other estimates \citep[\eg][]{Rieke1980,Rieke1988}.
\end{enumerate}
Existing evidence suggests, then, that the radio wavelength emission
from NGC\,253 is dominated by that from HII regions, radio supernovae,
and supernova remnants, and that the central source TH2 does not
appear to contain an LLAGN.

\cite{Ulvestad1997} classify the radio structure of the 2\,cm
continuum sources TH1, TH3, TH4, and TH6 as structurally resolved
with spectral indices $\alpha$ in the 1.3 to 6\,cm wavelength range of
$\sim -0.5$ to $+0.35$.  Each of these radio continuum sources is
associated with the dense gas GMC Regions 6, 5c, 5c, and 5a/5b,
respectively.  These sources have 2\,cm fluxes of 4 to 13\,mJy
\citep{Ulvestad1997}, and are prototypes of what appear to be HII
regions ionized by hot young stars.  Analyzing just the radio emission
properties of TH6, \cite{Ulvestad1997} calculate that there are at
least $5.2\times10^{51}$ ionizing photons per second produced from a
region $\sim 0.2\times0.1$\,arsec in size.  This ionizing photon flux
is equivalent to the emission produced by $\sim 100$ O5 stars in this HII
region, and corresponds to the upper end of the incident UV fields
modeled by \citet{Meijerink2011}; $10^5 G_0$, where $G_0 = 1.6\times10^{-3}$\,erg
  cm$^{-2}$ s$^{-1}$ (= one Habing).  This would make TH6 a slightly more
powerful version of the R136 cluster in the 30 Doradus region in the
Large Magellanic Cloud \citep{Kalari2018}.

Further evidence for a variety of heating processes is provided
  by near-infrared diagnostic probes such as Br$\gamma$, H$_2$, and
  [FeII].  \cite{Rosenberg2013} imaged the Br$\gamma$, H$_2$, and
  [FeII] emission toward NGC\,253, finding that all three trace the
  CMZ of NGC\,253, but with variations in intensity that suggest
  dominance of specific heating processes within specific regions of
  the NGC\,253 CMZ.  An example of the correlation between the
  millimeter continuum (dust) emission distribution from our Band 7
  measurements and the Br$\gamma$ emission imaged by
  \cite{Rosenberg2013} is shown in Figure~\ref{fig:Band7ContBrgamma}.
  Br$\gamma$ traces the emission from young massive stars and
  peaks near Region 4 in the dust and molecular emission.  This region
  also corresponds to the ``Infrared Core'', a region that dominates
  the emission at infrared wavelengths (see
  Section~\ref{vibHNCandCH3OH}).  The [FeII] emission, on the other
  hand, is stronger toward Regions 5, 6, and 7.  As [FeII] is a tracer
  of strong, grain-destroying, shocks (v$_{shock} \gtrsim 25$\,km/s,
  the [FeII] distribution suggests that shock heating is more
  prevalent toward Regions 5, 6, and 7.
It seems clear from these examples that
active massive star formation can provide the energy necessary to heat
the dense gas in the nucleus of NGC\,253 through a variety of physical
processes.

\subsection{Chemical Diagnostics of Dense Gas Heating}
\label{Chemical}

\begin{figure}
\centering
\includegraphics[scale=0.60]{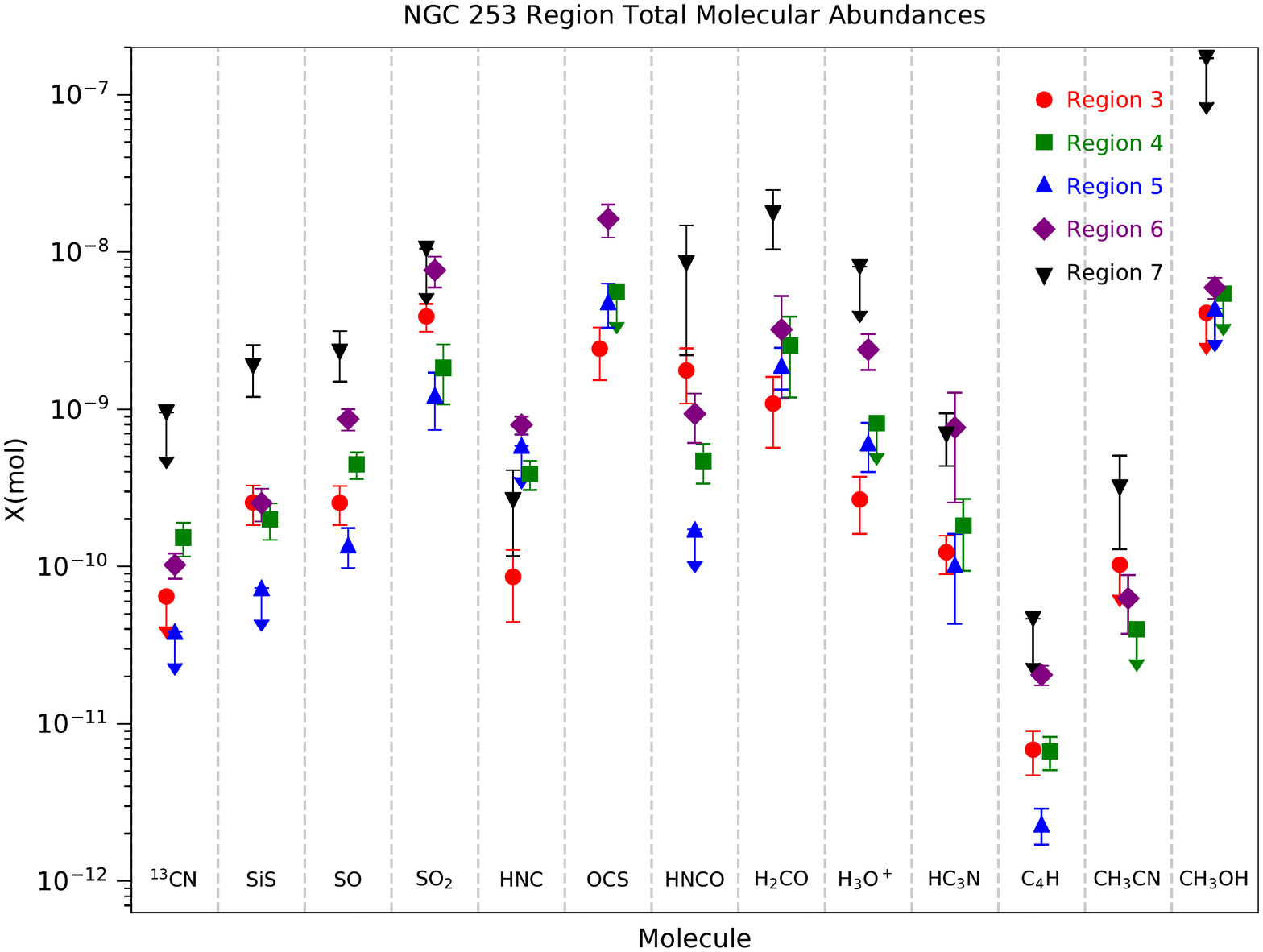}
\caption{Total molecular abundances, excluding $^{13}$CO and
  C$^{18}$O, derived from the total molecular column densities listed
  in Table~\ref{tab:ColumnDensity}.  Limits are shown with downward
  arrows.}
\label{fig:MolAbundances}
\end{figure}

In order to compare our molecular spectral line measurements of
the GMCs of NGC\,253 with molecular abundance predictions from
galactic starburst chemical models, we need to calculate the measured
molecular abundances ($X(mol) \equiv \frac{N^{thin}_{tot}(mol)}{N(H_2)}$)
using the total hydrogen and molecular column
densities listed in Tables~\ref{tab:DustColumnMassAverage} and
\ref{tab:ColumnDensity}.  Table~\ref{tab:MolAbundances} lists 
these calculated abundances, which are displayed in
Figure~\ref{fig:MolAbundances}.
Recall that in the calculations of the hydrogen column density and
mass presented in this section we have adopted the dust temperature
assumed by \cite{Leroy2015}; T$_d = 35$\,K.

\begin{deluxetable}{lccccc}
\tablewidth{0pt}
\tablecolumns{6}
\tablecaption{NGC\,253 Component Total Molecular Abundances\label{tab:MolAbundances}\tablenotemark{a}}
\tablehead{
& \multicolumn{5}{c}{$X_{tot}(\mathrm{Molecule})\times 10^{-10}$} \\
\cline{2-6}
\colhead{Molecule} & \colhead{Region 3} & \colhead{Region 4}
& \colhead{Region 5} & \colhead{Region 6} & \colhead{Region 7}}
\startdata
$^{13}$CN & (0.65) & $1.53\pm0.37$ & (0.38) & $1.02\pm0.19$ & (9.53) \\
SiS &  $2.55\pm0.72$ & $1.99\pm0.52$ & (0.73) & $2.53\pm0.60$ &
$18.84\pm6.86$ \\
SO & $2.54\pm0.70$ & $4.45\pm0.85$ & $1.37\pm0.39$ & $8.67\pm1.36$ &
$23.18\pm8.20$ \\
SO$_2$ & $38.94\pm7.76$ & $18.31\pm7.56$ & $12.26\pm4.86$ &
$76.55\pm17.05$ & (104.65) \\
HNC & $0.86\pm0.42$ & $3.89\pm0.82$ & (5.88) & $7.95\pm1.05$ &
$2.64\pm1.47$ \\
OCS & $24.26\pm8.93$ & (55.88) & $47.99\pm15.04$ & $162.14\pm38.44$ &
(22549.07) \\
HNCO & $17.65\pm6.78$ & $4.69\pm1.32$ & $1.72\pm0.64$ & $9.35\pm3.25$
& $84.81\pm62.73$ \\
H$_2$CO & $10.87\pm5.18$ & $25.36\pm13.50$ & $18.96\pm5.63$ &
$32.14\pm20.43$ & $175.81\pm72.00$ \\
H$_3$O$^+$ & $2.67\pm1.05$ & $8.18\pm2.94$ & $6.08\pm2.10$ &
$23.94\pm6.17$ & (80.70) \\
HC$_3$N & $1.23\pm0.34$ & $1.81\pm0.87$ & $1.02\pm0.59$ &
$7.65\pm5.09$ & $6.90\pm2.53$ \\
C$_4$H & $0.07\pm0.02$ & $0.07\pm0.02$ & $0.02\pm0.006$ &
$0.20\pm0.03$ & (0.47) \\
CH$_3$CN & (1.03) & (0.4) & \ldots & $0.63\pm0.25$ & $3.18\pm1.89$ \\
CH$_3$OH & (41.00) & (54.40) & (43.82) & $59.40\pm9.11$ & (1710.93) \\
\enddata
\tablenotetext{a}{Derived from measured total optically thin molecular
  abundances (Table~\ref{tab:ColumnDensity}) and dust-continuum derived
  H$_2$ column densities (Table~\ref{tab:DustColumnMassAverage}).
  Upper limits, listed in parentheses, are quoted as $3\sigma$.}
\end{deluxetable}

Some trends in our measured molecular abundances are apparent in
Figure~\ref{fig:MolAbundances}:
\begin{enumerate}
  \item For $^{13}$CN, SiS, SO, SO$_2$, HNCO, HC$_3$N, and C$_4$H,
    Region 5, the region associated with the central source, TH2,
    possesses the lowest abundance of the five regions studied.
  \item For SiS, SO, HNCO, and HC$_3$N, in addition to Region 5
    possessing the lowest abundance, Region 7, the region studied in
    this work that is the farthest from the NGC\,253 nucleus,
    possesses the highest abundance. 
  \item For those molecular species common to both studies, our
    derived molecular abundances are within the same range of
    $5\times10^{-9}$ to $5\times10^{-10}$ as those derived by
    \cite{Martin2006}.
\end{enumerate}
These molecular abundance patterns could be related to the spatial
distribution of energy sources, such as sources of CRs and
mechanical energy, within the NGC\,253 CMZ.  Past surveys designed to
sample the dominant chemical processes in NGC\,253
\citep[\eg][]{Martin2006,Aladro2015} have noted the influence of the
burst of star formation on the energetics within NGC\,253.  To
tease out the relative dominance of different dense gas heating
processes, models that predict the variation in molecular abundances
within starburst environments have been developed.  In general, these
models characterize the changes in molecular abundances as a function
of metallicity (z), volume density (n(H$_2$)), radiation field
intensity (G$_0$), CR rate ($\zeta$), and mechanical heating
($\Gamma_{mech}$).  These physical processes are then coupled to
chemical model networks to allow for the prediction of molecular
abundances as a function of varying physical conditions.  These
molecular abundance predictions can then be used to guide our
understanding of the abundance distributions measured in Galactic and
extragalactic star formation regions.  The adapted chemical PDR models
of \cite{Meijerink2011} and \cite{Bayet2011} are of this type and
have been specifically designed to predict molecular abundances in
extreme star formation environments such as starburst galaxies and
very high star formation rate ultraluminous infrared galaxies
(ULIRGs).  Even though these two models have been used to predict 
molecular abundances within somewhat different sets of physical
conditions, in general these models predict that when the CR
ionization rate $\zeta$ is increased, the abundances of molecular 
species other than simple ions such as OH$^+$, CO$^+$, CH$^+$, and
H$_2$O$^+$ (none of which are part of this study) decrease.  The range
of $\zeta$ modeled in these studies 
runs from the canonical Milky Way value of $2.6\pm0.8\times10^{-17}$
s$^{-1}$ \citep{vanderTak2000} to $5\times10^{-13}$\,s$^{-1}$.  This
upper limit to $\zeta$ is roughly 10 times the
CR ionization rate determined for NGC\,253 from ultra-high-energy CR
observations, which is 
$\zeta \simeq 3.6\times10^{-14}$\,s$^{-1}$ \citep{Acero2009}, and is
believed to be consistent with cosmic ray densities in ULIRGs
\citep{Papadopoulos2010}.  The supernova rate that corresponds to the
measured gamma-ray flux from NGC\,253 is $\sim 0.1$\,yr$^{-1}$
\citep{Acero2009}, which is most pronounced toward its
nucleus\footnote{Note, though, that the H.~E.~S.~S.~measurements from
  which the gamma-ray flux is derived have a spatial resolution of
  $\sim4.2$\,arcmin.}.

Given the existence of strong IR radiation fields in the NGC\,253 CMZ,
a note of caution is in order regarding the interpretation of
rotational transition intensities for molecules that might be
affected by radiative pumping of vibrational states that subsequently
experience radiative decay.  HCN, HNC, and HC$_3$N
(Section~\ref{vibHNCandCH3OH}) show vibrationally excited emission in
NGC\,253.  Rotational energy levels in other molecules, including
H$_2$CO \citep{Mangum1993}, can also be excited by far-infrared
emission, representing a potential unaccounted-for source of excitation of
these molecules.  Physical scenarios that attempt to describe
infrared excitation of molecular rotational energy levels in molecules
\citep{Carroll1981,Mangum1993} tend to require that the sources of
infrared emission be cospatial with the molecular distribution.  Our
measurements of vibrationally excited HNC and HC$_3$N emission
(Section~\ref{vibHNCandCH3OH}) suggest that this emission could
possess similar spatial distributions to their rotationally excited
counterparts.  We would expect that if infrared excitation through
vibrational transitions was significantly contributing to the
purely rotationally excited transitions that we measure, we would
expect to see local maxima in their spatial distributions that
correspond to peaks in the vibrationally excited emission from a given
molecule.  We have found that Region 6 represents a local maximum in
vibrationally excited HNC and HC$_3$N emission
(Section~\ref{vibHNCandCH3OH}), suggesting caution when interpreting
the rotationally excited emission from these molecules toward Region
6.

\begin{figure}
\centering
\includegraphics[scale=0.55]{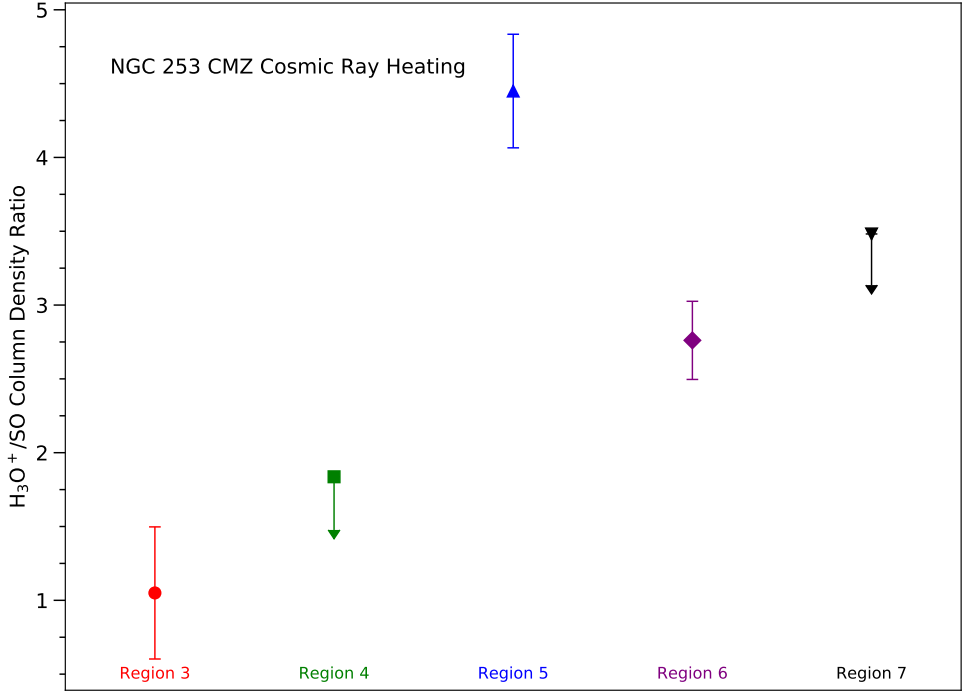}
\caption{Total H$_3$O$^+$/SO abundance ratio for each region in the
  NGC\,253 CMZ.  Derived from the total molecular column densities listed
  in Table~\ref{tab:ColumnDensity}.  Limits, shown with down arrows,
  have been calculated assuming three times the RMS of the undetected
  H$_3$O$^+$ column density.}
\label{fig:H3OpSO}
\end{figure}

Comparing our molecular abundance measurements to the model predictions of
\cite{Meijerink2011} and \cite{Bayet2011}:
\begin{enumerate}
  \item \cite{Bayet2011} and \cite{Meijerink2011} modeled H$_3$O$^+$
    and found that its abundance maintains a high level (of about
    $10^{-9}$ to $10^{-8}$, respectively) under a wide range of
    conditions, peaking at CR rates 
    of $\sim 10^{-15}$ to $10^{-13}$\,s$^{-1}$ at solar metallicity when the
    H$_2$ column density becomes high.  Our measured abundance of
    H$_3$O$^+$ ($\sim10^{-9}$) seems to be consistent with both the
    Bayet and Meijerink predictions.
  \item \cite{Bayet2011} modeled SO and found that it is destroyed by
    CRs starting at $10^{-16}$\,s$^{-1}$.  The regional
    SO abundance pattern we measure, with Region 5 showing the lowest
    abundance and Region 7 showing the highest, seems to be consistent
    with a higher concentration of CRs at the center of the
    NGC\,253 CMZ than in its outskirts.
  \item As H$_3$O$^+$ abundances are enhanced by CRs, while SO
    is destroyed by CRs, we have made a direct comparison of the
    abundances of these two direct CR tracers for the different
    regions (Figure~\ref{fig:H3OpSO}).  Region 5 has an H$_3$O$^+$/SO
    abundance ratio more than 30\% larger than that in Regions 3, 4,
    6, and 7, suggestive of enhanced CR heating near the center of
    NGC\,253. Note, though, that this abundance ratio assumes
    optically thin emission from both molecules.  Variations in the
    relative optical depth within the transitions measured to
    calculate this abundance ratio could at least partially explain
    this difference.  Furthermore, our H$_3$O$^+$ and SO
    measurements were made with different tunings of the ALMA
    receiver system, making their abundance ratio susceptible to our
    estimated absolute amplitude calibration uncertainties of 10\%
    and 15\% at Bands 6 and 7, respectively
    (Appendix~\ref{CalibrationDetails}).  Factor of two differences
    could be explained by these two effects.
  \item \cite{Meijerink2011} noted no obvious trends in HNC abundance
    as a function of CR rates.  We see no significant
    variation in X(HNC) amongst the various regions.  
  \item The abundance of H$_2$CO has been modeled by \cite{Bayet2011}
    and \cite{Meijerink2011}.  The changes in H$_2$CO abundance
    predicted by these models are largely consistent with those of
    other complex molecules and with our measurements.  H$_2$CO is
    destroyed by CRs at low densities.
  \item \cite{Meier2012} investigated shock chemistry influence on
    CH$_3$OH, HNCO, and SiO abundances.  They noted that CH$_3$OH and
    HNCO are possibly formed on grain mantles and hence simply require
    enough energy to liberate them into the gas phase.  This can be
    done with a shock that has 
    $10 \lesssim v_{shock} \lesssim 15$\,km/s (grain destruction
    happens for $v_{shock} \gtrsim 25$\,km/s).  \cite{Meier2012} also
    noted that the photodissociation rate for HNCO is twice that for
    CH$_3$OH and $\sim30$ times that of SiO. There are also
    large numbers of HII regions in Region 5 \citep{Ulvestad1997}.
    \cite{Leroy2018}, referring to the \cite{Gorski2017,Gorski2019}
    imaging of the relatively unattenuated $\sim$36\,GHz free-free
    emission from the GMCs in NGC\,253, have noted the utility of
    these continuum measurements as a sensitive measure of the
    free-free emission from young
    heavily embedded massive stars in NGC\,253.  Region 5 is the most
    intense source of $\sim$36\,GHz continuum emission in the NGC\,253
    CMZ.  These embedded HII regions could explain our nondetection
    of HNCO in Region 5.
  \item Note also that CRs are toxic to many molecules
    \citep{Bayet2011,Meijerink2011}, where abundances of molecules
    such as \eg CN, SO, HNC, HCN, OCS, and H$_2$CO are shown to
    decrease by upward of $10^3$ when the CR rate increases from
    $10^{-16}$ to $10^{-14}$\,s$^{-1}$.  Low abundances of many
    molecules in Region 5 might be due to a higher CR rate in
    this  Region.
  \item Region 7 is the farthest away from the main source of
    CRs.  Its higher level of complex molecular abundances
    relative to regions closer to the center of the galaxy is
    consistent with a lower level of CR heating.
\end{enumerate}

To identify possible
sources of CRs, we note that \cite{Ulvestad1997} calculated
spectral indices for all of the radio sources they detected.  In those
measurements spectral indices in the range $\alpha < -0.4$ are
more common in the Region 5 area (where TH2 through TH6 are located)
than elsewhere in the NGC\,253 CMZ.  Since spectral indices in this
range would imply synchrotron emission, which can be generated by
supernova remnants, and which produce CRs, one might expect a
larger flux of CRs in Region 5 than within the other regions.
CR heating, then, appears to be a plausible mechanism by which
the GMCs in the CMZ of NGC\,253 are heated to the high kinetic
temperatures that we measure.  Variations in the molecular abundances
in these GMCs support this CR-heating-dominated scenario, but
they also do not rule out a significant influence due to mechanical
heating. 

The recent dust and molecular spectral line study of the $\sim
40$\,arcsec-scale structures within the NGC\,253 CMZ by
\cite{PerezBeaupuits2018} concludes that mechanical heating is
responsible for the highest kinetic temperatures measured within
NGC\,253.  Analysis of the submillimeter dust and CO SEDs indicates
that mechanical heating drives the high 
kinetic temperatures for the higher-excitation (J$_u > 13$) CO
transitions, but not the lower-excitation transitions.
\cite{PerezBeaupuits2018} note also that the effects and role of 
CRs in the dense gas heating process within the NGC\,253 CMZ cannot be
assessed through their measurements.

While we conclude that CRs may be responsible for the high observed
temperatures in NGC\,253's CMZ, \cite{Ginsburg2016} reached a somewhat
different conclusion for the Milky Way's CMZ.  Their inference was
based on the mismatch between dust and gas temperature at moderately
high density ($\sim10^{4-5}$), which is difficult to explain by CR
heating and is better explained by mechanical (turbulent) heating.
\cite{Ginsburg2016} notably found no regional variance, instead
finding that the elevated T$_{gas}$/T$_{dust}$ was relatively uniform
across the CMZ.  By contrast, we have found that there are significant
regional variations in molecular abundances and that these variations
are correlated with the locations of likely supernova remnants.  It
therefore seems that the temperature structure in NGC\,253 is more
heavily influenced by CRs than the Milky Way CMZ because of its higher
star formation (and therefore supernova) rate.

\section{Vibrationally Excited Molecules and Possible Non-LTE Methanol Emission}
\label{vibHNCandCH3OH}

\begin{figure}
\centering
\includegraphics[trim=0mm 0mm 0mm 0mm,clip,scale=0.40]{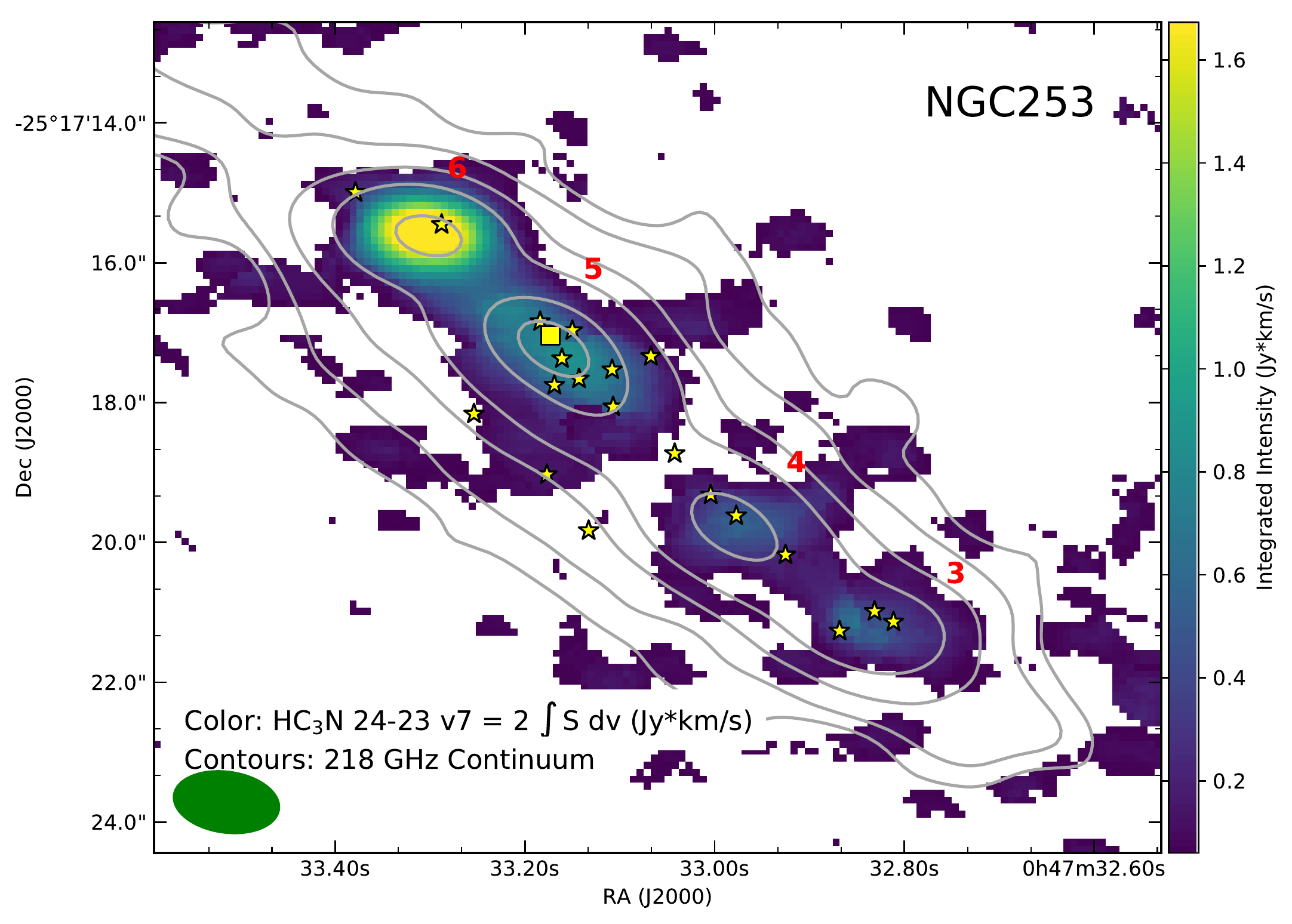}
\includegraphics[trim=0mm 0mm 0mm 0mm,clip,scale=0.38]{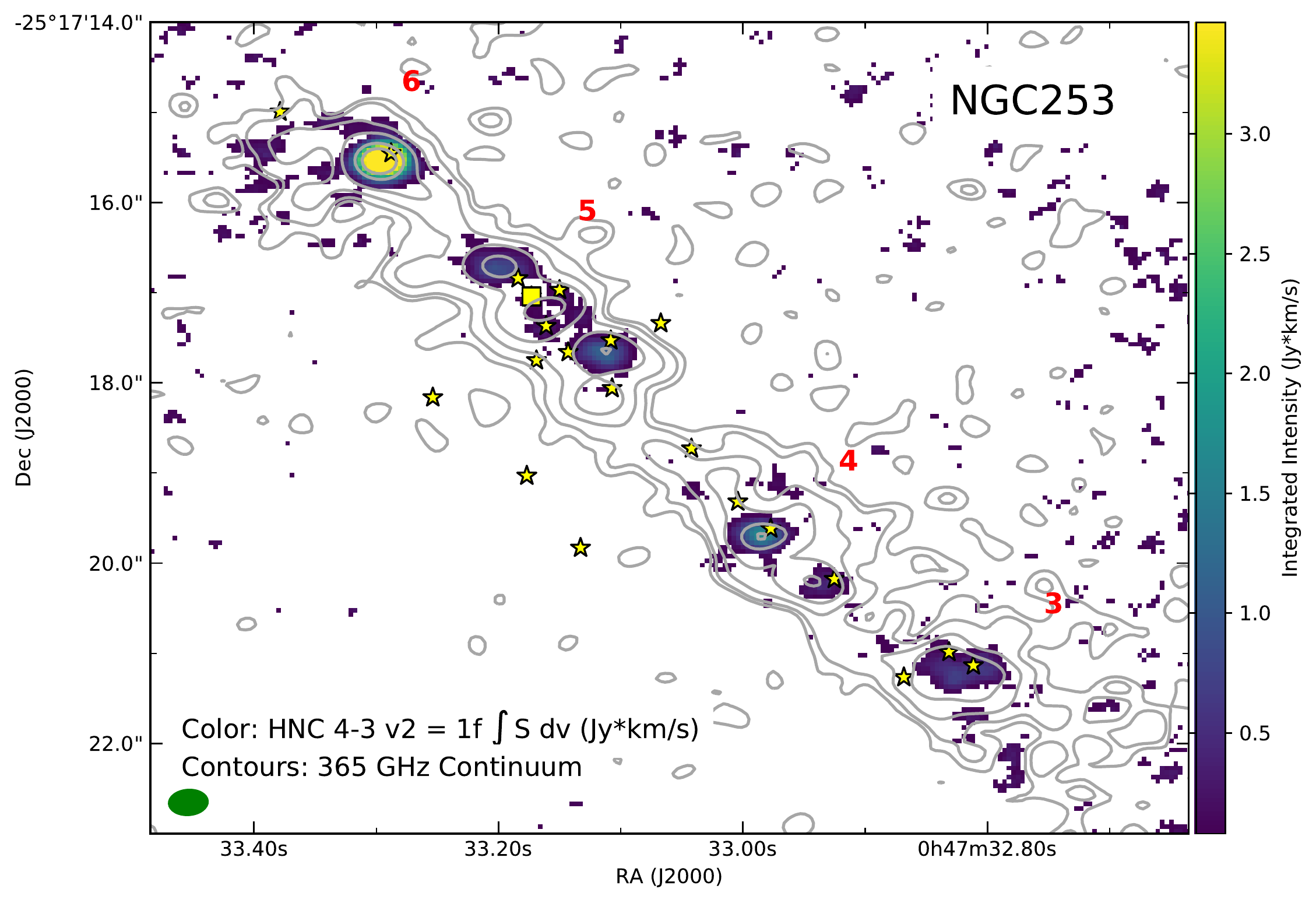} 
\includegraphics[trim=0mm 0mm 0mm 0mm,clip,scale=0.38]{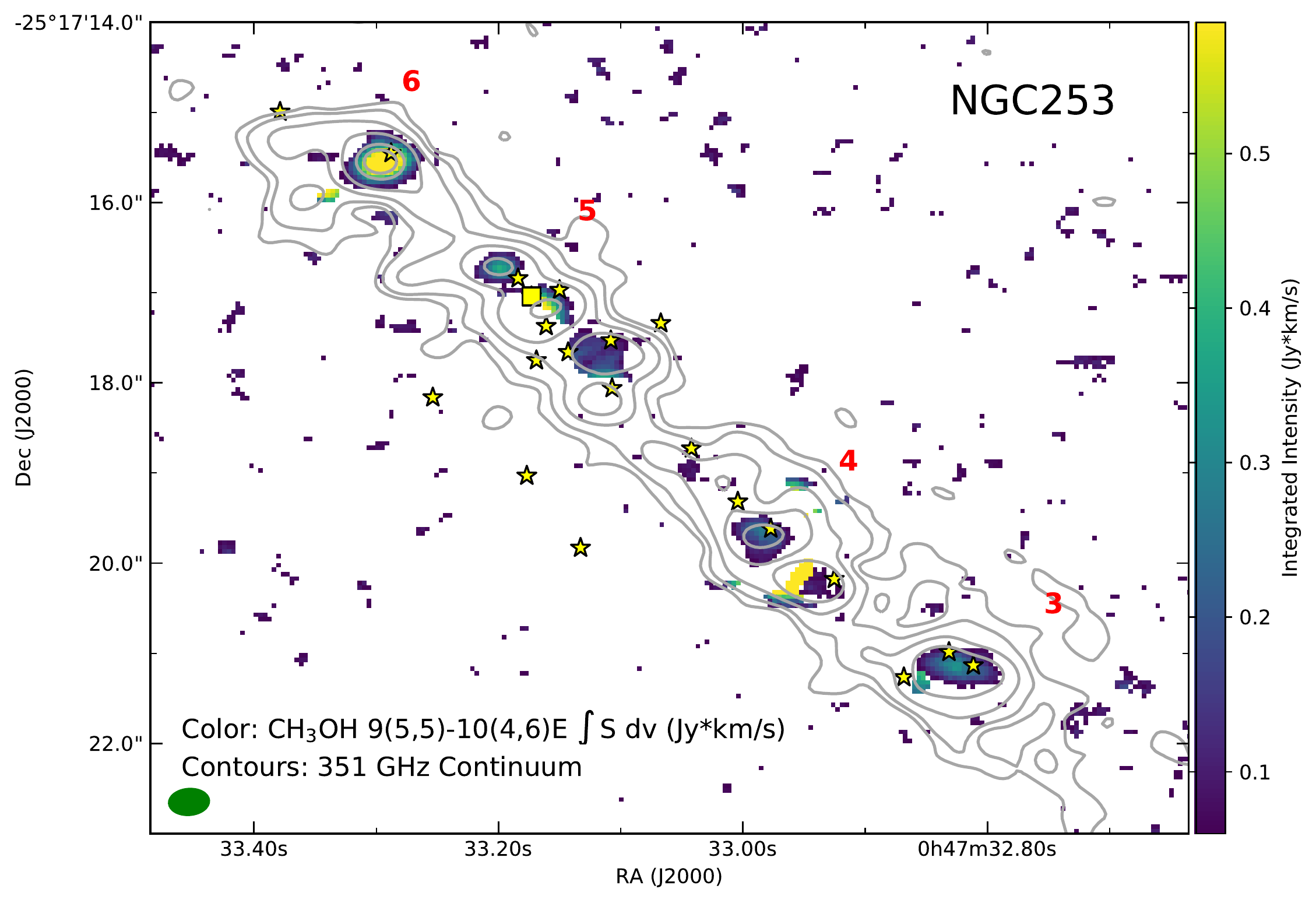}
\caption{Integrated intensity images for the vibrationally excited
  HC$_3$N $24-23~\nu_7 = 2$ and HNC $4-3~\nu_2 = 1f$ transitions and the
  CH$_3$OH $9_{5}-10_{4}E$ transition.  The synthesized beam is
  displayed in the lower left of each panel.  Red numbers indicate the locations of 
  the dense molecular transition regions identified by
  \citet[][Table~\ref{tab:ContPositions}]{Leroy2015}.  Yellow markers
  locate the positions of the 2\,cm radio continuum emission peaks
  \citep{Ulvestad1997}, with a square indicating the position of the
  strongest radio continuum peak identified by \cite{Turner1985} (TH2:
  RA(J2000) = 00$^h$ 47$^m$ 33$^s$.18, Dec(J2000) = $-25^\circ$
  17$^\prime$ 16$^{\prime\prime}$.93).  Contour levels are 3, 5, 10,
  20, 50, and 75\,mJy/beam (218\,GHz continuum); 5, 10, 20, 50, and 75
  \,mJy/beam (365\,GHz continuum); and 3, 5, 10, 20, 50, and
  75\,mJy/beam (351\,GHz continuum).}
\label{fig:vibandCH3OHImages}
\end{figure}

The molecular emission from vibrationally excited HC$_3$N $24-23~\nu_7 =
2$ and HNC $4-3~\nu_2 = 1f$ transitions at 219675.114 and 365147.495\,GHz,
respectively, are detected toward Regions 3 through 7 (see
Figure~\ref{fig:vibandCH3OHImages}).  With the exception of Region 3 in the
HC$_3$N $24-23~\nu_7 = 2$ transition, vibrationally excited integrated
intensities are at greater than twice the integrated intensity RMS
noise levels in our measurements.  Contrary to the integrated emission
from transitions within the vibrational ground states of the molecules
we study, though, the spatial distribution for HC$_3$N $24-23~\nu_7 =
2$ and HNC $4-3~\nu_2 = 1f$ is strongly peaked toward Region 6.  As
noted by \cite{Ando2017}, who also reported the detection of HNC
$4-3~\nu_2 = 1f$, this is the third detection of vibrationally excited
HNC toward an external galaxy, the others being the luminous infrared
galaxies (LIRGs) NGC\,4418 \citep{Costagliola2013,Costagliola2015} and
IRAS\,20551$-$4250 \citep{Imanishi2016}.

A similar spatial distribution is measured from the CH$_3$OH
$9_{5}-10_{4}E$ transition at 351.236\,GHz (Figure~\ref{fig:vibandCH3OHImages})
and the CH$_3$OH $13_{0}-12_{1}A^+$ transition near 355.6\,GHz
\citep{Ando2017}.  For both transitions the strongest emission
emanates from Region 6.
For reference, the 36.2\,GHz
$4_{-1}-3_0 E$ CH$_3$OH maser emission sources in NGC\,253
\citep{Ellingsen2014,Chen2018} emanate from \cite{Leroy2015} Regions
1, 7, and 8, on the far edges of the CMZ imaged in our measurements.
Region 1 is also the source of HC$_3$N $4-3$ maser emission
\citep{Ellingsen2017}. As the emission distributions for
vibrationally-exited HNC, HC$_3$N, and the CH$_3$OH $9_{5}-10_{4}E$
(this work) and $13_{0}-12_{1}A^+$ \citep{Ando2017} transitions
show such spatial similarities, and the CH$_3$OH molecule possesses
numerous inverted (potentially masing) transitions \citep{Mueller2004}, we have
investigated the possibility that the excitation of these transitions
shares a common origin. 

To investigate the potential for inverted level populations in the
CH$_3$OH $9_{5}-10_{4}E$ and $13_{0}-12_{1}A^+$ transitions, we
have run LVG models
\citep[RADEX\footnote{http://var.sron.nl/radex/radex.php};][]{vanderTak2007}
over representative ranges in n(H$_2$), N(CH$_3$OH), and T$_K$:
\begin{enumerate}
  \item n(H$_2$) = $10^4$ to $10^7$\,cm$^{-3}$
  \item N(CH$_3$OH) = $10^{13}$ to $10^{16}$\, cm$^{-2}$
  \item T$_K$ = 50 to 300\,K
  \item FWHM line width = 50\,km/s (typical value from our spectral
    extraction process (Section~\ref{Extraction}))
\end{enumerate}
Over these ranges in volume density, CH$_3$OH column density, and
kinetic temperature we could not find \textit{any} set of physical
conditions where the populations of the CH$_3$OH $9_{5}-10_{4}E$ and
$13_{0}-12_{1}A^+$ transitions were \textit{not} inverted.
Excitation temperatures for these two transitions range from
T$_{ex}$(CH$_3$OH-A) = $-8$ to $-32$\,K and from T$_{ex}$(CH$_3$OH-E) =
$-21$ to $-200$\,K within our LVG models.  It appears that
amplification of a background continuum source through these two
CH$_3$OH transitions is a plausible explanation for the anomalous
spatial distributions measured.

Region 6 contains a rather strong \citep[S(2\,cm) =
  5.59\,mJy;][]{Ulvestad1997} radio continuum source.  As the vibrational
transitions from HNC and HC$_3$N are excited by mid-infrared emission
at 21\,$\mu$m \citep[HNC $4-3~\nu_2 = 1f$;][]{Aalto2007} and 34\,$\mu$m
for \citep[HC$_3$N $24-23~\nu_7 = 2$;][]{Wyrowski1999}, we have investigated the 
potential sources of mid-infrared emission from Region 6.
\cite{Gunthardt2015} imaged the mid-infrared emission from NGC\,253,
finding numerous compact sources along the circumnuclear disk.  Region
6 corresponds to source A3 in Table 2 of \cite{Gunthardt2015}, which
has a flux of 45\,mJy in the infrared Qa band ($\lambda =
18.3$\,$\mu$m).  In the infrared Q-band ($\lambda = 18.7$\,$\mu$m)
images of \cite{FernandezOntiveros2009}, a strong unnamed infrared
source exists that is within ($\Delta$RA,$\Delta$Dec) = (1.0,0.2) arcsec
of Region 6.  Finally, note that the Qa-band sources A2 and A4 in
\cite{Gunthardt2015}, the brightest mid-infrared sources in NGC\,253
with about 1.2\,Jy of Qa-band flux (called the ``Infrared Core'', or
``IRC'' in many references that describe the infrared properties of
NGC\,253), are about ($\Delta$RA,$\Delta$Dec) = ($-4$,$-3$) arcsec
away from Region 6, located near Regions 3 and 4.

As there is a strong source of mid-infrared
emission associated with the submillimeter continuum and molecular
spectral line peaks of Region 6, it seems plausible to suggest that the
source(s) of these radio, millimeter, submillimeter, and infrared
continuum components may provide the sources of amplification for the 
vibrationally excited HC$_3$N $24-23~\nu_7 = 2$ and HNC $4-3~\nu_2 = 1f$
and rotational level inverted CH$_3$OH $9_{5}-10_{4}E$ and
$13_{0}-12_{1}A^+$ transitions.  This suggests that caution is in order
when using these CH$_3$OH transitions to investigate molecular
abundance variations within the NGC\,253 CMZ.

\section{Conclusions}
\label{Conclusions}

Combined ALMA 12m Array and ACA imaging at ALMA
frequency Band 6 (217.3 to 220.8\,GHz) and Band 7 (350.8 to
352.7\,GHz; 362.5 to 364.4 and 363.9 to 365.8\,GHz) of the CMZ within
NGC\,253 have been used to characterize the dense gas kinetic
temperature structure and its relationship to potential sources of
heating in the GMCs within the NGC\,253 nucleus.  These measurements
have yielded the following conclusions:
\begin{enumerate}
  \item Continuum images extracted from our spectral imaging have been
    used to identify five continuum/gas structures, two with
    identifiable substructure, within the NGC\,253 CMZ.  The
    components identified within our measurements correspond to those
    identified previously
    \citep{Turner1985,Sakamoto2011,Leroy2015,Meier2015,Ando2017},
    and appear to be analogs to GMCs in our Galaxy.
  \item Our dust continuum measurements have been used to calculate
    H$_2$ column densities and masses for all measured GMCs.  N(H$_2$)
    and M(H$_2$) range from $1.29-68.09\times10^{23}$\,cm$^{-2}$ and
    $0.19-23.63\times10^{6}$\,\msun, respectively.
  \item Using the selected continuum component positions, we have
    identified 15 molecular species/isotopologues represented by a
    total of 29 molecular transitions.  Integrated
    intensities have been extracted from our spectral imaging, which
    were subsequently used to derive molecular column densities and
    abundances within the GMCs that compose the NGC\,253 CMZ.
  \item Using 10 transitions from the H$_2$CO molecule, we derive the
    kinetic temperature within the $\sim 0.^{\prime\prime}5$ to
    $5^{\prime\prime}$ GMCs.  On $\sim 5^{\prime\prime}$ ($\sim
    80$\,pc) scales we measure $T_K \gtrsim 50$\,K, while on size
    scales $\lesssim 1^{\prime\prime}$ ($\lesssim 16$\,pc) we measure
    $T_K \gtrsim 300$\,K, which surpasses the limits of our H$_2$CO kinetic
    temperature measurement technique.  These kinetic temperature
    measurements are consistent with those derived from previous lower
    spatial resolution studies
    \citep{Mangum2008,Mangum2013,Mangum2013b,Gorski2017}. 
  \item Further evidence for the influence of different heating
    mechanisms within the GMCs of NGC\,253 has been identified by
    comparing the relative abundances of 13 molecular species
    within these GMCs to those derived from modified PDR model
    predictions \citep{Bayet2011,Meijerink2011}.  A general gradient with
    increasing abundances of CR- and mechanical-heating-sensitive
    molecules from the nucleus of NGC\,253 to GMCs farther 
    out in the nuclear disk is observed.  These variations in the
    molecular abundances among the GMCs in NGC\,253 support a
    CR-heating-dominated scenario, but also do not rule out a 
    significant influence due to mechanical heating. 
  \item The simultaneous use of a number of molecular species, as
    opposed to a single chemical diagnostic, has proven to be an
    effective indicator of heating processes within the NGC\,253 CMZ.
    These chemical indicators point to CR and/or mechanical
    heating as plausible mechanisms by which the GMCs in the CMZ of
    NGC\,253 are heated to the high kinetic temperatures that we
    measure.
  \item While investigating the spatial variations among our derived
    GMC molecular column densities, we noted that
    vibrationally excited transitions from HNC and HC$_3$N are
    strongly peaked toward an off-nuclear (Region 6) GMC with an
    associated strong mid-infrared source.  
  \item The Region 6 GMC is also the source of the strongest CH$_3$OH
    emission within the NGC\,253 CMZ.  As numerous CH$_3$OH
    transitions are known to have inverted energy level populations,
    which present as maser emission (\ie\ 36.2\,GHz $4_{-1}-3_0 E$), we
    investigated the potential for our measurements of the CH$_3$OH
    $9_{5}-10_{4}E$ and the \cite{Ando2017} measurements of the
    CH$_3$OH $13_{0}-12_{1}A^+$ transition to possess inverted
    level populations for physical conditions that are representative
    of the NGC\,253 CMZ.  LVG models over 
    representative ranges of n(H$_2$), N(CH$_3$OH), and T$_K$ indicate
    that \textit{all} investigated physical conditions produced
    inverted CH$_3$OH level populations.  Amplification of a
    background continuum source through these two CH$_3$OH transitions
    is a plausible explanation for the anomalous spatial distributions
    measured.
  \item A review of the radio through infrared measurements of the
    center of the NGC\,253 nucleus suggests that there is insufficient
    evidence to support the existence of a low-luminosity AGN at the
    center of NGC\,253.  The radio 
    wavelength emission from NGC\,253 is dominated by that from HII
    regions, radio supernovae, and supernova remnants.
  \item A search for the sources of the high kinetic temperatures in
    the NGC\,253 GMCs encompassed a study of existing high spatial
    resolution radio through infrared imaging.  Potential sources of
    CRs and mechanical heating were determined by noting the
    radio spectral indices from sources in each NGC\,253 GMC
    \citep{Ulvestad1997}.  Radio sources with spectral indices in the
    range $\alpha < -0.4$ are more numerous in the central GMC (Region
    5, which contains the \cite{Turner1985} sources TH2 through TH6)
    than elsewhere in the NGC\,253 CMZ.  Radio spectral indices in
    this range are indicative of synchrotron emission, which can be
    generated by supernova remnants, which produce CRs and
    inject mechanical energy into dense gas.
\end{enumerate}

\acknowledgments

J.G.M.~thanks Ed Fomalont and Bill Cotton for providing invaluable
insight into the physical underpinnings of the ALMA 12m Array and ACA data
combination.  Adam Leroy also provided valuable insight regarding the
gas-to-dust ratio in star formation regions.  This research has made
use of NASA's Astrophysics Data 
System.  This research has also made use of the NASA/IPAC
Extragalactic Database (NED), which is operated by the Jet Propulsion
Laboratory, California Institute 
of Technology, under contract with the National Aeronautics and Space
Administration. This article makes use of the following ALMA data:
ADS/JAO.ALMA\#2013.1.00099.S and ADS/JAO.ALMA\#2015.1.00476.S. ALMA is
a partnership of ESO (representing its member states), NSF (USA) and
NINS (Japan), together with NRC (Canada), NSC and ASIAA (Taiwan), and
KASI (Republic of Korea), in cooperation with the Republic of
Chile. The Joint ALMA Observatory is operated by ESO, AUI/NRAO, and NAOJ.
The National Radio Astronomy Observatory is a facility of the National
Science Foundation operated under cooperative agreement by Associated
Universities, Inc.

\facility{ALMA}
\software{CASA, spectral-cube \citep{Robitaille2016}, PySpecKit
  \citep{Ginsburg2011b}, Astropy \citep{astropy2018}}

\appendix

\section{Data Calibration Details}
\label{CalibrationDetails}

Table~\ref{tab:calibrators} lists the measured and assumed fluxes and
uncertainties associated with all of the calibration measurements
acquired with our NGC\,253 Band 6 and 7 ALMA data.  This information
is meant to support our estimates of the flux uncertainties associated
with the measurements presented.

\startlongtable
\begin{deluxetable*}{llll}
\tablewidth{0pt}
\tablecolumns{4}
\tablecaption{Gain, Flux, and Bandpass Calibrator Fluxes\label{tab:calibrators}}
\tablehead{
\colhead{Measurement} &
\colhead{Calibrator} &
\colhead{Derived/Assumed Flux \tablenotemark{a}} &
\colhead{Flux Uncertainty (\%)}\\
\colhead{} &
\colhead{} &
\colhead{(GA,BP:mJy; FL,PT:Jy)} &
\colhead{}
}
\startdata
\multicolumn{4}{l}{NGC\,253} \\
\tableline
\multicolumn{4}{l}{Band 6 12m Array Measurement on 2014-12-28} \\
GA & J0038$-$2459 & 341.1$\pm$1.8, 339.6$\pm$1.9, 330.4$\pm$2.1 & 0.5, 0.6, 0.6 \\
BP & J2258$-$2758 & 453.4$\pm$1.7, 449.5$\pm$1.7, 424.2$\pm$2.0 & 0.4, 0.4, 0.5 \\
FL & Uranus & 30.9, 31.3, 34.8 & 5\tablenotemark{b} \\
PT & J0108$+$0135 & 900$\pm$130 & 14 \\
\tableline
\multicolumn{4}{l}{Band 6 12m Array Measurement on 2015-05-02} \\
GA & J0038$-$2459 & 263.8$\pm$3.9, 262.4$\pm$3.9, 252.5$\pm$4.3 & 1.5, 1.5, 1.7 \\
BP & J0334$-$4008 & 688.2$\pm$4.8, 685.8$\pm$5.5, 653.3$\pm$6.5 & 0.7 0.8, 1.0 \\
FL & Mars & 78.5, 79.7, 89.5 & 10\tablenotemark{b} \\
PT & J0238$+$1636 & 1550$\pm$50 & 3 \\
\tableline
\multicolumn{4}{l}{Band 6 ACA Measurement on 2014-06-04} \\
GA & J0038$-$2459 & 458.4$\pm$4.6, 453.5$\pm$4.6, 422.4$\pm$5.2 & 1.0, 1.0, 1.2 \\
BP & J2258$-$2758 & 570.7$\pm$2.3, 569.3$\pm$2.6, 521.5$\pm$3.4 & 0.4, 0.4, 0.6 \\
FL & Neptune & 13.0, 13.1, 13.6 & 10\tablenotemark{b} \\
\tableline
\multicolumn{4}{l}{Band 6 ACA Measurement on 2014-06-04} \\
GA & J0038$-$2459 & 450.6$\pm$4.0, 448.0$\pm$5.8, 425.1$\pm$4.8 & 0.9, 1.3, 1.1 \\
BP & J2258$-$2758 & 570.7$\pm$2.4, 567.7$\pm$2.7, 538.6$\pm$3.0 & 0.4, 0.5, 0.5 \\
FL & Uranus & 28.9, 29.3, 32.6 & 5\tablenotemark{b} \\
\tableline
\multicolumn{4}{l}{Band 6 ACA Measurement on 2014-06-04} \\
GA & J0038$-$2459 & 445.5$\pm$4.6, 442.4$\pm$5.3, 427.7$\pm$4.0 & 1.0, 1.0, 1.2 \\
BP & J2258$-$2758 & 559.4$\pm$2.7, 554.2$\pm$2.6, 528.3$\pm$2.4 & 0.4, 0.4, 0.6 \\
FL & Uranus & 28.9, 29.3, 32.6 & 5\tablenotemark{b} \\
\tableline
\multicolumn{4}{l}{Band 7 12m Array Measurement on 2014-05-19} \\
GA & J0038$-$2459 & 290.0$\pm$12.5, 287.5$\pm$16.6, 289.2$\pm$18.5 & 4.3, 5.8, 6.4 \\
BP & J0006$-$0623 & 2092.47$\pm$34.6, 2074.5$\pm$45.7, 2066.6$\pm$48.4 & 1.7, 2.2, 2.3 \\
FL & J2258$-$2758 & 0.4$\pm$0.1\tablenotemark{c} & 15\tablenotemark{c} \\
\tableline
\multicolumn{4}{l}{Band 7 ACA Measurement on 2014-05-19} \\
GA & J0038$-$2459 & 274.7$\pm$8.2, 279.9$\pm$7.5, 284.1$\pm$11.2 & 3.0, 2.7, 3.9 \\
BP & J0006$-$0623 & 2045.2$\pm$14.0, 2080.7$\pm$10.6, 2066.0$\pm$10.8 & 0.7, 0.5, 0.5 \\
FL & Neptune & 25.8, 27.7, 27.8 & 10\tablenotemark{b} \\
\tableline
\multicolumn{4}{l}{Band 7 ACA Measurement on 2014-06-08} \\
GA & J0038$-$2459 & 324.3$\pm$7.7, 324.9$\pm$9.3, 320.3$\pm$6.1 & 3.0, 2.7, 3.9 \\
BP & J0006$-$0623 & 2243.8$\pm$15.4, 2214.2$\pm$17.5, 2226.3$\pm$15.6 & 0.7, 0.8, 0.7 \\
FL & Uranus & 63.8, 66.3, 66.6 & 5\tablenotemark{b} \\
\enddata
\tablenotetext{a}{Derived (GA and BP) or assumed (FL and PT) flux,
  where GA $\equiv$ gain amplitude, BP $\equiv$ bandpass, FL $\equiv$
  flux, and PT $\equiv$ pointing calibration.}
\tablenotetext{b}{See discussion in Section~\ref{Observations}.}
\tablenotetext{c}{Assumed flux from ALMA calibrator catalog.}
\end{deluxetable*}

\section{Spectral Properties}
\label{Spectral}

Table~\ref{tab:SpectralProperties} lists the frequency, upper-state
energy, dipole moment, line strength, transition degeneracies, and
rotational partition function values at kinetic temperatures of 50,
150, and 300\,K for each measured transition.

\begin{longrotatetable}
\begin{deluxetable}{lccccccc}
\tablewidth{0pt}
\tablecolumns{8}
\tablecaption{Measured Transition Spectral Properties\label{tab:SpectralProperties}}
\tablehead{
\colhead{Molecule} &
\colhead{Transition} & 
\colhead{Frequency (MHz)} &
\colhead{$E_u$ (K)} &
\colhead{$\mu$ (Debye)} &
\colhead{S\tablenotemark{a}} &
\colhead{$g_J$/$g_K$/$g_I$} &
\colhead{Q$_{rot}$(50/150/300)}
}
\startdata
$^{13}$CO & $2-1$ & 220398.684 & 15.8662 & 0.11046 & $\frac{2}{5}$ &
5/1/1 & 19.22/56.97/113.86 \\
C$^{18}$O & $2-1$ & 219560.358 & 15.8059 & 0.11079 & $\frac{2}{5}$ &
5/1/1 & 19.31/57.19/114.29 \\
$^{13}$CN\tablenotemark{b} & F$_1=3-2$ & 217467.150 & 15.684 & 1.45 & $\frac{2}{5}$ &
5/1/1 & 19.47/57.74/115.42 \\
SiS & $12-11$ & 217817.663 & 67.954 & 1.730 & $\frac{12}{25}$ & 25/1/1 &
114.99/344.17/689.59 \\
SO & $6_5-5_4$ & 219949.442 & 34.9847 & 1.55 & $\frac{5}{11}$ & 13/1/0.5 &
41.83/137.83/283.33 \\
SO$_2$ & $4_{22}-3_{13}$ & 235151.72 & 19.0298 & 1.6331 & 0.1906 &
9/1/1 & 401.17/2084.56/5896.02 \\
SO$_2$ & $5_{33}-4_{22}$ & 351257.224 & 35.88646 & 1.6331 & 0.2495 &
11/1/1 & 401.17/2084.56/5896.02 \\
SO$_2$ & $14_{4,10}-14_{3,11}$ & 351873.873 & 135.87076 &
1.6331 & 0.2538 & 29/1/1 & 401.17/2084.56/5896.02 \\
HNC & $4-3$ & 362630.30 & 43.5097 & 3.05 & $\frac{4}{9}$ & 9/1/1 &
23.29/69.18/138.31 \\
OCS & $30-29$ & 364748.960 & 271.379 & 0.7152 & $\frac{30}{61}$ &
61/1/1 & 171.46/513.44/1028.60 \\
HNCO & $10_0 - 9_0$ & 219798.27 & 58.0194 & 1.602 & $\frac{10}{21}$ &
21/1/1 & 181.60/945.23/2695.60 \\
HNCO & $10_1 - 9_1$ & 218981.02 & 101.079 & 1.602 & $\frac{99}{210}$ &
21/1/1 & 181.60/945.23/2695.60 \\
HNCO & $10_2-9_2$ & 219737.19 & 228.2851 & 1.602 & $\frac{48}{105}$ &
21/1/1 & 181.60/945.23/2695.60 \\
H$_2$CO & $3_{03}-2_{02}$ & 218222.192 & 20.957 & 2.332
& $\frac{9}{21}$ & 7/1/0.25 & 50.14/258.97/731.44 \\
H$_2$CO & $3_{21}-2_{20}$ & 218760.071 & 68.112 &
2.332 & $\frac{5}{21}$ & 7/1/0.25 & 50.14/258.97/731.44 \\
H$_2$CO & $5_{05}-4_{04}$ & 362736.048 & 52.313 & 2.332 &
$\frac{25}{55}$ & 11/1/0.25 & 50.14/258.97/731.44 \\
H$_2$CO & $5_{24}-4_{23}$ & 363945.894 & 99.539 & 2.332 &
$\frac{21}{55}$ & 11/1/0.25 & 50.14/258.97/731.44 \\
H$_2$CO & $5_{23}-4_{22}$ & 365363.428 & 99.658 & 2.332 &
$\frac{21}{55}$ & 11/1/0.25 & 50.14/258.97/731.44 \\
H$_2$CO & $5_4-4_4$ & 364103.249 & 240.730 & 2.332 & $\frac{9}{55}$ &
11/1/0.25 & 50.14/258.97/731.44 \\
H$_2$CO & $5_{15}-4_{14}$ & 351768.645 & 64.453 & 2.332 &
$\frac{24}{55}$ & 11/1/0.75 & 50.14/258.97/731.44 \\
H$_2$CO & $5_3-4_3$ & 364288.884 & 158.424 & 2.332 & $\frac{16}{55}$ &
11/1/0.75 & 50.14/258.97/731.44 \\
H$_3$O$^+$ & $3_2-2_2$ & 364797.427 & 139.338 & 1.44 &
$\frac{5}{21}$ & 7/2/0.25 & 4.39/22.79/64.46 \\
HC$_3$N & $24-23$ v$_0$ & 218324.72 & 130.982 &
3.73172 & $\frac{24}{49}$ & 49/1/1 & 229.10/686.25/1374.92 \\
HC$_3$N & $40-39$ & 363785.40 & 357.9731 &
3.73172 & $\frac{40}{81}$ & 81/1/1 & 229.10/686.25/1374.92 \\
C$_4$H & N=23-22 & 218857.0 & 126.05 & 0.90 & 23.0 & 47/1/1 &
221.05/439.60/1315.80 \\
CH$_3$CN & $12-11,K=0-3$ & 220747.259 & 68.8657 &
3.92197 & $\frac{12}{25}$ & 25/1/0.5 & 167.17/876.86/2500.55 \\
CH$_3$OH & $9_{55}$-$10_{46}$E & 351236.343 & 240.50455 & 1.412 &
0.6908 & 19/1/0.25 & 72.65/377.50/1067.74 \\
\enddata
\tablenotetext{a}{Line strengths calculated using \cite{Mangum2015}
  excluding the following: \cite[SO:][]{Tiemann1974},
  \cite[SO$_2$:][]{Lovas1985}, \cite[CH$_3$OH:][]{Xu1997}.}
\tablenotetext{b}{Strongest component of the hyperfine group with
  N=$2-1$, J=$\frac{5}{2}-\frac{3}{2}$.}
\end{deluxetable}
\end{longrotatetable}

\section{Spectral Line Integrated Intensity Images}
\label{IntegInt}

\begin{figure}
\centering
\includegraphics[trim=20mm 0mm 20mm 10mm, scale=0.65]{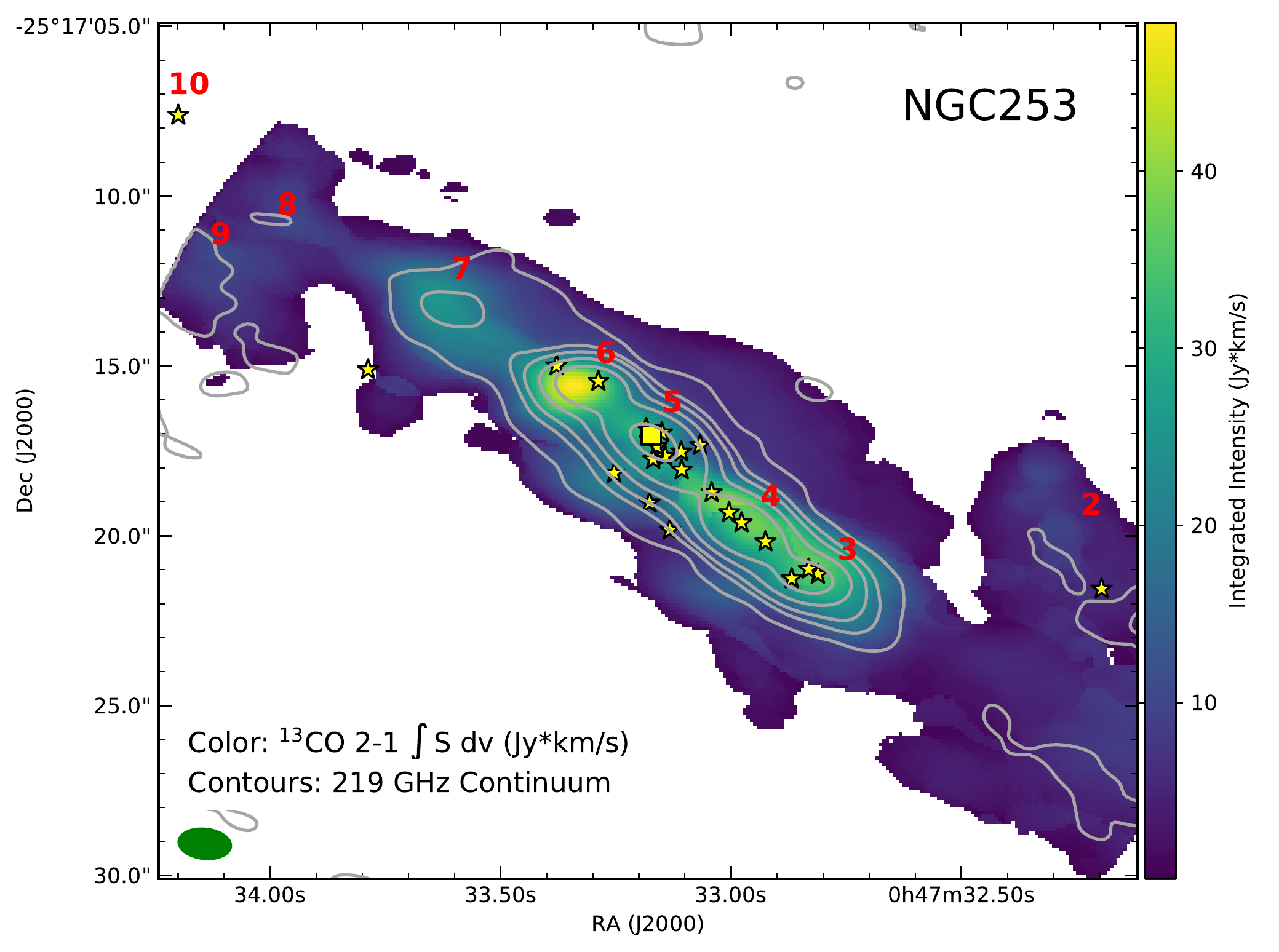}\\
\includegraphics[trim=20mm 0mm 20mm 0mm, scale=0.65]{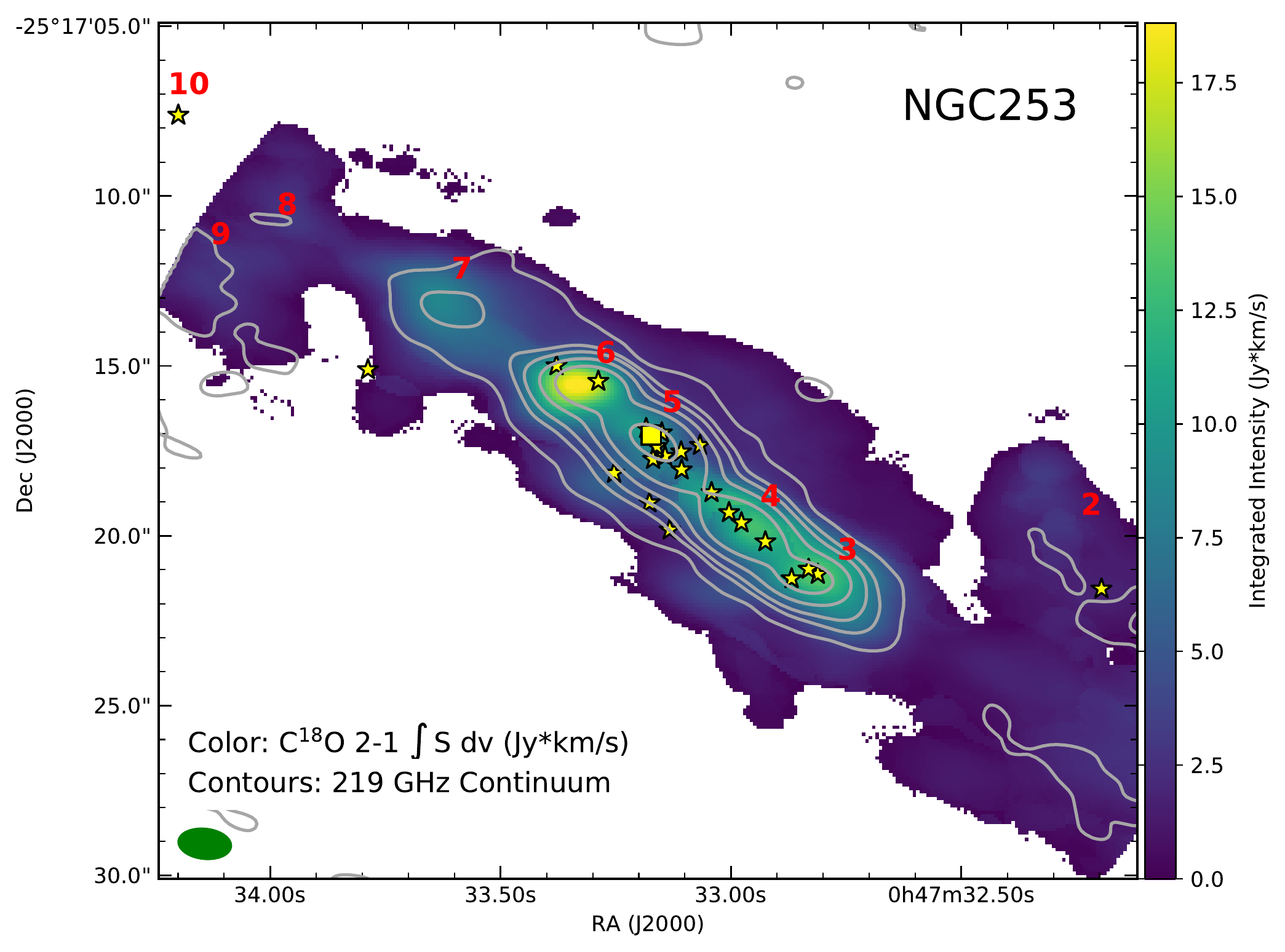}
\caption{$^{13}$CO $2-1$ (top) and C$^{18}$O $2-1$ (bottom) integrated intensity.}
\label{fig:IntegIntCO}
\end{figure}

\begin{figure}
\centering
\includegraphics[trim=20mm 0mm 20mm 0mm, scale=0.65]{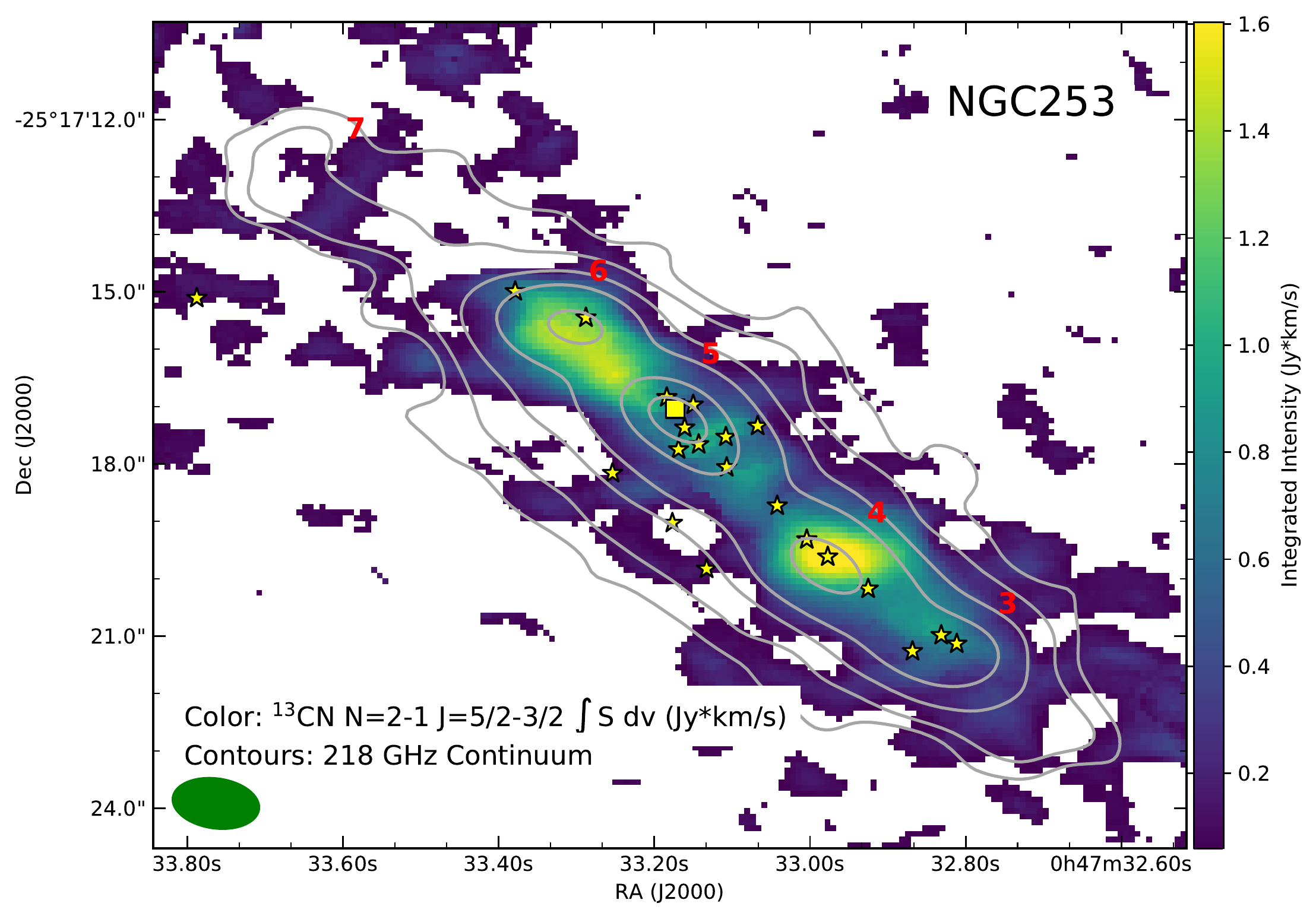}
\caption{$^{13}$CN F$_1$=$3-2$ integrated intensity.}
\label{fig:IntegIntCN}
\end{figure}

\begin{figure}
\centering
\includegraphics[trim=20mm 0mm 20mm 0mm, scale=0.65]{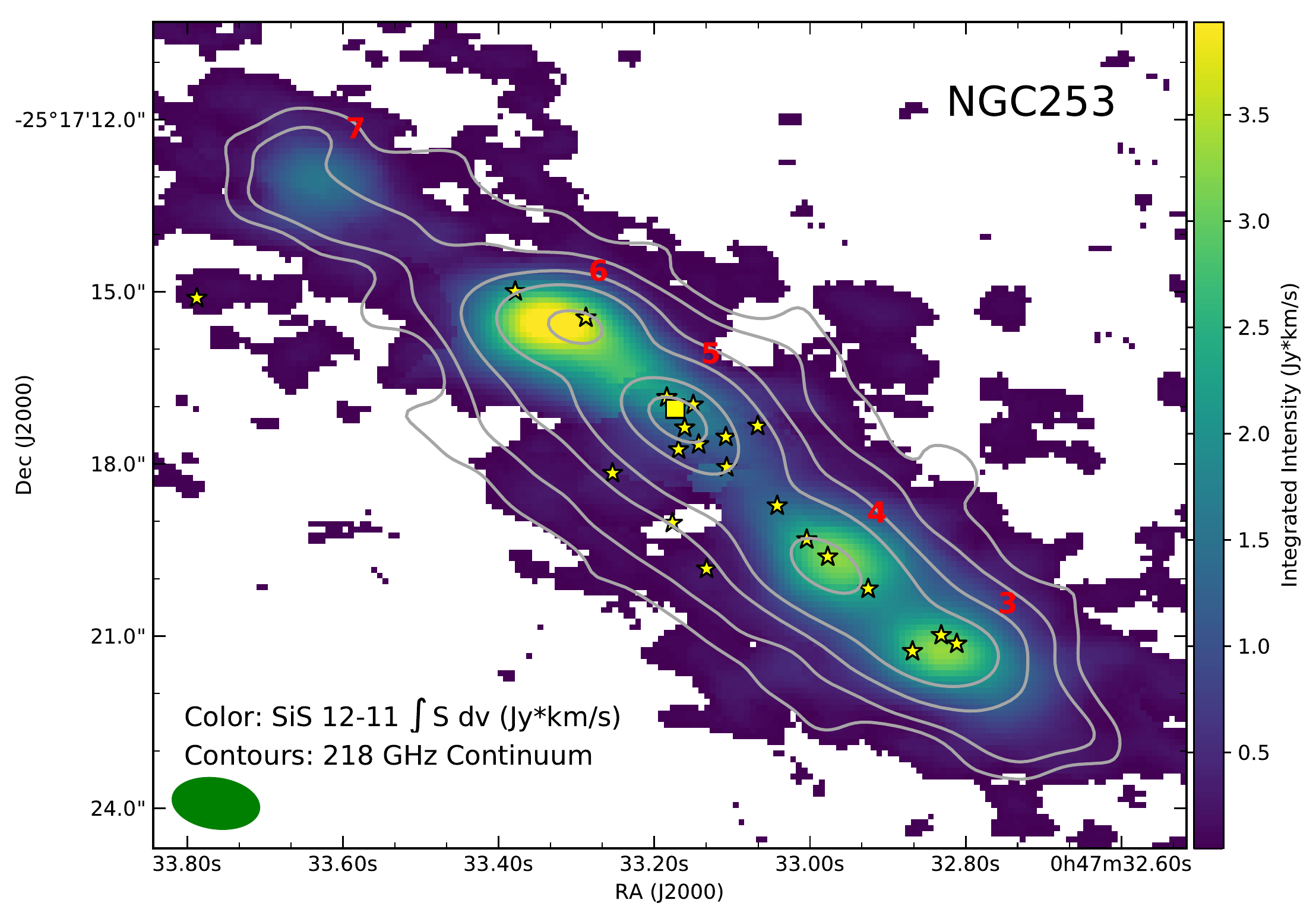}
\caption{SiS $12-11$ integrated intensity.}
\label{fig:IntegIntSiS}
\end{figure}

\begin{figure}
\centering
\includegraphics[trim=20mm 0mm 20mm 0mm, scale=0.65]{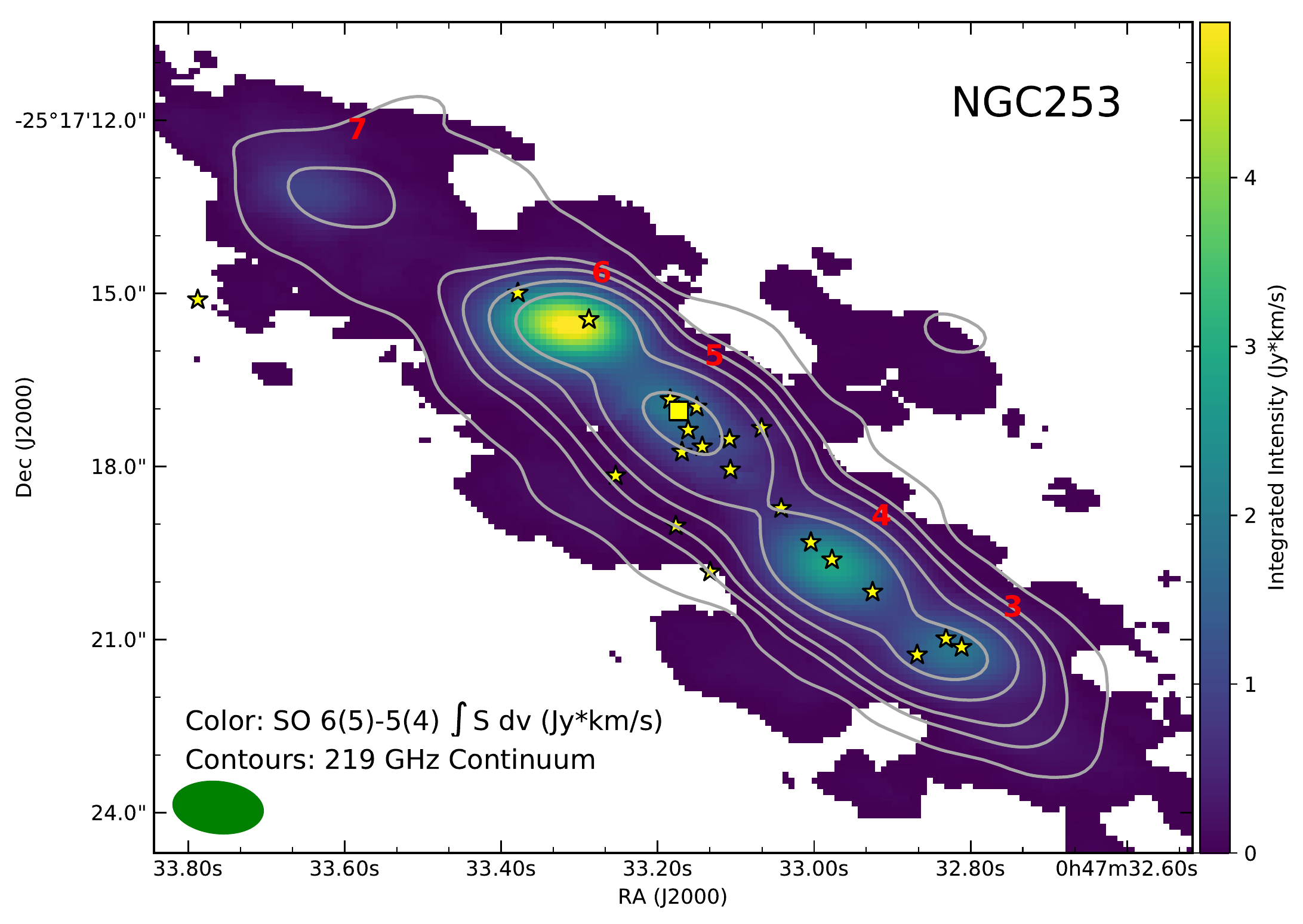}
\caption{SO $6_5-5_4$ integrated intensity.}
\label{fig:IntegIntSO}
\end{figure}

\begin{figure}
\centering
\includegraphics[trim=20mm 0mm 22mm 0mm, scale=0.43]{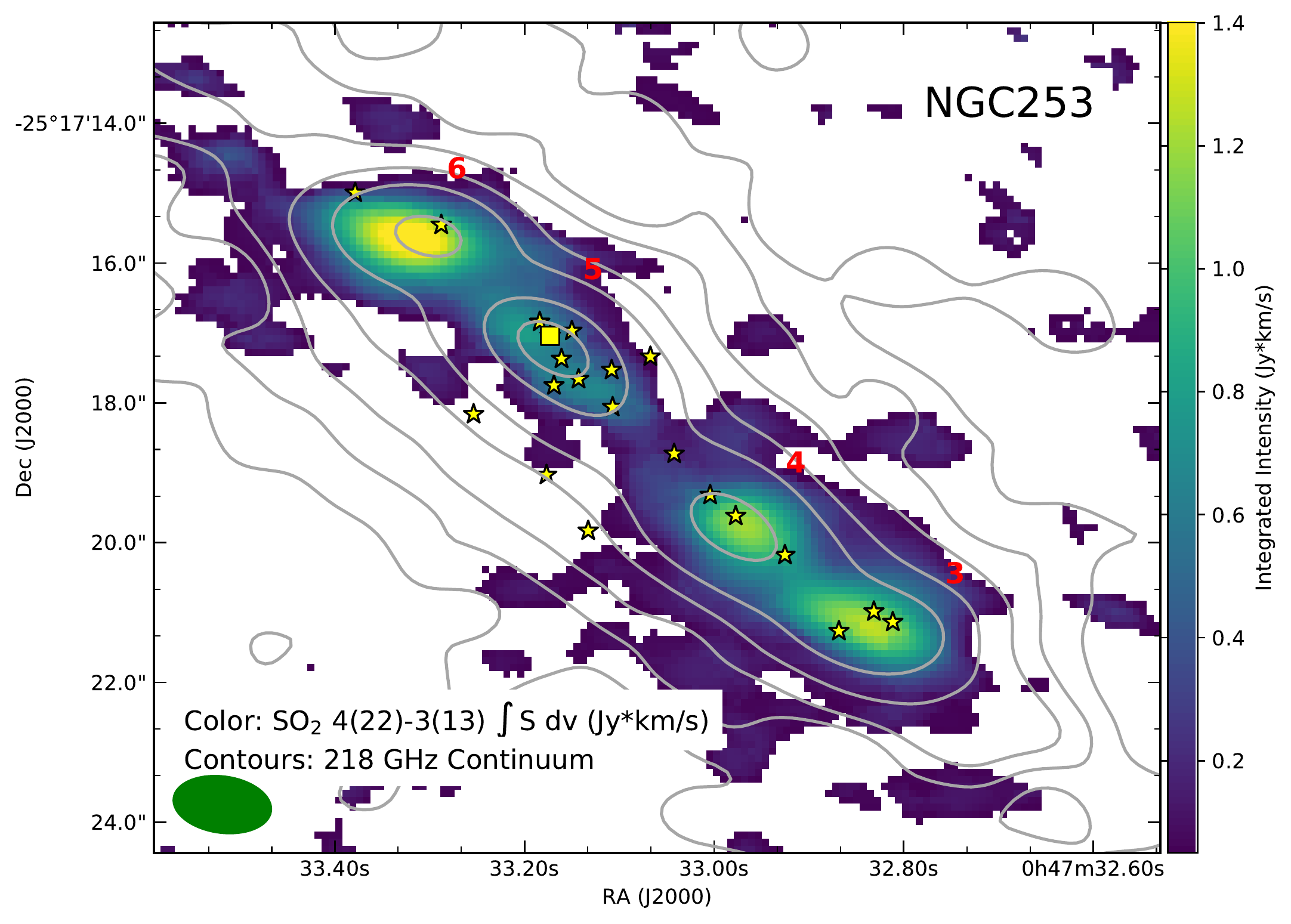}\\
\includegraphics[trim=20mm 0mm 22mm 0mm, scale=0.41]{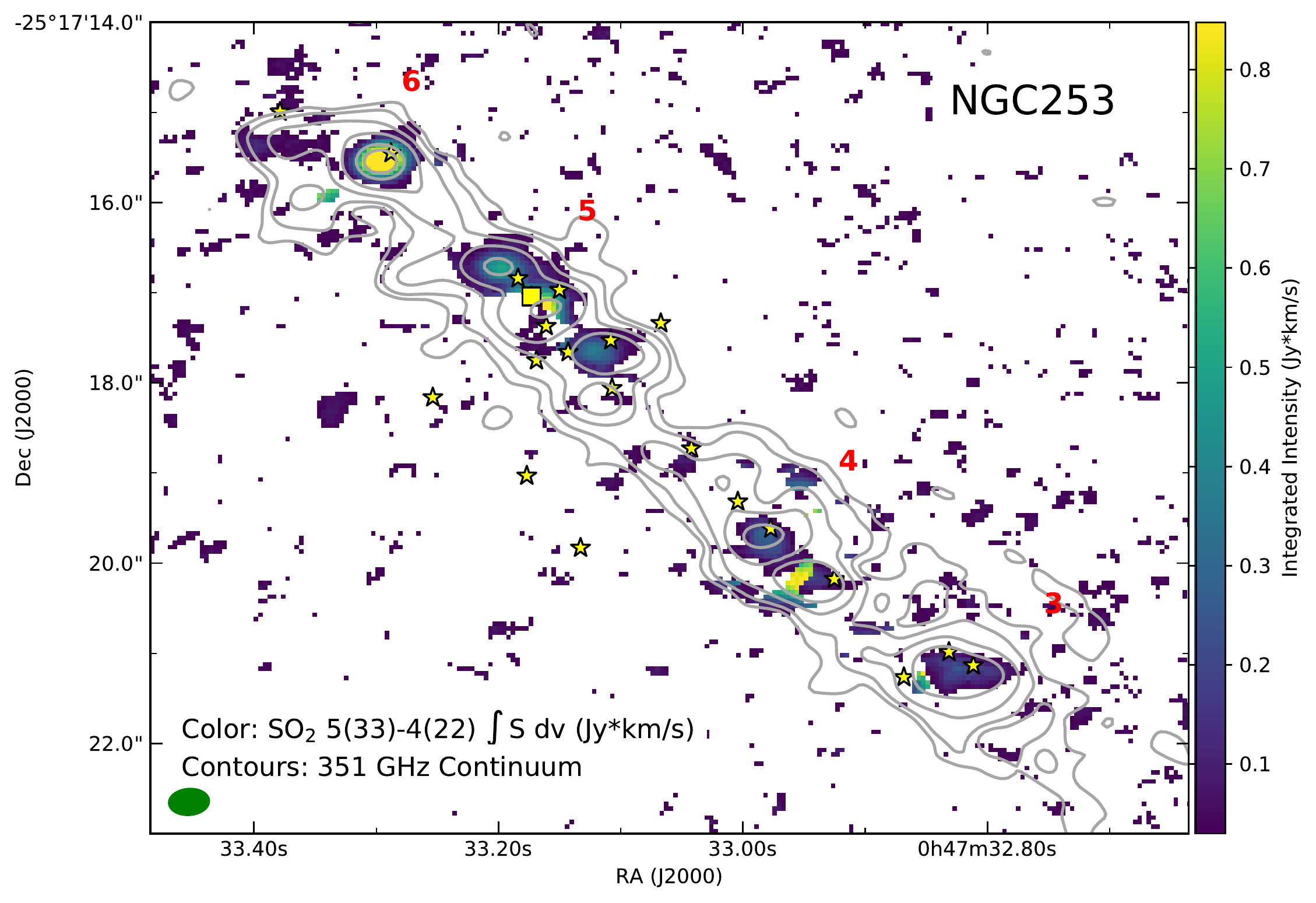}\\
\includegraphics[trim=20mm 0mm 22mm 0mm, scale=0.41]{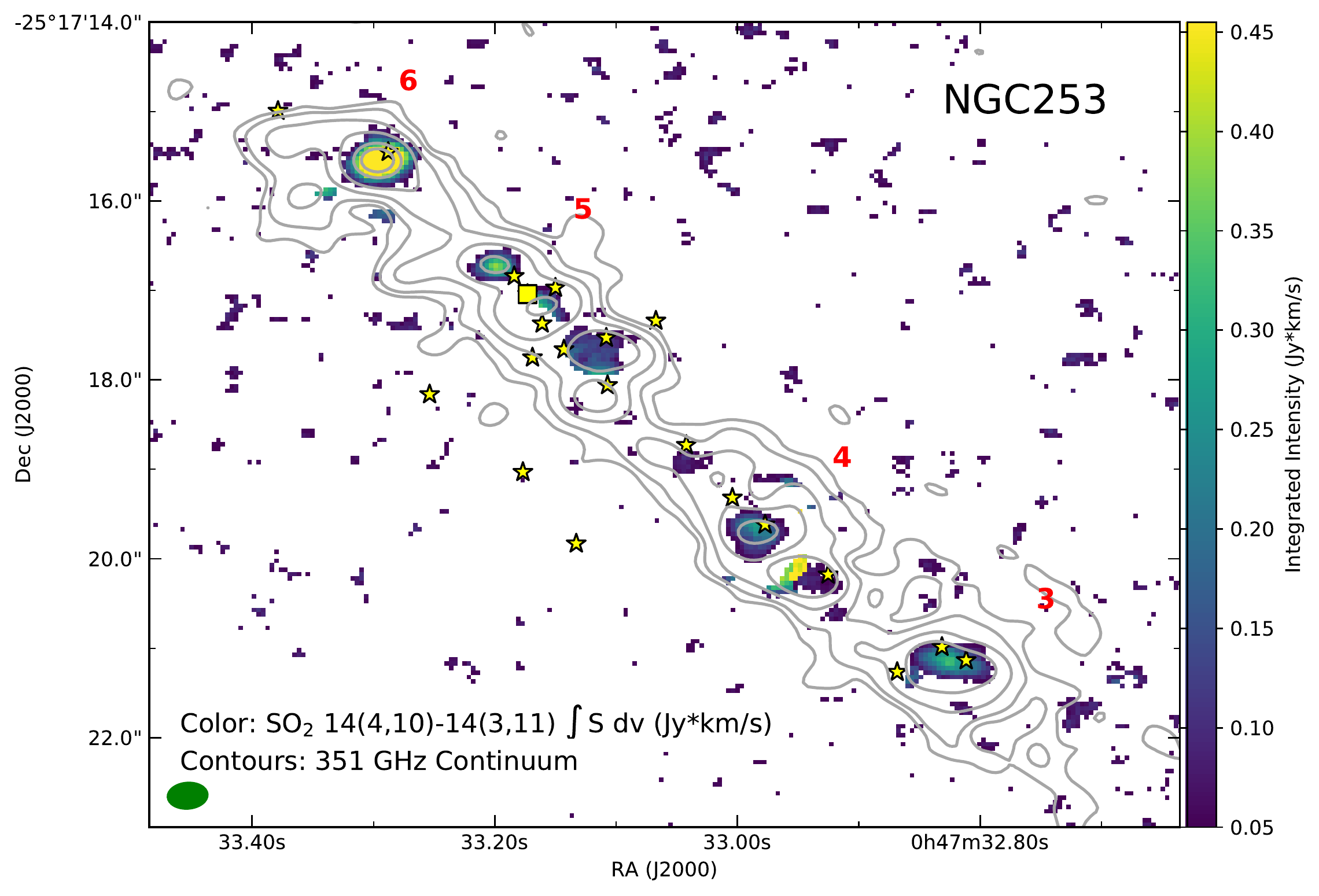}
\caption{SO$_2$ $4_{22}-3_{13}$ (top), $5_{33}-4_{22}$ (middle), and
  $14_{4,10}-14_{3,11}$ (bottom) integrated intensity.}
\label{fig:IntegIntSO2}
\end{figure}

\begin{figure}
\centering
\includegraphics[trim=20mm 0mm 20mm 0mm, scale=0.65]{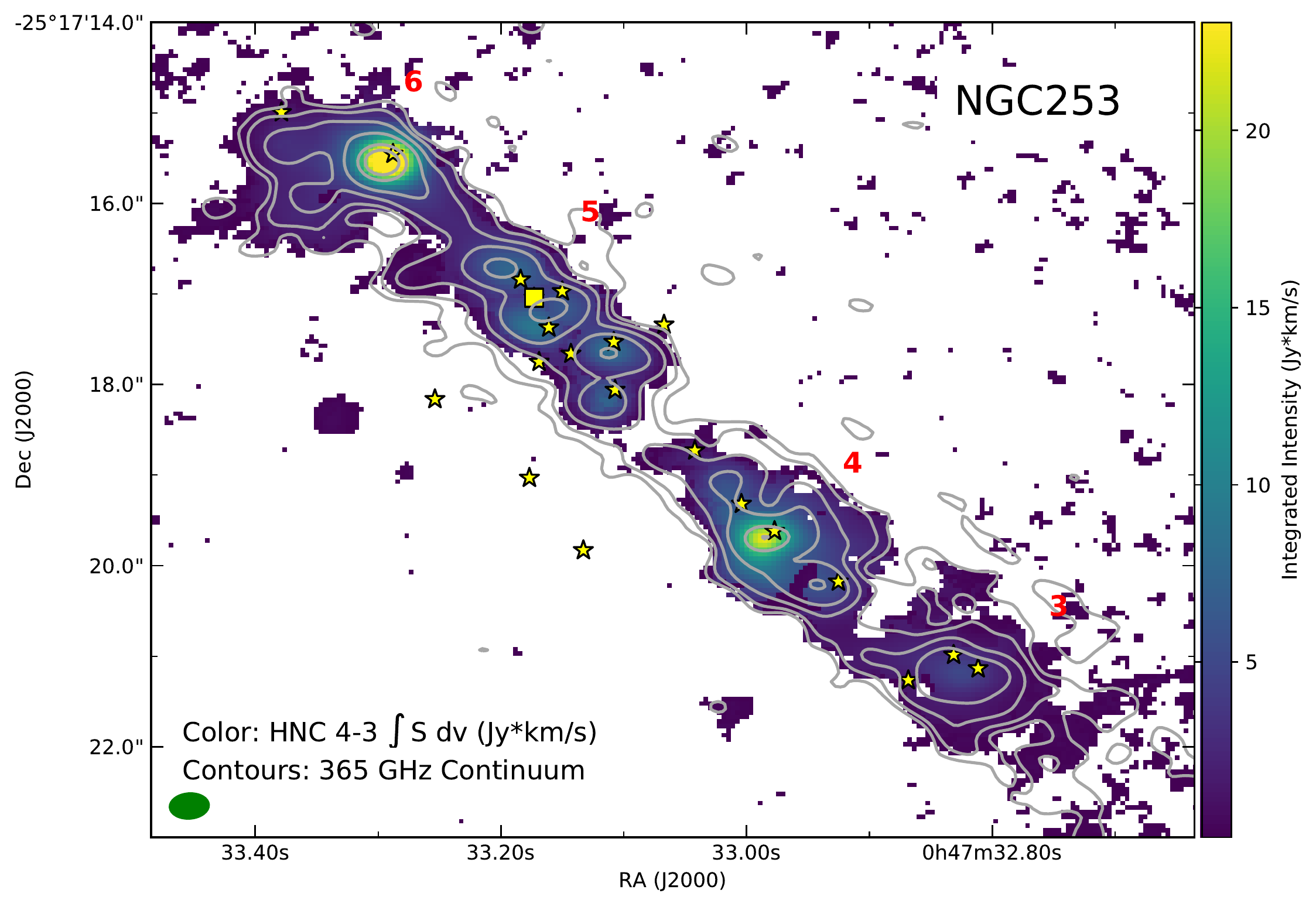}\\
\includegraphics[trim=20mm 0mm 20mm 0mm, scale=0.65]{NGC253-HNC43v21fInt.pdf}
\caption{HNC $4-3$ (top) and $4-3~v_2=1f$ (bottom) integrated intensity.}
\label{fig:IntegIntHNC}
\end{figure}

\begin{figure}
\centering
\includegraphics[trim=20mm 0mm 20mm 0mm, scale=0.66]{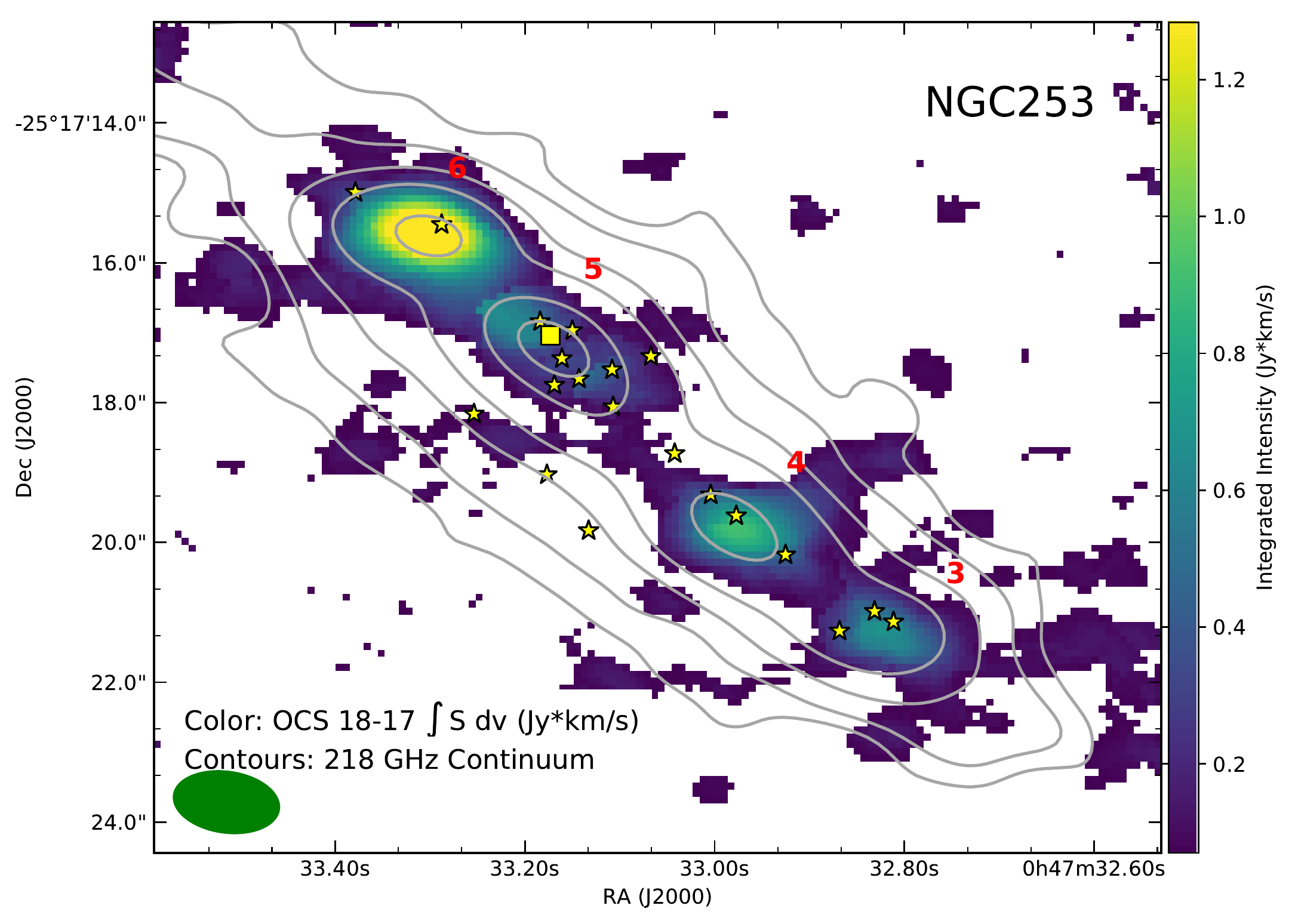}\\
\includegraphics[trim=20mm 0mm 20mm 0mm, scale=0.64]{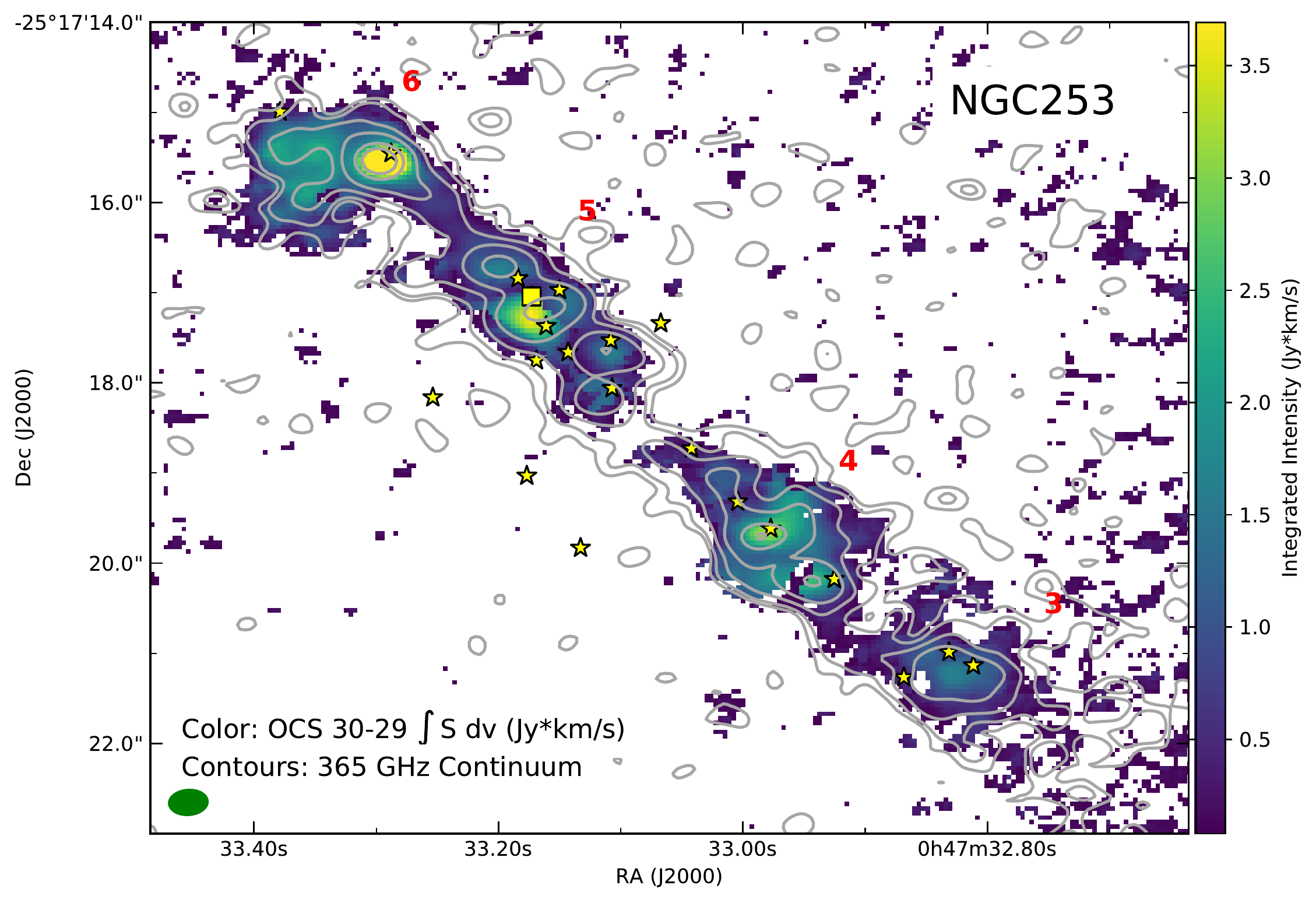}
\caption{OCS $18-17$ (top) and $30-29$ (bottom) integrated intensity.}
\label{fig:IntegIntOCS}
\end{figure}

\begin{figure}
\centering
\includegraphics[trim=20mm 0mm 20mm 0mm, scale=0.45]{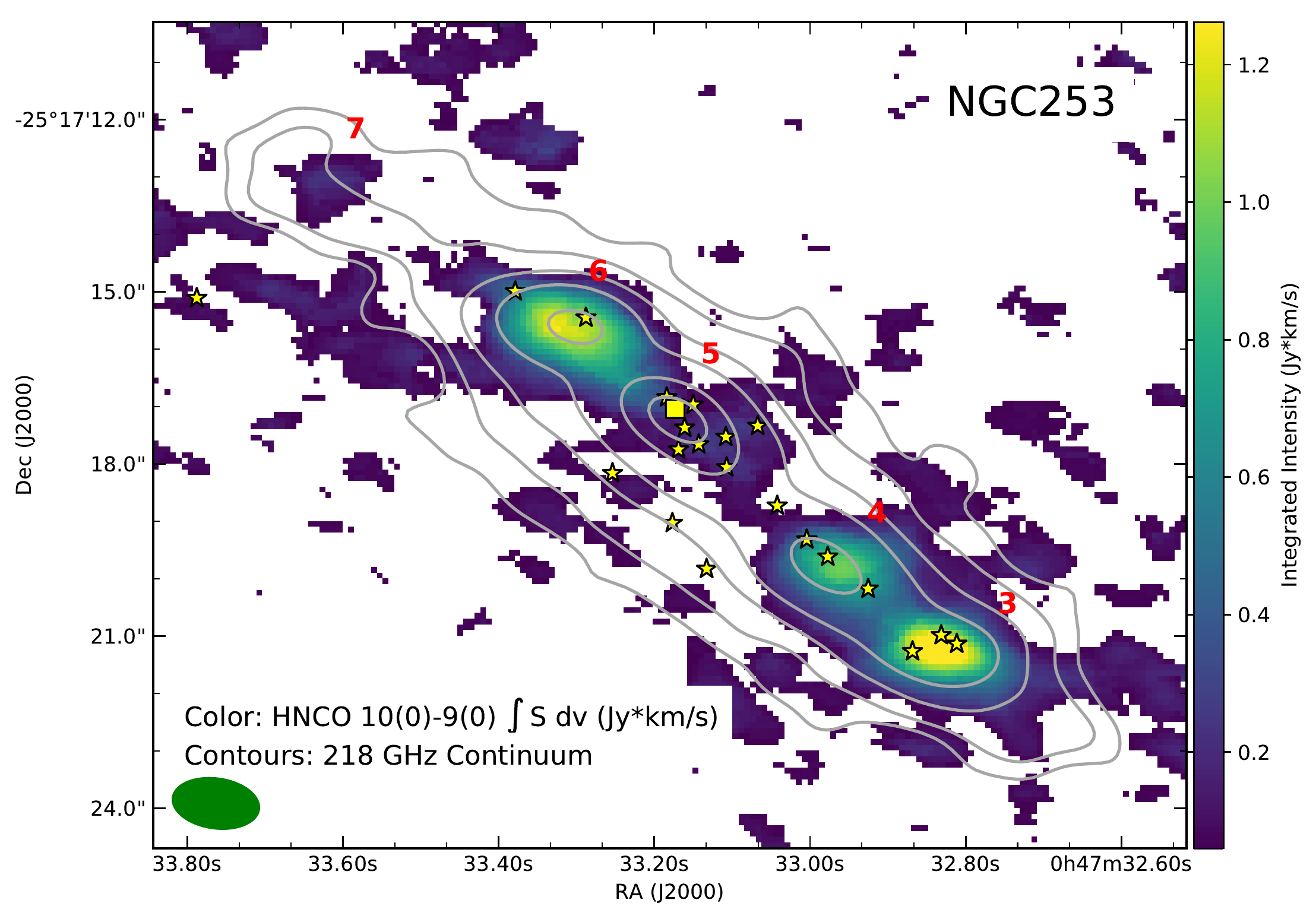}\\
\includegraphics[trim=20mm 0mm 20mm 0mm, scale=0.45]{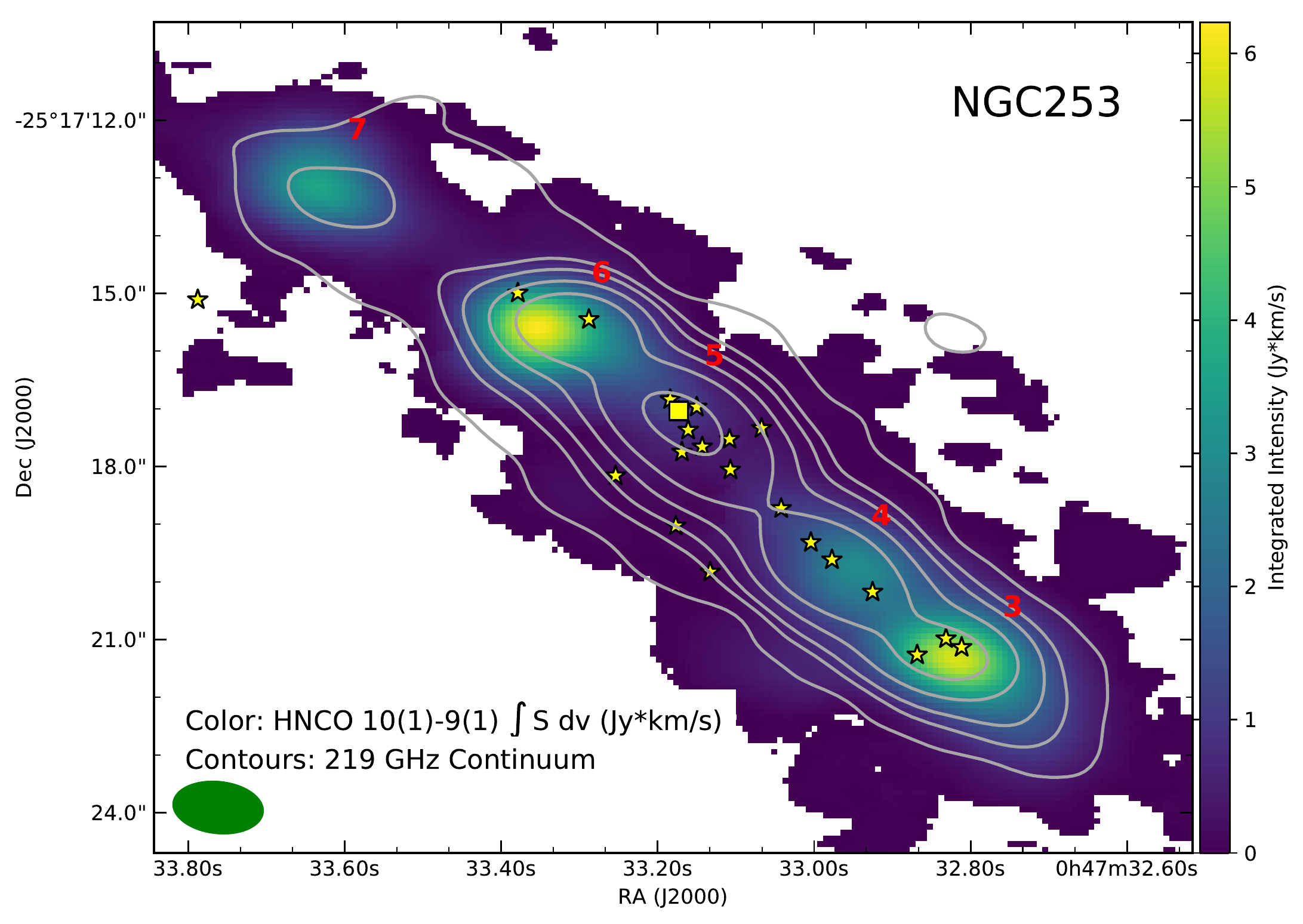}\\
\includegraphics[trim=20mm 0mm 20mm 0mm, scale=0.45]{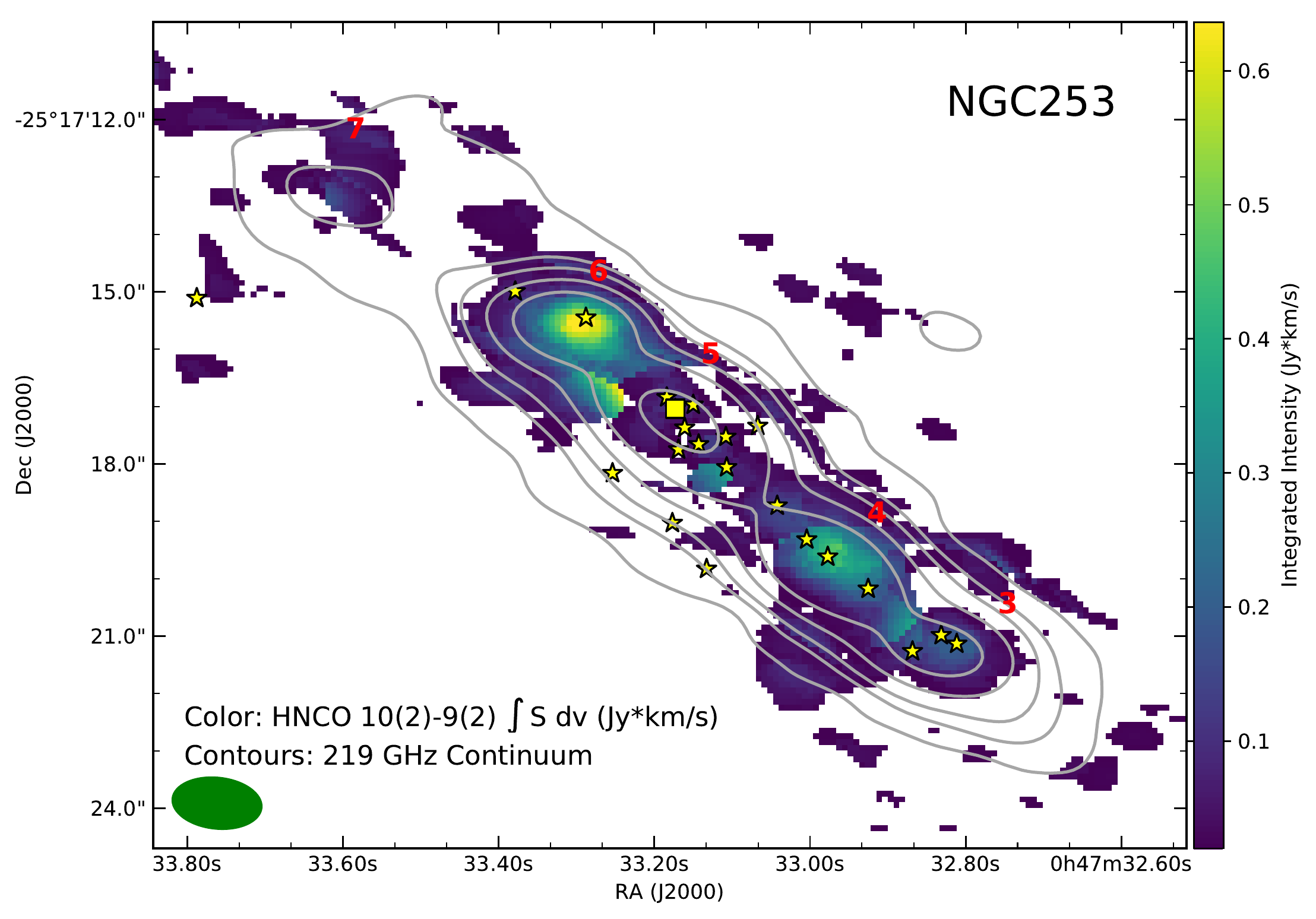}
\caption{HNCO $10_0-9_0$ (top), $10_1-9_1$ (middle), $10_2-9_2$
  (bottom) integrated intensity.}
\label{fig:IntegIntHNCO}
\end{figure}

\begin{figure}
\centering
\includegraphics[trim=20mm 0mm 20mm 0mm, scale=0.65]{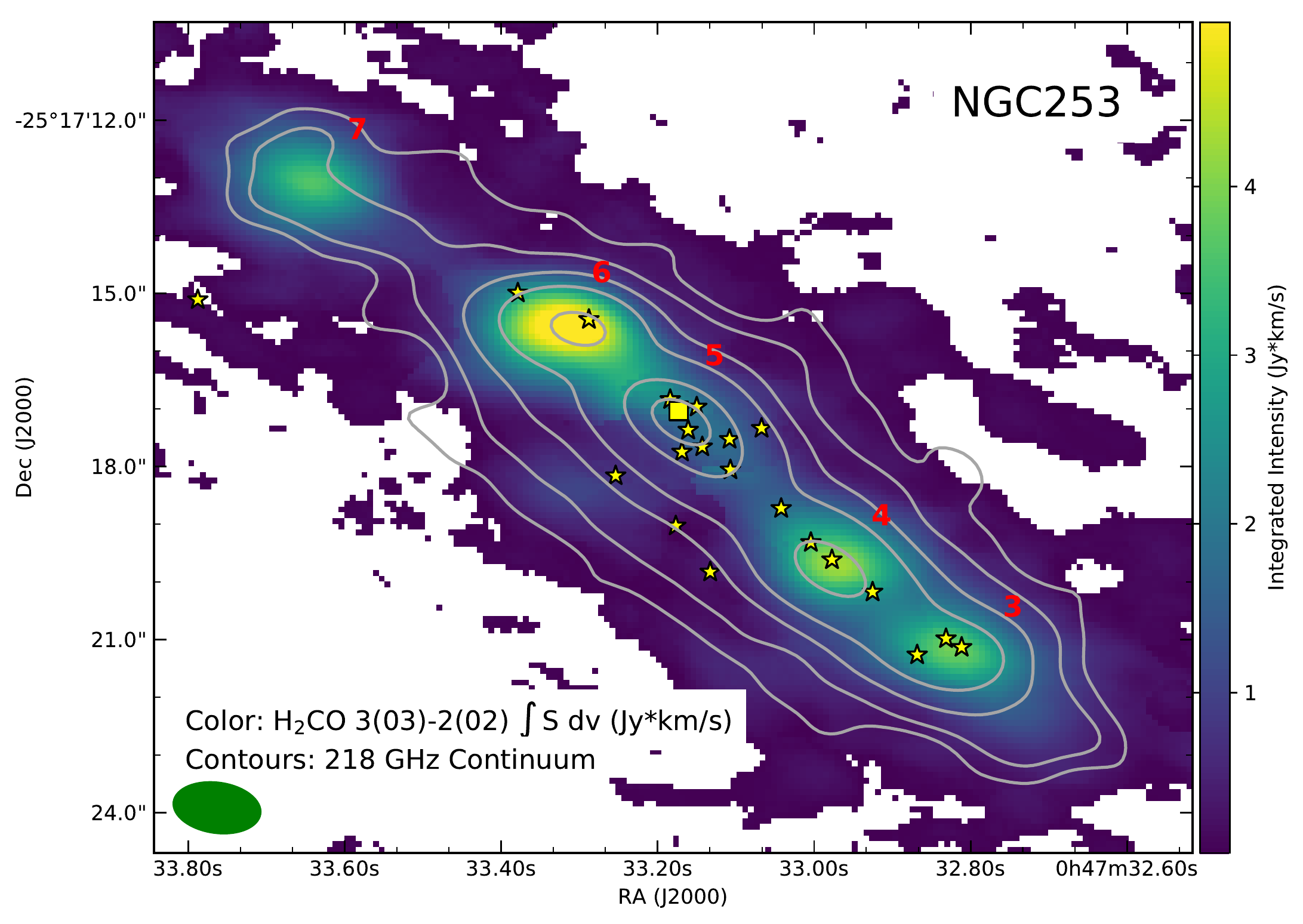}\\
\includegraphics[trim=20mm 0mm 20mm 0mm, scale=0.65]{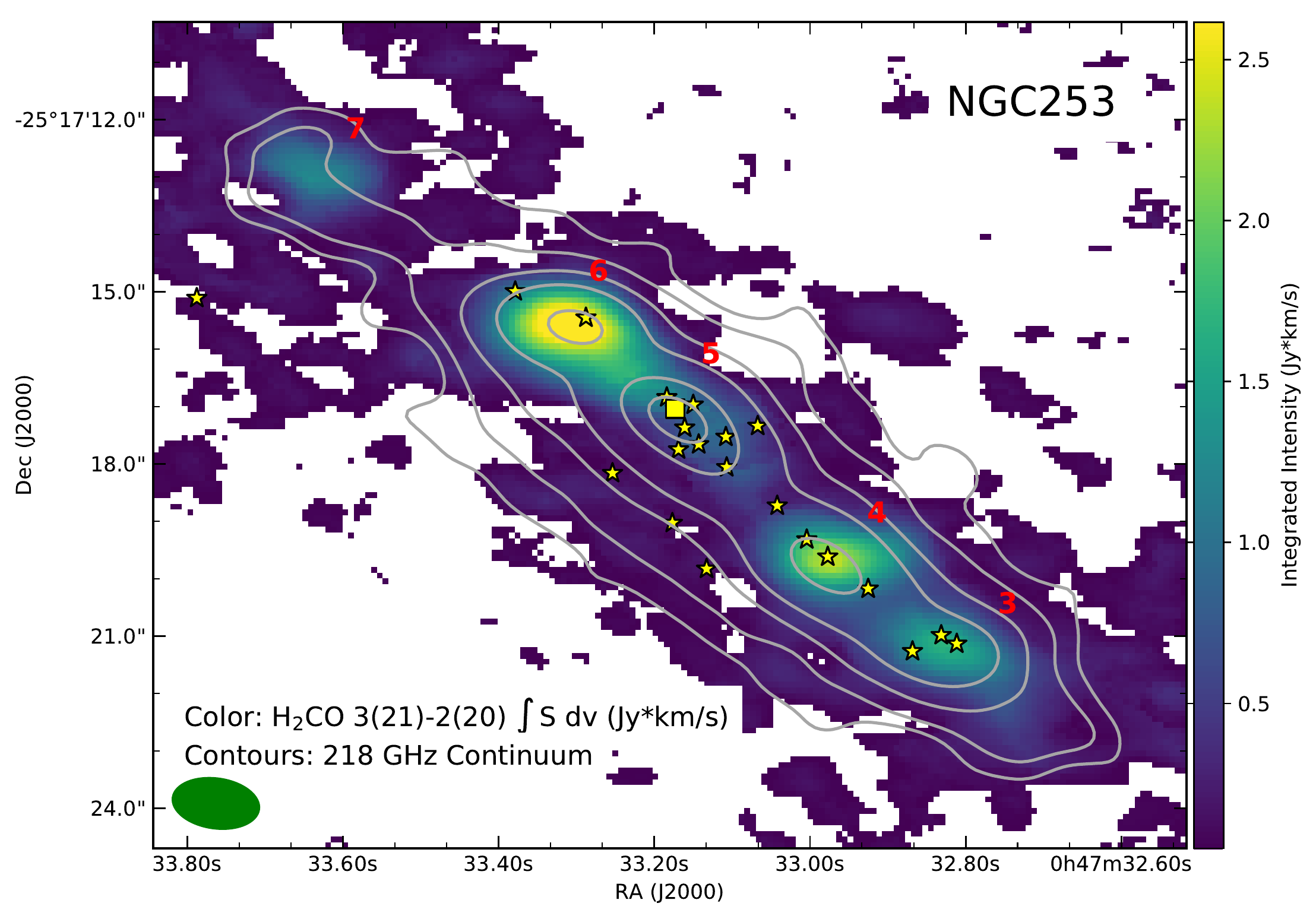}
\caption{Para H$_2$CO $3_{03}-2_{02}$ (top) and $3_{21}-2_{20}$ (bottom) integrated intensity.}
\label{fig:IntegIntpH2COlow}
\end{figure}

\begin{figure}
\centering
\includegraphics[trim=5mm 0mm 0mm 0mm, scale=0.32]{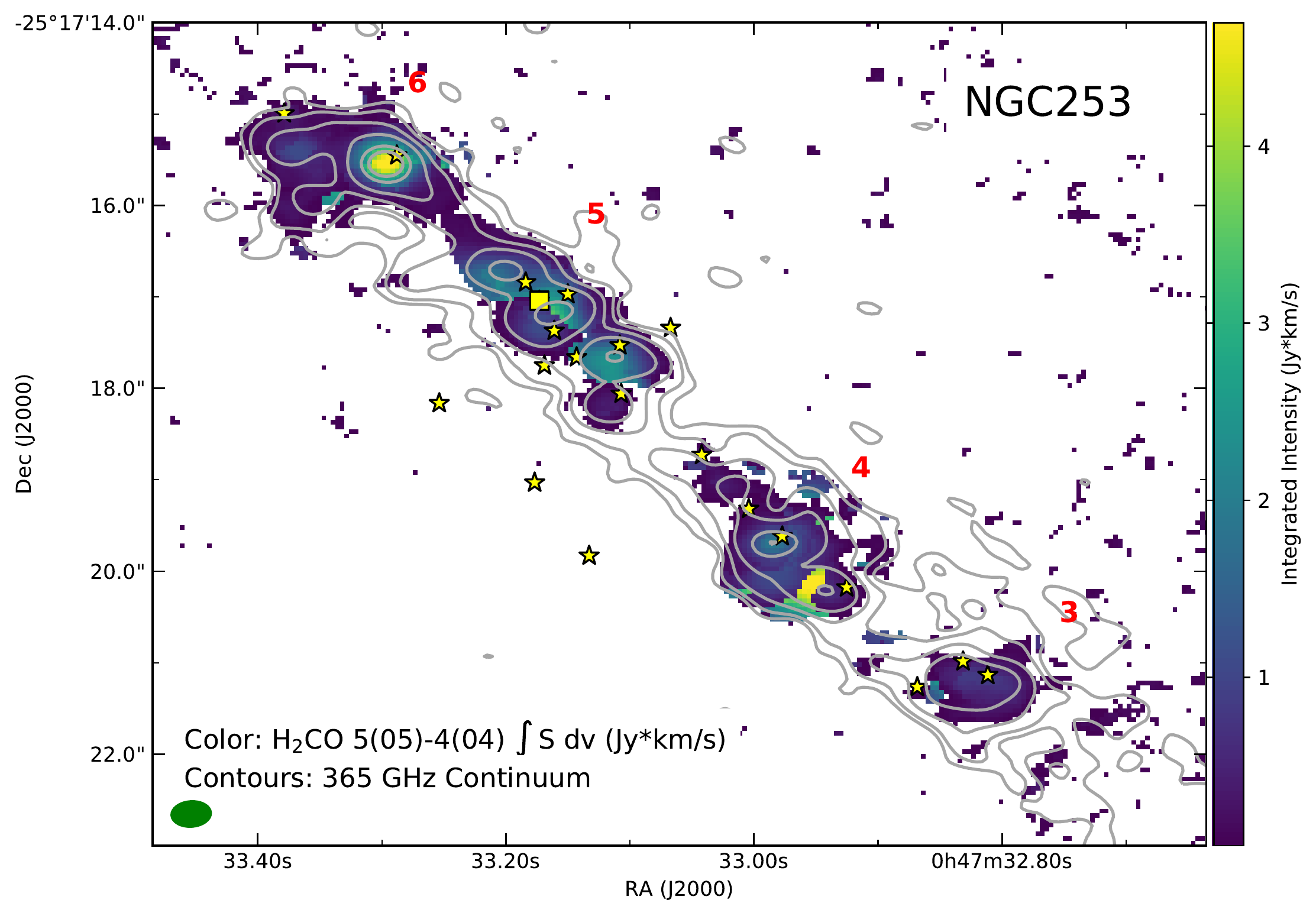}
\includegraphics[trim=5mm 0mm 0mm 0mm, scale=0.32]{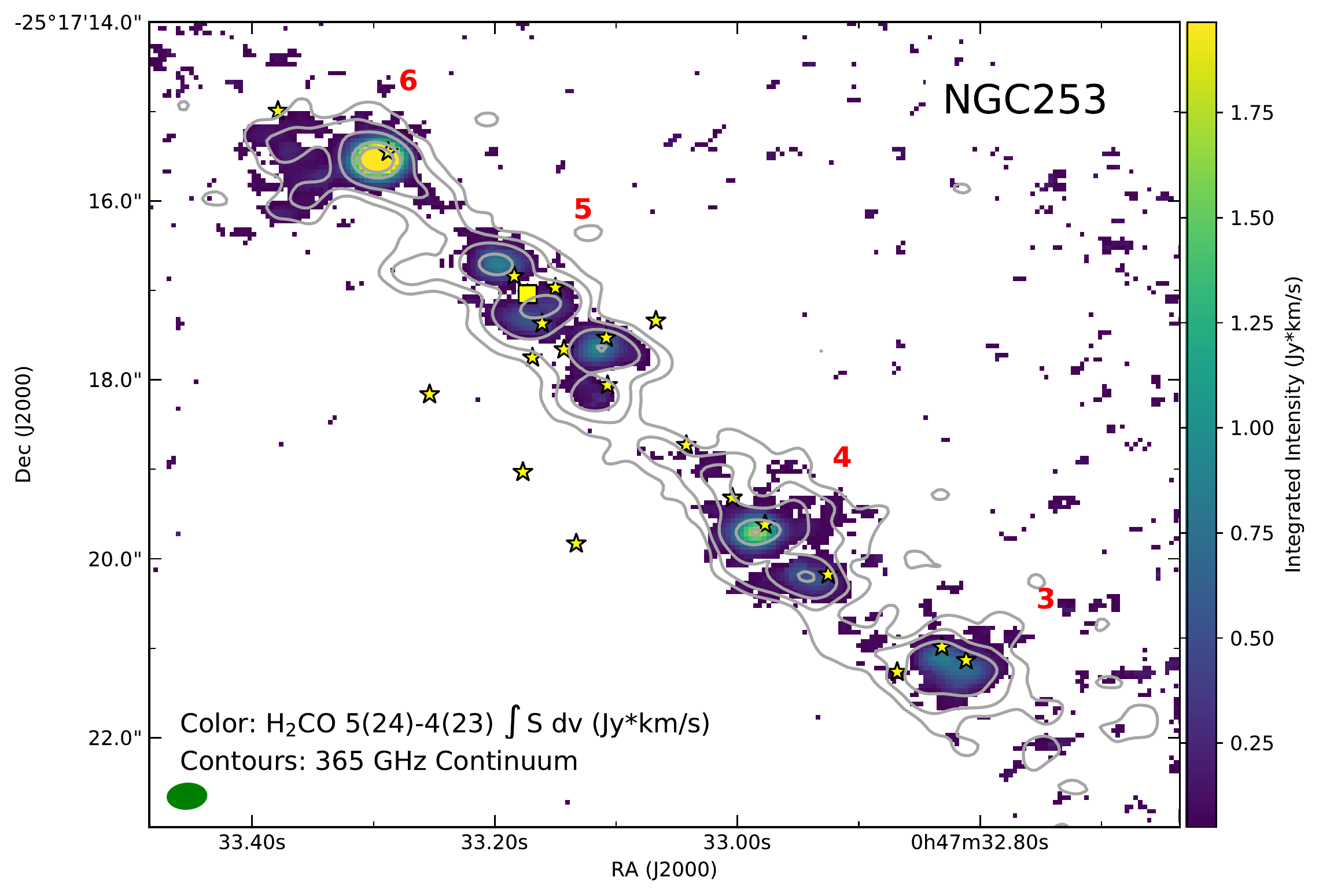}\\
\includegraphics[trim=5mm 0mm 0mm 0mm, scale=0.32]{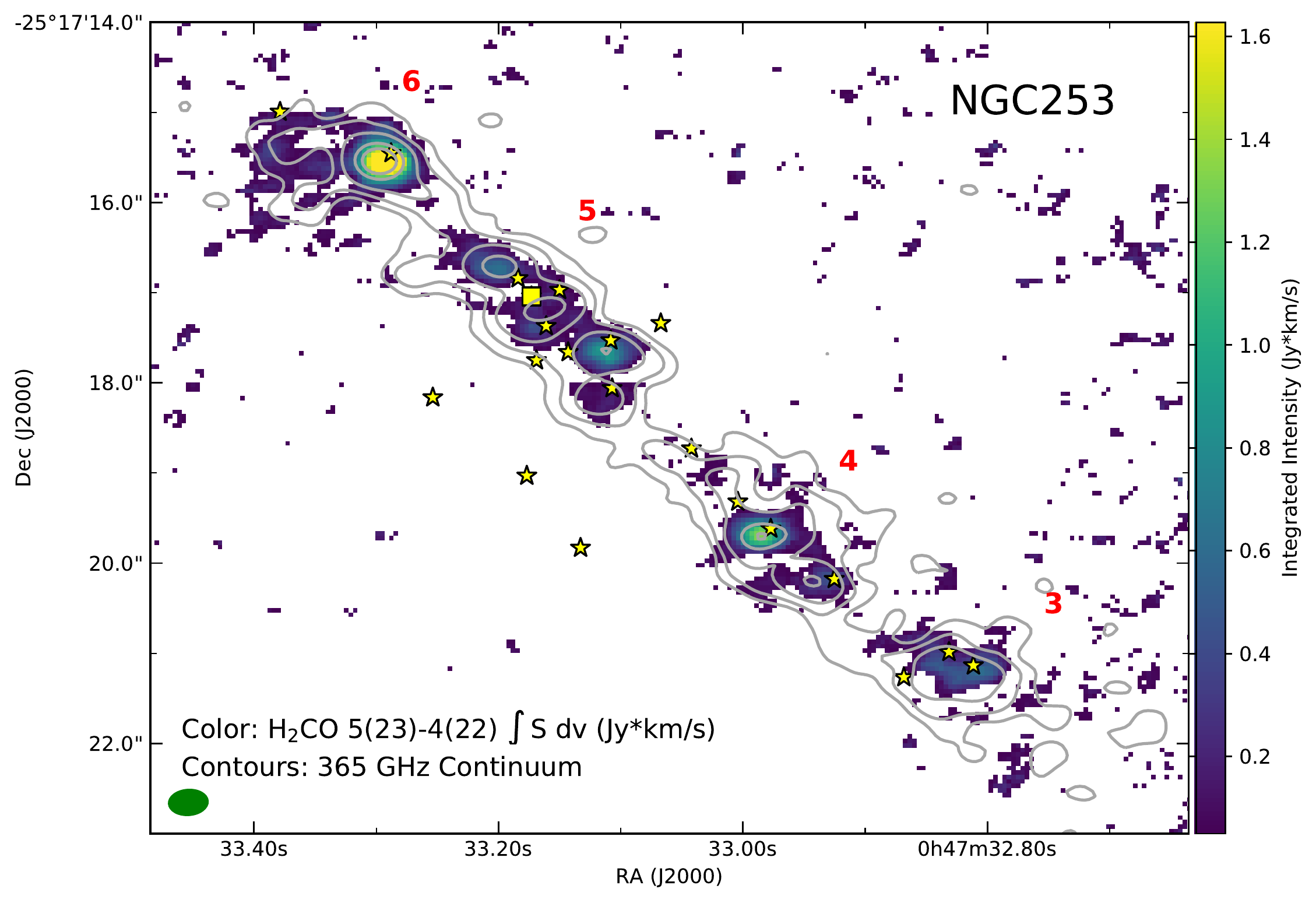}
\includegraphics[trim=5mm 0mm 0mm 0mm, scale=0.32]{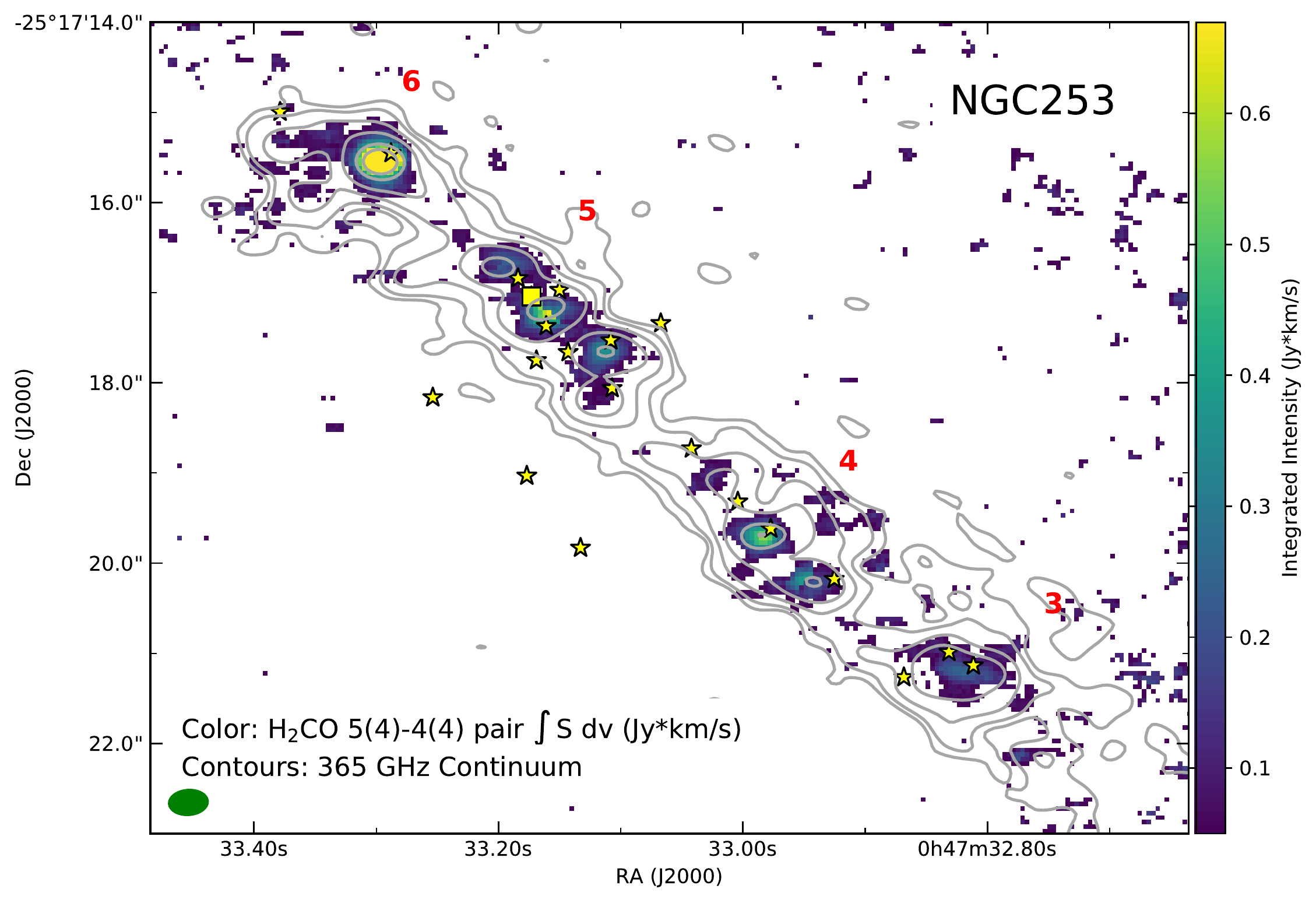}
\caption{Para H$_2$CO $5_{05}-4_{04}$ (top left), $5_{24}-4_{23}$ (top right),
  $5_{23}-4_{22}$ (bottom left), and $5_{4}-4_{4}$ (bottom right) integrated intensity.}
\label{fig:IntegIntpH2COhigh}
\end{figure}

\begin{figure}
\centering
\includegraphics[trim=20mm 0mm 20mm 0mm, scale=0.65]{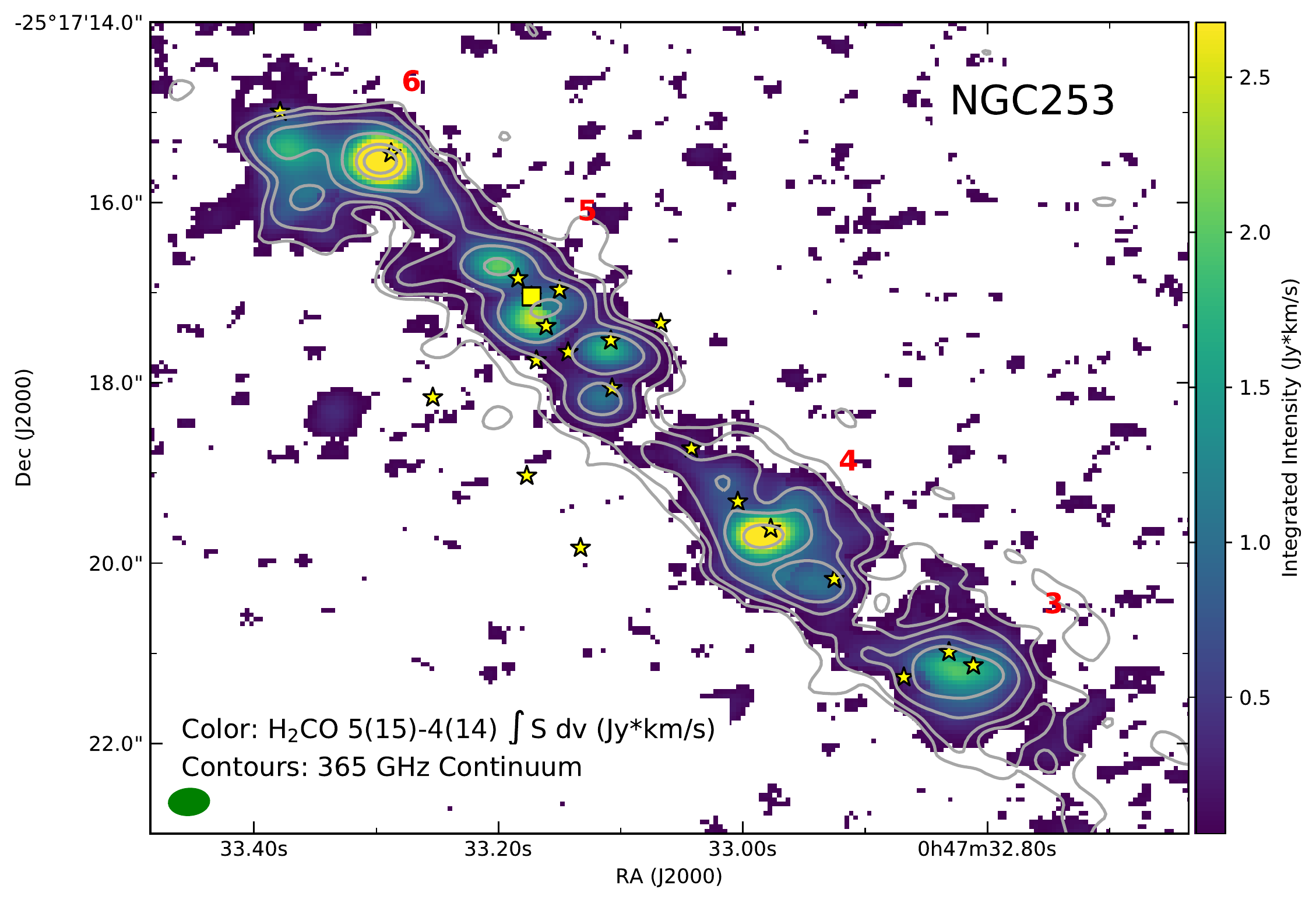}\\
\includegraphics[trim=20mm 0mm 20mm 0mm, scale=0.65]{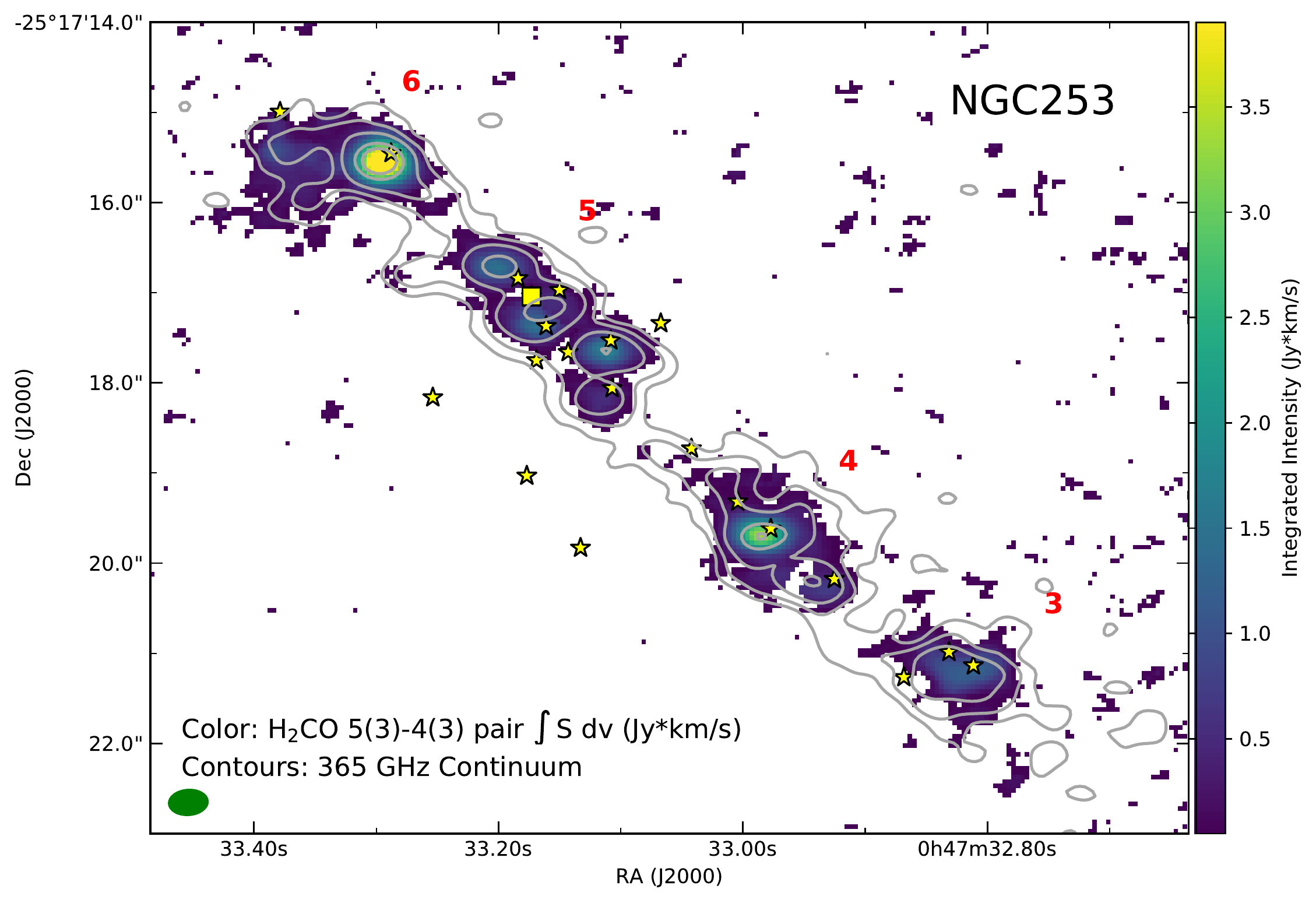}
\caption{Ortho H$_2$CO $5_{15}-4_{14}$ (top) and $5_{3}-4_{3}$ (bottom) integrated intensity.}
\label{fig:IntegIntoH2CO}
\end{figure}

\begin{figure}
\centering
\includegraphics[trim=20mm 0mm 20mm 0mm, scale=0.65]{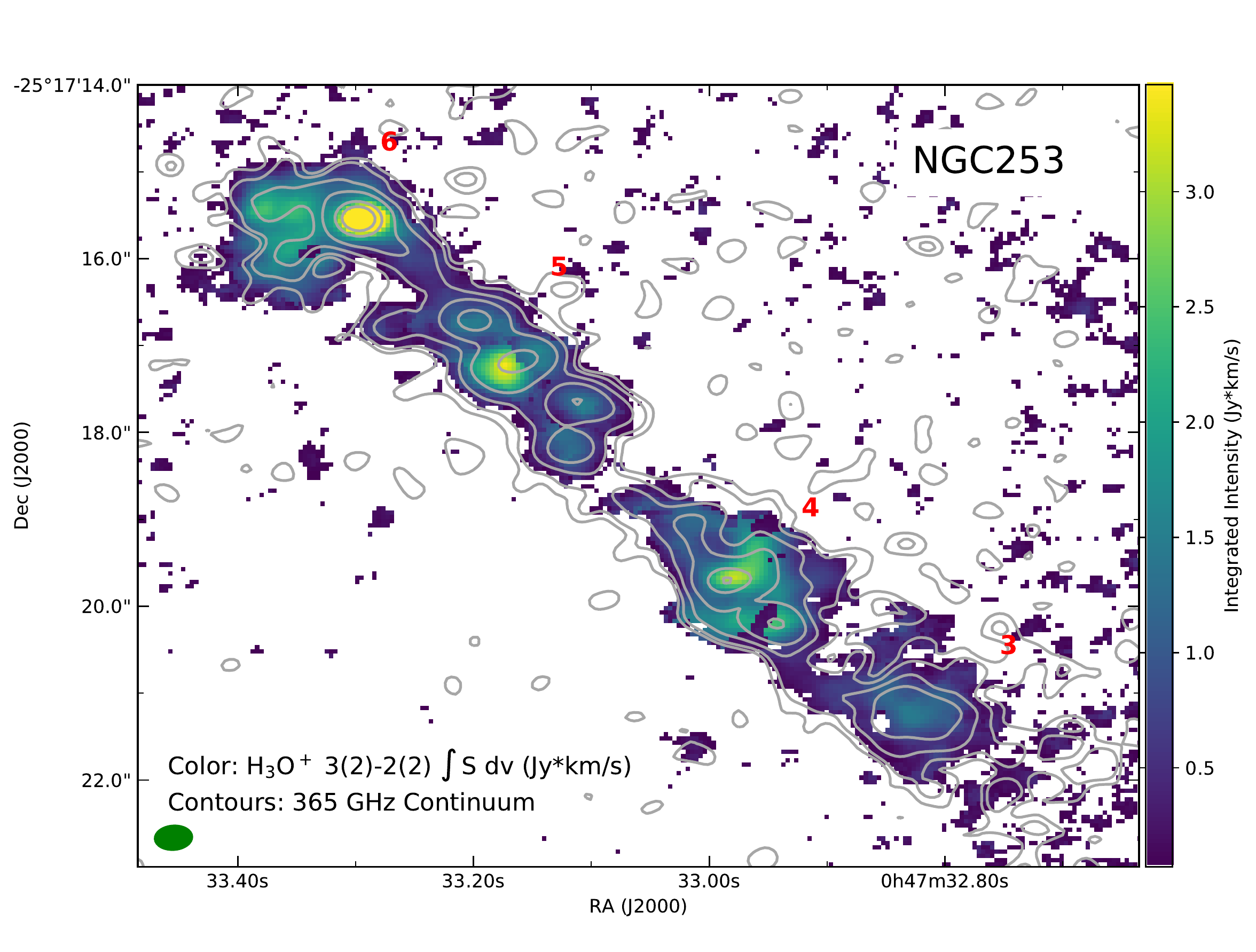}
\caption{H$_3$O$^+$ $3_{2}-2_{2}$ integrated intensity.}
\label{fig:IntegIntH3Op}
\end{figure}

\begin{figure}
\centering
\includegraphics[trim=20mm 0mm 20mm 0mm, scale=0.45]{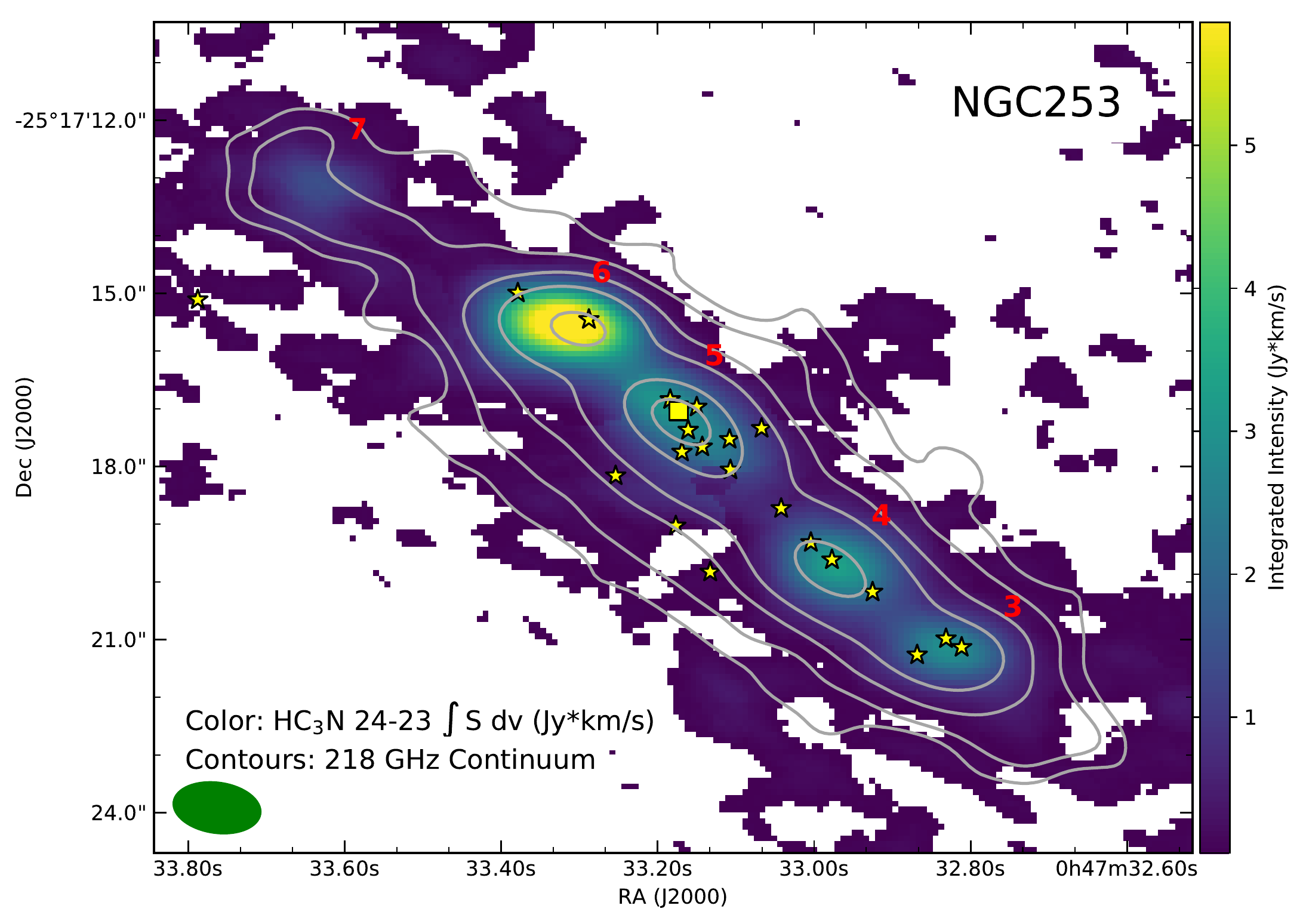}\\
\includegraphics[trim=20mm 0mm 20mm 0mm, scale=0.45]{NGC253-HC3N2423v7Int.pdf}\\
\includegraphics[trim=20mm 0mm 20mm 0mm, scale=0.43]{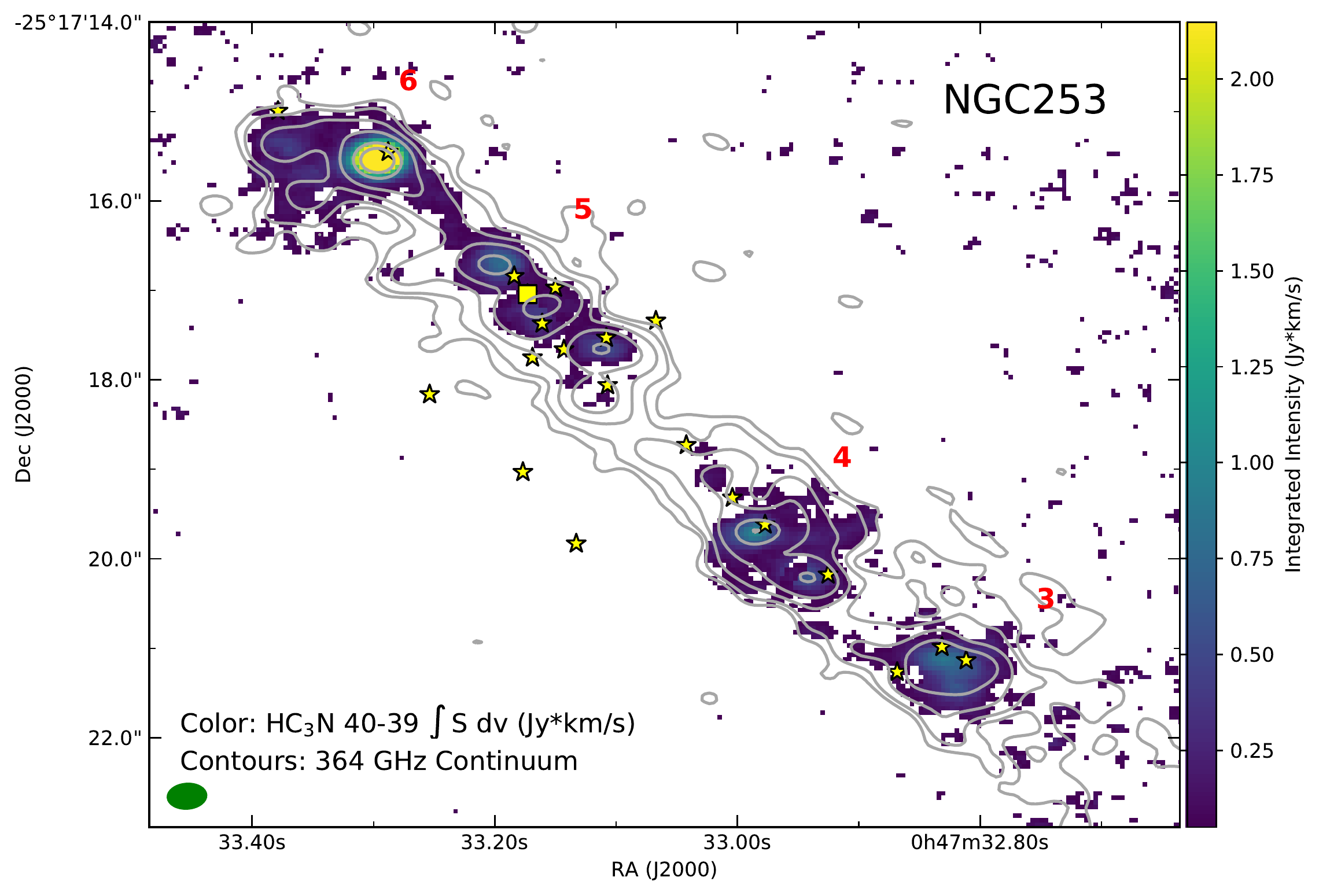}
\caption{HC$_3$N $24-23$ (top), $24-23 \nu_7 = 2$ (middle), and $40-39$ (bottom) integrated intensity.}
\label{fig:IntegIntHC3N}
\end{figure}

\begin{figure}
\centering
\includegraphics[trim=20mm 0mm 20mm 0mm, scale=0.65]{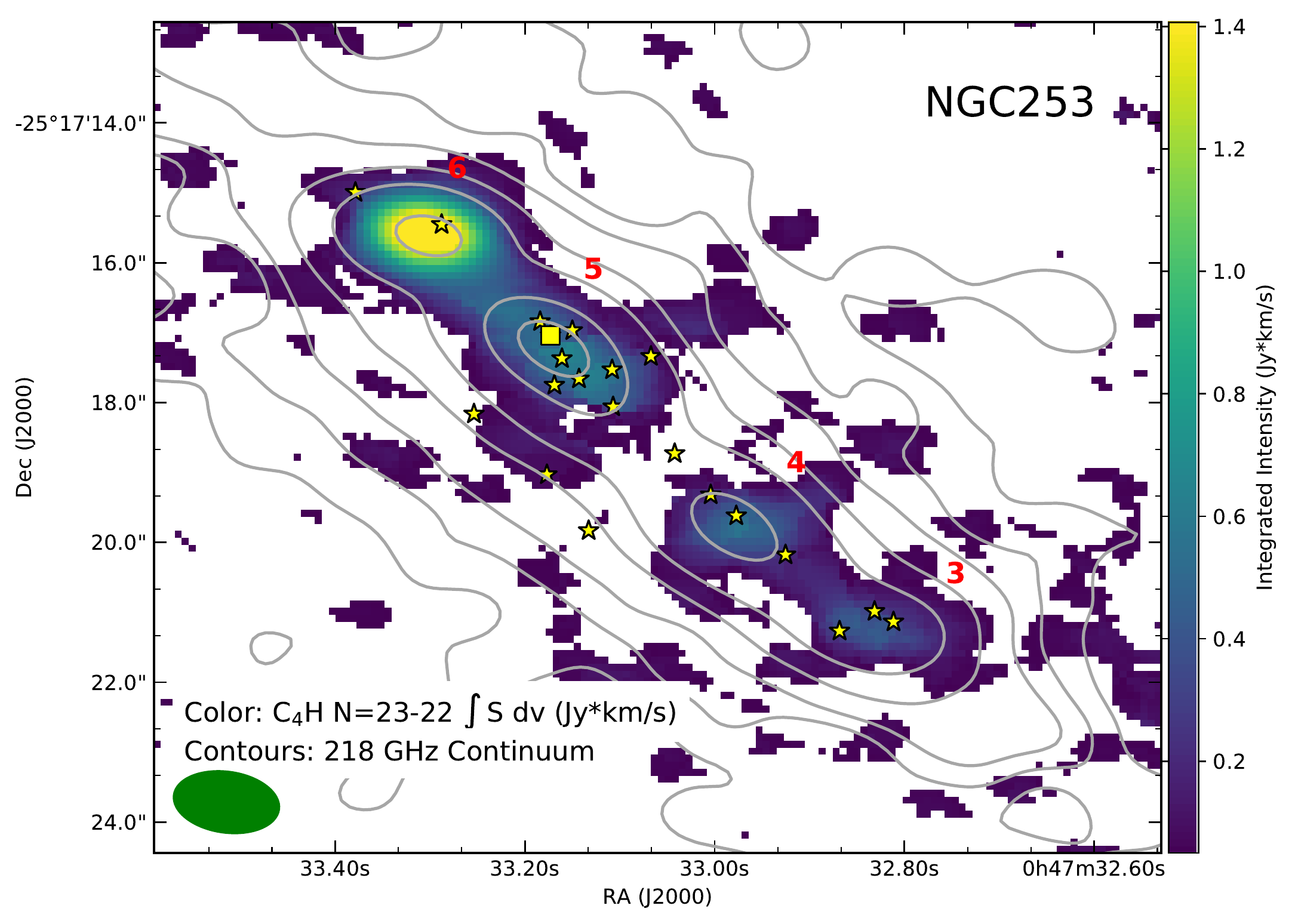}
\caption{C$_4$H N=$23-22$ integrated intensity.}
\label{fig:IntegIntC4H}
\end{figure}

\begin{figure}
\centering
\includegraphics[trim=20mm 0mm 20mm 0mm, scale=0.65]{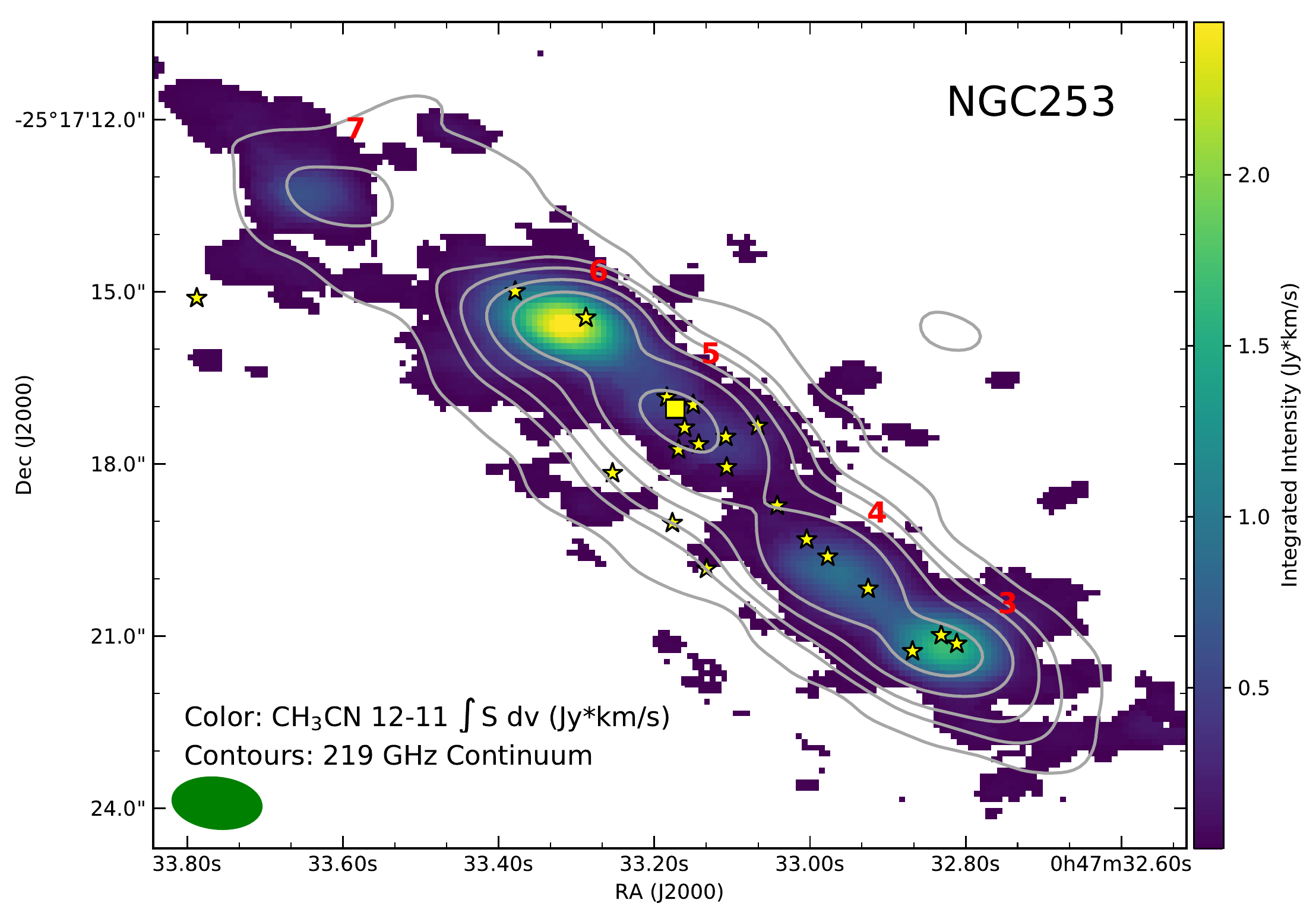}
\caption{CH$_3$CN $12-11$ integrated intensity.}
\label{fig:IntegIntCH3CN}
\end{figure}

\begin{figure}
\centering
\includegraphics[trim=20mm 0mm 20mm 0mm, scale=0.63]{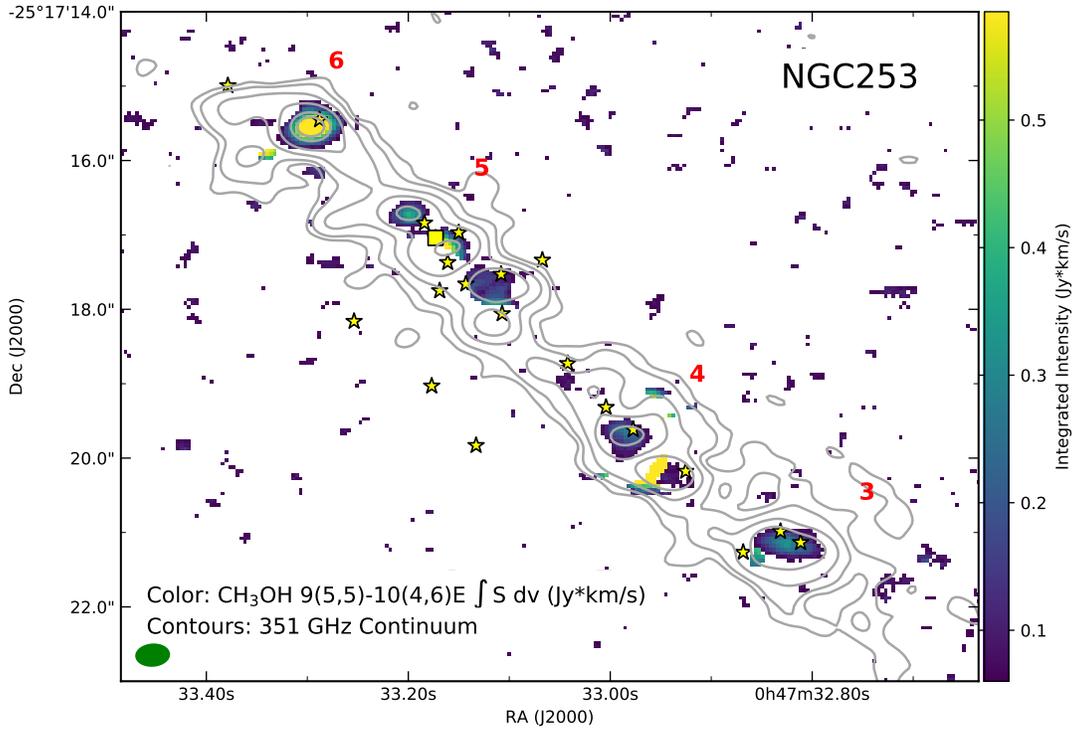}
\caption{CH$_3$OH $9_{5}-10_{4}E$ integrated intensity.}
\label{fig:IntegIntCH3OH}
\end{figure}

\begin{figure}
\centering
\includegraphics[trim=5mm 0mm 0mm 0mm, scale=0.30]{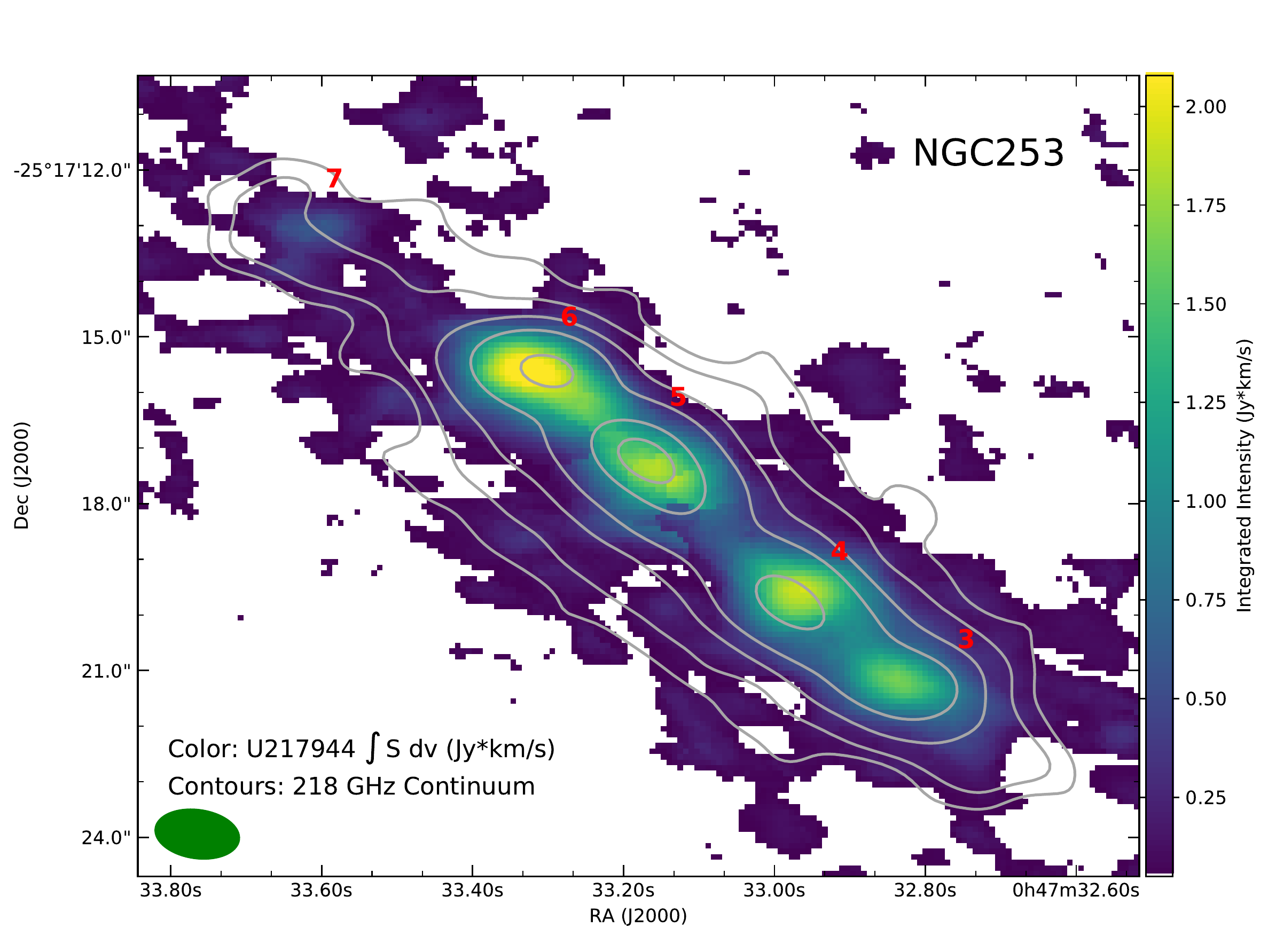}
\includegraphics[trim=5mm 0mm 0mm 0mm, scale=0.32]{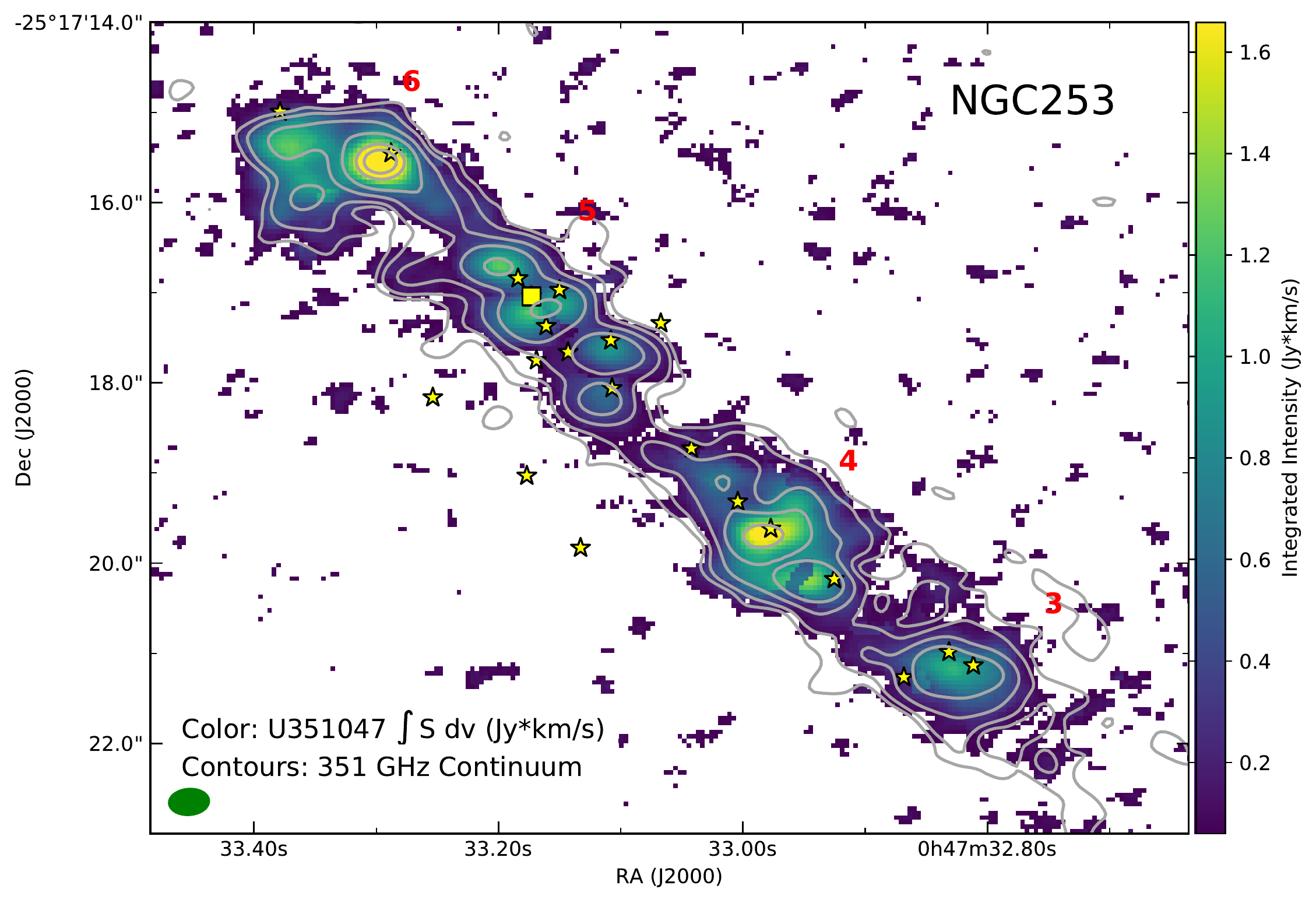}\\
\includegraphics[trim=5mm 0mm 0mm 0mm, scale=0.32]{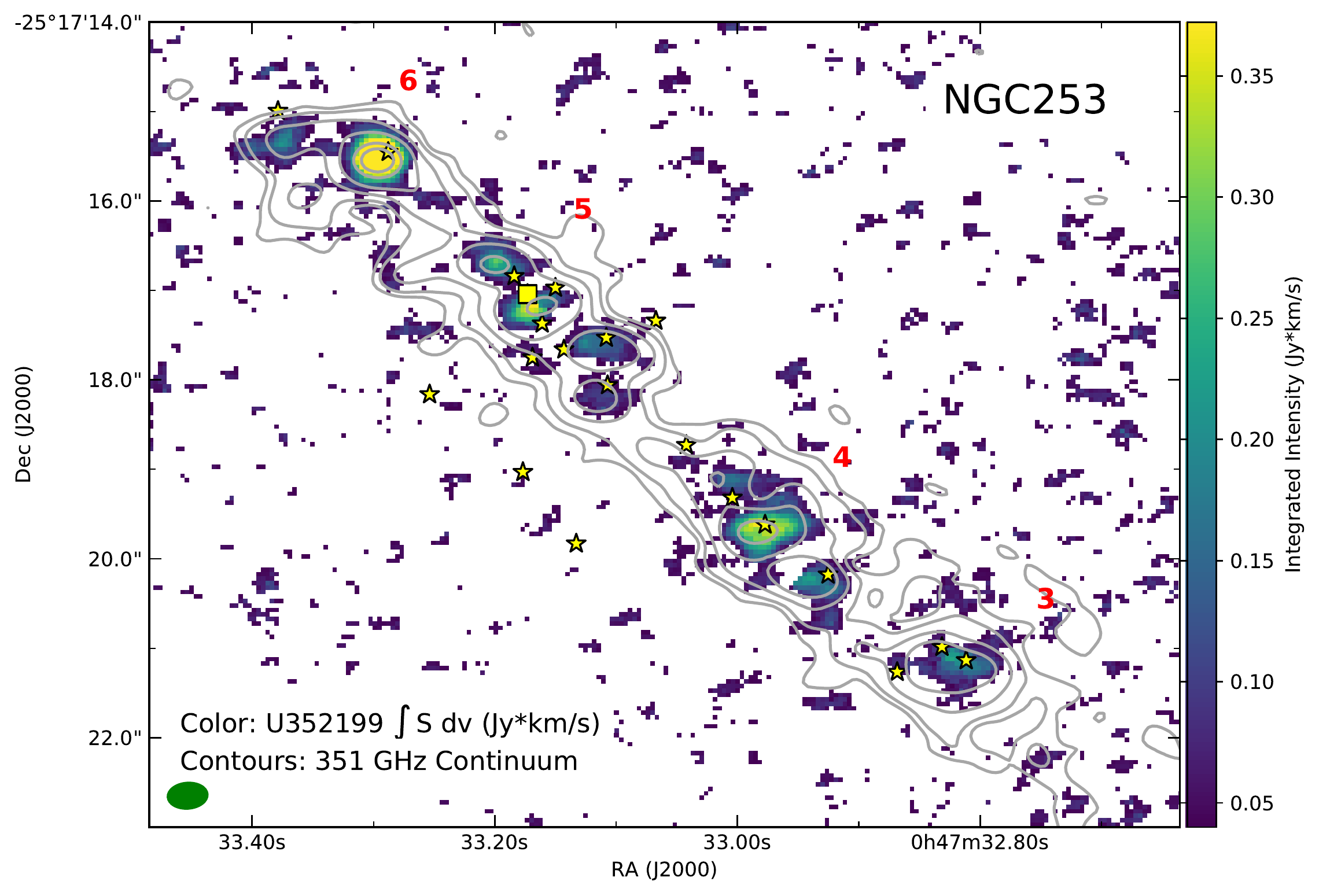}
\includegraphics[trim=5mm 0mm 0mm 0mm, scale=0.32]{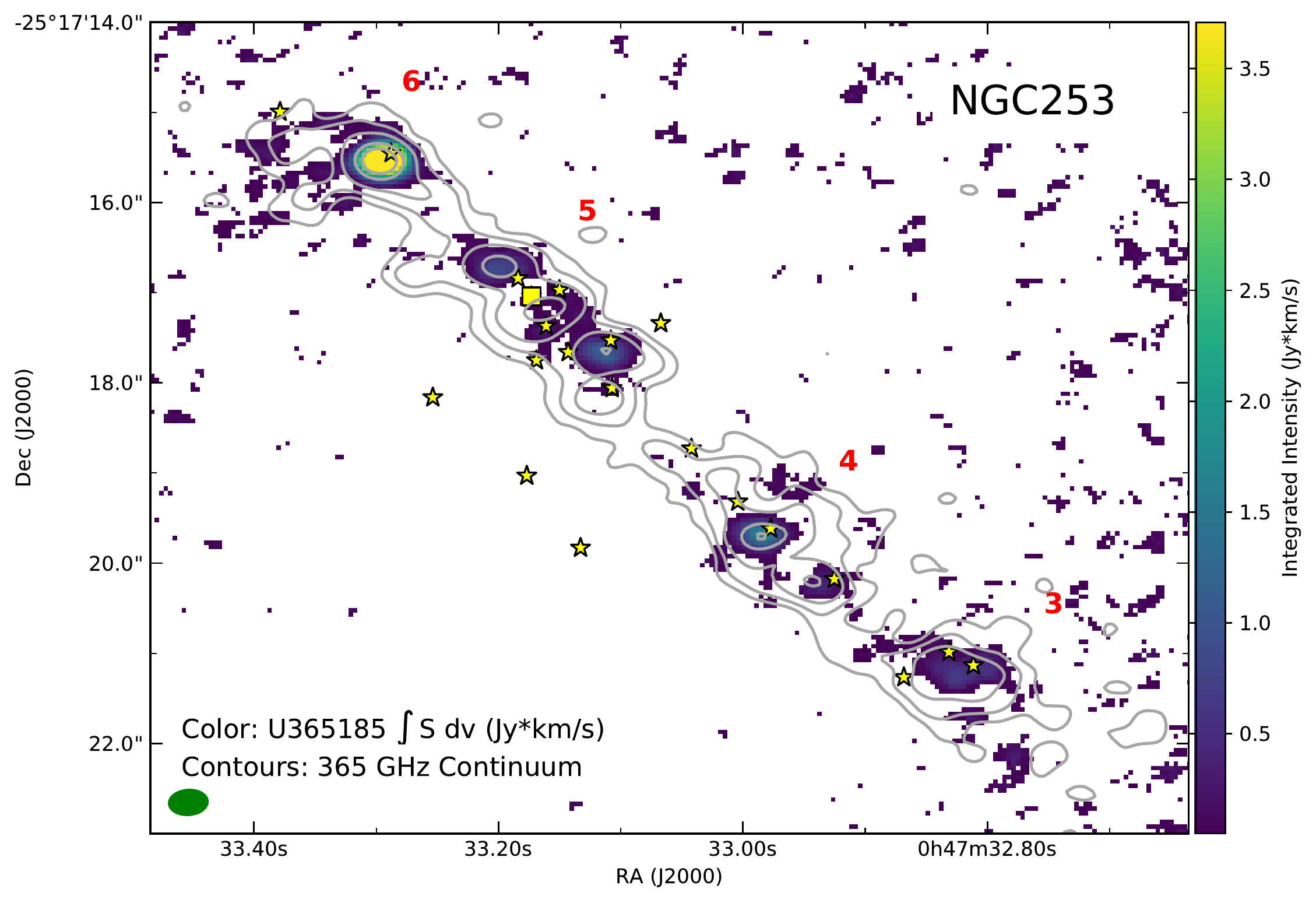}
\caption{U217944 (top left), U351047 (top right), U352199 (bottom left),
  and U365185 (bottom right) integrated intensity.}
\label{fig:IntegIntUnid}
\end{figure}

Figures~\ref{fig:IntegIntCO} through \ref{fig:IntegIntUnid} show the
spectral line integrated intensities derived from our ALMA Band 6 and
7 imaging of NGC\,253.  As was shown in the other integrated intensity
figures in this article, red numbers indicate the locations of the
dense molecular transition regions identified by
\citet[][Table~\ref{tab:ContPositions}]{Leroy2015}, while yellow markers
locate the positions of the 2\,cm radio continuum emission peaks (with
a square indicating the position of the strongest radio continuum
peak) identified by \cite{Turner1985} (TH2: RA(J2000) = 00$^h$ 47$^m$
33$^s$.18, Dec(J2000) = $-25^\circ$ 17$^\prime$
16$^{\prime\prime}$.93).  Contour levels shown are 3, 5, 10, 20, 50,
and 75\,mJy/beam (218\,GHz continuum); 1, 3, 5, 10, 20, 50, and
75\,mJy/beam (219\,GHz continuum); 3, 5, 10, 20, 50, and 75\,mJy/beam (351\,GHz
continuum); and 5, 10, 20, 50, and 75\,mJy/beam (365\,GHz continuum).
Note that the spatial distribution is limited  
by the primary beam of our measurements for our $^{13}$CO and C$^{18}$O
$3-2$ integrated intensity images (Figure~\ref{fig:IntegIntCO}).  For
this reason we consider our measurements of 
these two integrated intensity distributions to not be a completely
accurate representation of the spatial distribution of the $^{13}$CO
and C$^{18}$O distributions in NGC\,253.

\bibliographystyle{apj}
\bibliography{NGC253H2COALMACycle2PaperFinal.bib}

\begin{thebibliography}{}
\expandafter\ifx\csname natexlab\endcsname\relax\def\natexlab#1{#1}\fi

\bibitem[{{Aalto} {et~al.}(2007){Aalto}, {Spaans}, {Wiedner}, \&
  {H{\"u}ttemeister}}]{Aalto2007}
{Aalto}, S., {Spaans}, M., {Wiedner}, M.~C., \& {H{\"u}ttemeister}, S. 2007,
  \aap, 464, 193

\bibitem[{{Acero} {et~al.}(2009){Acero}, {Aharonian}, {Akhperjanian}, {Anton},
  {Barres de Almeida}, {Bazer-Bachi}, {Becherini}, {Behera}, {Bernl{\"o}hr},
  {Bochow}, {Boisson}, {Bolmont}, {Borrel}, {Brucker}, {Brun}, {Brun},
  {B{\"u}hler}, {Bulik}, {B{\"u}sching}, {Boutelier}, {Chadwick},
  {Charbonnier}, {Chaves}, {Cheesebrough}, {Chounet}, {Clapson}, {Coignet},
  {Dalton}, {Daniel}, {Davids}, {Degrange}, {Deil}, {Dickinson},
  {Djannati-Ata{\"i}}, {Domainko}, {Drury}, {Dubois}, {Dubus}, {Dyks}, {Dyrda},
  {Egberts}, {Emmanoulopoulos}, {Espigat}, {Farnier}, {Fegan}, {Feinstein},
  {Fiasson}, {F{\"o}rster}, {Fontaine}, {F{\"u}{\ss}ling}, {Gabici}, {Gallant},
  {G{\'e}rard}, {Gerbig}, {Giebels}, {Glicenstein}, {Gl{\"u}ck}, {Goret},
  {G{\"o}ring}, {Hauser}, {Hauser}, {Heinz}, {Heinzelmann}, {Henri}, {Hermann},
  {Hinton}, {Hoffmann}, {Hofmann}, {Hofverberg}, {Hoppe}, {Horns},
  {Jacholkowska}, {de Jager}, {Jahn}, {Jung}, {Katarzy{\'n}ski}, {Katz},
  {Kaufmann}, {Kerschhaggl}, {Khangulyan}, {Kh{\'e}lifi}, {Keogh}, {Klochkov},
  {Klu{\'z}niak}, {Kneiske}, {Komin}, {Kosack}, {Kossakowski}, {Lamanna},
  {Lenain}, {Lohse}, {Marandon}, {Martineau-Huynh}, {Marcowith}, {Masbou},
  {Maurin}, {McComb}, {Medina}, {M{\'e}hault}, {Moderski}, {Moulin},
  {Naumann-Godo}, {de Naurois}, {Nedbal}, {Nekrassov}, {Nicholas}, {Niemiec},
  {Nolan}, {Ohm}, {Olive}, {Wilhelmi}, {Orford}, {Ostrowski}, {Panter},
  {Arribas}, {Pedaletti}, {Pelletier}, {Petrucci}, {Pita}, {P{\"u}hlhofer},
  {Punch}, {Quirrenbach}, {Raubenheimer}, {Raue}, {Rayner}, {Reimer}, {Renaud},
  {Rieger}, {Ripken}, {Rob}, {Rosier-Lees}, {Rowell}, {Rudak}, {Rulten},
  {Ruppel}, {Sahakian}, {Santangelo}, {Schlickeiser}, {Sch{\"o}ck}, {Schwanke},
  {Schwarzburg}, {Schwemmer}, {Shalchi}, {Sikora}, {Skilton}, {Sol}, {Stawarz},
  {Steenkamp}, {Stegmann}, {Stinzing}, {Superina}, {Szostek}, {Tam},
  {Tavernet}, {Terrier}, {Tibolla}, {Tluczykont}, {van Eldik}, {Vasileiadis},
  {Venter}, {Venter}, {Vialle}, {Vincent}, {Vivier}, {V{\"o}lk}, {Volpe},
  {Wagner}, {Ward}, {Zdziarski}, \& {Zech}}]{Acero2009}
{Acero}, F., {Aharonian}, F., {Akhperjanian}, A.~G., {et~al.} 2009, Science,
  326, 1080

\bibitem[{{Aladro} {et~al.}(2015){Aladro}, {Mart{\'\i}n}, {Riquelme}, {Henkel},
  {Mauersberger}, {Mart{\'\i}n-Pintado}, {Wei{\ss}}, {Lefevre}, {Kramer},
  {Requena- Torres}, \& {Armijos-Abenda{\~n}o}}]{Aladro2015}
{Aladro}, R., {Mart{\'\i}n}, S., {Riquelme}, D., {et~al.} 2015, \aap, 579, A101

\bibitem[{{Ando} {et~al.}(2017){Ando}, {Nakanishi}, {Kohno}, {Izumi},
  {Mart{\'{\i}}n}, {Harada}, {Takano}, {Kuno}, {Nakai}, {Sugai}, {Sorai},
  {Tosaki}, {Matsubayashi}, {Nakajima}, {Nishimura}, \& {Tamura}}]{Ando2017}
{Ando}, R., {Nakanishi}, K., {Kohno}, K., {et~al.} 2017, \apj, 849, 81

\bibitem[{{Ao} {et~al.}(2013){Ao}, {Henkel}, {Menten}, {Requena-Torres},
  {Stanke}, {Mauersberger}, {Aalto}, {M{\"u}hle}, \& {Mangum}}]{Ao2013}
{Ao}, Y., {Henkel}, C., {Menten}, K.~M., {et~al.} 2013, \aap, 550, A135

\bibitem[{{Astropy Collaboration} {et~al.}(2018){Astropy Collaboration},
  {Price-Whelan}, {Sip{\'{o}}cz}, {G{\"u}nther}, {Lim}, {Crawford}, {Conseil},
  {Shupe}, {Craig}, {Dencheva}, {Ginsburg}, {VanderPlas}, {Bradley},
  {P{\'e}rez-Su{\'a}rez}, {de Val- Borro}, {Aldcroft}, {Cruz}, {Robitaille},
  {Tollerud}, {Ardelean}, {Babej}, {Bach}, {Bachetti}, {Bakanov}, {Bamford},
  {Barentsen}, {Barmby}, {Baumbach}, {Berry}, {Biscani}, {Boquien}, {Bostroem},
  {Bouma}, {Brammer}, {Bray}, {Breytenbach}, {Buddelmeijer}, {Burke},
  {Calderone}, {Cano Rodr{\'\i}guez}, {Cara}, {Cardoso}, {Cheedella}, {Copin},
  {Corrales}, {Crichton}, {D'Avella}, {Deil}, {Depagne}, {Dietrich}, {Donath},
  {Droettboom}, {Earl}, {Erben}, {Fabbro}, {Ferreira}, {Finethy}, {Fox},
  {Garrison}, {Gibbons}, {Goldstein}, {Gommers}, {Greco}, {Greenfield},
  {Groener}, {Grollier}, {Hagen}, {Hirst}, {Homeier}, {Horton}, {Hosseinzadeh},
  {Hu}, {Hunkeler}, {Ivezi{\'c}}, {Jain}, {Jenness}, {Kanarek}, {Kendrew},
  {Kern}, {Kerzendorf}, {Khvalko}, {King}, {Kirkby}, {Kulkarni}, {Kumar},
  {Lee}, {Lenz}, {Littlefair}, {Ma}, {Macleod}, {Mastropietro}, {McCully},
  {Montagnac}, {Morris}, {Mueller}, {Mumford}, {Muna}, {Murphy}, {Nelson},
  {Nguyen}, {Ninan}, {N{\"o}the}, {Ogaz}, {Oh}, {Parejko}, {Parley}, {Pascual},
  {Patil}, {Patil}, {Plunkett}, {Prochaska}, {Rastogi}, {Reddy Janga},
  {Sabater}, {Sakurikar}, {Seifert}, {Sherbert}, {Sherwood-Taylor}, {Shih},
  {Sick}, {Silbiger}, {Singanamalla}, {Singer}, {Sladen}, {Sooley},
  {Sornarajah}, {Streicher}, {Teuben}, {Thomas}, {Tremblay}, {Turner},
  {Terr{\'o}n}, {van Kerkwijk}, {de la Vega}, {Watkins}, {Weaver}, {Whitmore},
  {Woillez}, {Zabalza}, \& {Astropy Contributors}}]{astropy2018}
{Astropy Collaboration}, {Price-Whelan}, A.~M., {Sip{\'{o}}cz}, B.~M., {et~al.}
  2018, \aj, 156, 123

\bibitem[{{Bayet} {et~al.}(2011){Bayet}, {Williams}, {Hartquist}, \&
  {Viti}}]{Bayet2011}
{Bayet}, E., {Williams}, D.~A., {Hartquist}, T.~W., \& {Viti}, S. 2011, \mnras,
  414, 1583

\bibitem[{{Brunthaler} {et~al.}(2009){Brunthaler}, {Castangia}, {Tarchi},
  {Henkel}, {Reid}, {Falcke}, \& {Menten}}]{Brunthaler2009}
{Brunthaler}, A., {Castangia}, P., {Tarchi}, A., {et~al.} 2009, \aap, 497, 103

\bibitem[{{Carroll} \& {Goldsmith}(1981)}]{Carroll1981}
{Carroll}, T.~J., \& {Goldsmith}, P.~F. 1981, \apj, 245, 891

\bibitem[{{Chen} {et~al.}(2018){Chen}, {Ellingsen}, {Shen}, {McCarthy},
  {Zhong}, \& {Deng}}]{Chen2018}
{Chen}, X., {Ellingsen}, S.~P., {Shen}, Z.-Q., {et~al.} 2018, \apjl, 856, L35

\bibitem[{{Condon}(1992)}]{Condon1992}
{Condon}, J.~J. 1992, \araa, 30, 575

\bibitem[{{Costagliola} {et~al.}(2013){Costagliola}, {Aalto}, {Sakamoto},
  {Mart{\'\i}n}, {Beswick}, {Muller}, \& {Kl{\"o}ckner}}]{Costagliola2013}
{Costagliola}, F., {Aalto}, S., {Sakamoto}, K., {et~al.} 2013, \aap, 556, A66

\bibitem[{{Costagliola} {et~al.}(2015){Costagliola}, {Sakamoto}, {Muller},
  {Mart{\'\i}n}, {Aalto}, {Harada}, {van der Werf}, {Viti}, {Garcia-Burillo},
  \& {Spaans}}]{Costagliola2015}
{Costagliola}, F., {Sakamoto}, K., {Muller}, S., {et~al.} 2015, \aap, 582, A91

\bibitem[{{Cotton}(2017)}]{Cotton2017}
{Cotton}, W.~D. 2017, Publications of the Astronomical Society of the Pacific,
  129, 094501

\bibitem[{{Draine}(2011)}]{Draine2011}
{Draine}, B.~T. 2011, {Physics of the Interstellar and Intergalactic Medium}
  (Physics of the Interstellar and Intergalactic Medium by Bruce
  T.~Draine.~Princeton University Press, 2011.~ISBN: 978-0-691-12214-4)

\bibitem[{{Ellingsen} {et~al.}(2017){Ellingsen}, {Chen}, {Breen}, \&
  {Qiao}}]{Ellingsen2017}
{Ellingsen}, S.~P., {Chen}, X., {Breen}, S.~L., \& {Qiao}, H.-h. 2017, \apjl,
  841, L14

\bibitem[{{Ellingsen} {et~al.}(2014){Ellingsen}, {Chen}, {Qiao}, {Baan}, {An},
  {Li}, \& {Breen}}]{Ellingsen2014}
{Ellingsen}, S.~P., {Chen}, X., {Qiao}, H.-H., {et~al.} 2014, \apjl, 790, L28

\bibitem[{{Fern{\'a}ndez-Ontiveros} {et~al.}(2009){Fern{\'a}ndez-Ontiveros},
  {Prieto}, \& {Acosta-Pulido}}]{FernandezOntiveros2009}
{Fern{\'a}ndez-Ontiveros}, J.~A., {Prieto}, M.~A., \& {Acosta-Pulido}, J.~A.
  2009, \mnras, 392, L16

\bibitem[{{Ginsburg} \& {Mirocha}(2011)}]{Ginsburg2011b}
{Ginsburg}, A., \& {Mirocha}, J. 2011, {PySpecKit: Python Spectroscopic
  Toolkit}, Astrophysics Source Code Library, ascl:1109.001

\bibitem[{{Ginsburg} {et~al.}(2016){Ginsburg}, {Henkel}, {Ao}, {Riquelme},
  {Kauffmann}, {Pillai}, {Mills}, {Requena-Torres}, {Immer}, {Testi}, {Ott},
  {Bally}, {Battersby}, {Darling}, {Aalto}, {Stanke}, {Kendrew}, {Kruijssen},
  {Longmore}, {Dale}, {Guesten}, \& {Menten}}]{Ginsburg2016}
{Ginsburg}, A., {Henkel}, C., {Ao}, Y., {et~al.} 2016, \aap, 586, A50

\bibitem[{{Gorski} {et~al.}(2017){Gorski}, {Ott}, {Rand}, {Meier}, {Momjian},
  \& {Schinnerer}}]{Gorski2017}
{Gorski}, M., {Ott}, J., {Rand}, R., {et~al.} 2017, \apj, 842, 124

\bibitem[{{Gorski} {et~al.}(2019){Gorski}, {Ott}, {Rand}, {Meier}, {Momjian},
  {Schinnerer}, \& {Ellingsen}}]{Gorski2019}
{Gorski}, M.~D., {Ott}, J., {Rand}, R., {et~al.} 2019, \mnras, 483, 5434

\bibitem[{{Green}(1991)}]{Green1991}
{Green}, S. 1991, \apjs, 76, 979

\bibitem[{{G{\"u}nthardt} {et~al.}(2015){G{\"u}nthardt}, {Ag{\"u}ero},
  {Camperi}, {D{\'{\i}}az}, {Gomez}, {Bosch}, \& {Schirmer}}]{Gunthardt2015}
{G{\"u}nthardt}, G.~I., {Ag{\"u}ero}, M.~P., {Camperi}, J.~A., {et~al.} 2015,
  \aj, 150, 139

\bibitem[{{Hildebrand}(1983)}]{Hildebrand1983}
{Hildebrand}, R.~H. 1983, \qjras, 24, 267

\bibitem[{{Imanishi} {et~al.}(2016){Imanishi}, {Nakanishi}, \&
  {Izumi}}]{Imanishi2016}
{Imanishi}, M., {Nakanishi}, K., \& {Izumi}, T. 2016, \apj, 825, 44

\bibitem[{{Jenkins}(2009)}]{Jenkins2009}
{Jenkins}, E.~B. 2009, \apj, 700, 1299

\bibitem[{{Kalari} {et~al.}(2018){Kalari}, {Rubio}, {Elmegreen}, {Guzm{\'a}n},
  {Zinnecker}, \& {Herrera}}]{Kalari2018}
{Kalari}, V.~M., {Rubio}, M., {Elmegreen}, B.~G., {et~al.} 2018, \apj, 852, 71

\bibitem[{{Leroy} {et~al.}(2015){Leroy}, {Bolatto}, {Ostriker}, {Rosolowsky},
  {Walter}, {Warren}, {Donovan Meyer}, {Hodge}, {Meier}, {Ott}, {Sandstrom},
  {Schruba}, {Veilleux}, \& {Zwaan}}]{Leroy2015}
{Leroy}, A.~K., {Bolatto}, A.~D., {Ostriker}, E.~C., {et~al.} 2015, \apj, 801,
  25

\bibitem[{{Leroy} {et~al.}(2018){Leroy}, {Bolatto}, {Ostriker}, {Walter},
  {Gorski}, {Ginsburg}, {Krieger}, {Levy}, {Meier}, {Mills}, {Ott},
  {Rosolowsky}, {Thompson}, {Veilleux}, \& {Zschaechner}}]{Leroy2018}
---. 2018, \apj, 869, 126

\bibitem[{{Lovas}(1985)}]{Lovas1985}
{Lovas}, F.~J. 1985, Journal of Physical and Chemical Reference Data, 14, 395

\bibitem[{{Lovas}(1992)}]{Lovas1992}
---. 1992, Journal of Physical and Chemical Reference Data, 21, 181

\bibitem[{{Mangum} {et~al.}(2013{\natexlab{a}}){Mangum}, {Darling}, {Henkel},
  \& {Menten}}]{Mangum2013}
{Mangum}, J.~G., {Darling}, J., {Henkel}, C., \& {Menten}, K.~M.
  2013{\natexlab{a}}, \apj, 766, 108

\bibitem[{{Mangum} {et~al.}(2013{\natexlab{b}}){Mangum}, {Darling}, {Henkel},
  {Menten}, {MacGregor}, {Svoboda}, \& {Schinnerer}}]{Mangum2013b}
{Mangum}, J.~G., {Darling}, J., {Henkel}, C., {et~al.} 2013{\natexlab{b}},
  \apj, 779, 33

\bibitem[{{Mangum} {et~al.}(2008){Mangum}, {Darling}, {Menten}, \&
  {Henkel}}]{Mangum2008}
{Mangum}, J.~G., {Darling}, J., {Menten}, K.~M., \& {Henkel}, C. 2008, \apj,
  673, 832

\bibitem[{{Mangum} \& {Shirley}(2015)}]{Mangum2015}
{Mangum}, J.~G., \& {Shirley}, Y.~L. 2015, \pasp, 127, 266

\bibitem[{{Mangum} \& {Wootten}(1993)}]{Mangum1993}
{Mangum}, J.~G., \& {Wootten}, A. 1993, \apjs, 89, 123

\bibitem[{{Mart{\'{\i}}n} {et~al.}(2006){Mart{\'{\i}}n}, {Mauersberger},
  {Mart{\'{\i}}n-Pintado}, {Henkel}, \& {Garc{\'{\i}}a-Burillo}}]{Martin2006}
{Mart{\'{\i}}n}, S., {Mauersberger}, R., {Mart{\'{\i}}n-Pintado}, J., {Henkel},
  C., \& {Garc{\'{\i}}a-Burillo}, S. 2006, \apjs, 164, 450

\bibitem[{{Mason} \& {Brogan}(2013)}]{Mason2013}
{Mason}, B.~S., \& {Brogan}, C. 2013, {Relative Integration Times for the ALMA
  Cycle 1 12-m, 7-m, and Total Power Arrays}, Tech. Rep. 598, National Radio
  Astronomy Observatory, Charlottesville, VA 22903, USA

\bibitem[{{McCormick} {et~al.}(2013){McCormick}, {Veilleux}, \&
  {Rupke}}]{McCormick2013}
{McCormick}, A., {Veilleux}, S., \& {Rupke}, D. S.~N. 2013, \apj, 774, 126

\bibitem[{{Meier} \& {Turner}(2012)}]{Meier2012}
{Meier}, D.~S., \& {Turner}, J.~L. 2012, \apj, 755, 104

\bibitem[{{Meier} {et~al.}(2015){Meier}, {Walter}, {Bolatto}, {Leroy}, {Ott},
  {Rosolowsky}, {Veilleux}, {Warren}, {Wei{\ss}}, {Zwaan}, \&
  {Zschaechner}}]{Meier2015}
{Meier}, D.~S., {Walter}, F., {Bolatto}, A.~D., {et~al.} 2015, \apj, 801, 63

\bibitem[{{Meijerink} {et~al.}(2011){Meijerink}, {Spaans}, {Loenen}, \& {van
  der Werf}}]{Meijerink2011}
{Meijerink}, R., {Spaans}, M., {Loenen}, A.~F., \& {van der Werf}, P.~P. 2011,
  \aap, 525, A119

\bibitem[{{Mills} \& {Morris}(2013)}]{Mills2013}
{Mills}, E.~A.~C., \& {Morris}, M.~R. 2013, \apj, 772, 105

\bibitem[{{M{\"u}ller} {et~al.}(2004){M{\"u}ller}, {Menten}, \&
  {M{\"a}der}}]{Mueller2004}
{M{\"u}ller}, H.~S.~P., {Menten}, K.~M., \& {M{\"a}der}, H. 2004, \aap, 428,
  1019

\bibitem[{{M{\"u}ller} {et~al.}(2001){M{\"u}ller}, {Thorwirth}, {Roth}, \&
  {Winnewisser}}]{Muller2001}
{M{\"u}ller}, H.~S.~P., {Thorwirth}, S., {Roth}, D.~A., \& {Winnewisser}, G.
  2001, \aap, 370, L49

\bibitem[{{M{\"u}ller} {et~al.}(2016){M{\"u}ller}, {Balog}, {Nielbock},
  {Moreno}, {Klaas}, {Mo{\'o}r}, {Linz}, \& {Feuchtgruber}}]{Mueller2016}
{M{\"u}ller}, T.~G., {Balog}, Z., {Nielbock}, M., {et~al.} 2016, \aap, 588,
  A109

\bibitem[{{Orton} {et~al.}(2014){Orton}, {Fletcher}, {Moses}, {Mainzer},
  {Hines}, {Hammel}, {Martin-Torres}, {Burgdorf}, {Merlet}, \&
  {Line}}]{Orton2014a}
{Orton}, G.~S., {Fletcher}, L.~N., {Moses}, J.~I., {et~al.} 2014, \icarus, 243,
  494

\bibitem[{{Papadopoulos}(2010)}]{Papadopoulos2010}
{Papadopoulos}, P.~P. 2010, \apj, 720, 226

\bibitem[{{P{\'e}rez-Beaupuits} {et~al.}(2018){P{\'e}rez-Beaupuits},
  {G{\"u}sten}, {Harris}, {Requena-Torres}, {Menten}, {Wei{\ss}},
  {Polehampton}, \& {van der Wiel}}]{PerezBeaupuits2018}
{P{\'e}rez-Beaupuits}, J.~P., {G{\"u}sten}, R., {Harris}, A., {et~al.} 2018,
  \apj, 860, 23

\bibitem[{{Perley} \& {Butler}(2013)}]{Perley2013}
{Perley}, R.~A., \& {Butler}, B.~J. 2013, \apjs, 204, 19

\bibitem[{{Rekola} {et~al.}(2005){Rekola}, {Richer}, {McCall}, {Valtonen},
  {Kotilainen}, \& {Flynn}}]{Rekola2005}
{Rekola}, R., {Richer}, M.~G., {McCall}, M.~L., {et~al.} 2005, \mnras, 361, 330

\bibitem[{{Rieke} {et~al.}(1980){Rieke}, {Lebofsky}, {Thompson}, {Low}, \&
  {Tokunaga}}]{Rieke1980}
{Rieke}, G.~H., {Lebofsky}, M.~J., {Thompson}, R.~I., {Low}, F.~J., \&
  {Tokunaga}, A.~T. 1980, \apj, 238, 24

\bibitem[{{Rieke} {et~al.}(1988){Rieke}, {Lebofsky}, \& {Walker}}]{Rieke1988}
{Rieke}, G.~H., {Lebofsky}, M.~J., \& {Walker}, C.~E. 1988, \apj, 325, 679

\bibitem[{{Robitaille} {et~al.}(2016){Robitaille}, {Ginsburg}, {Beaumont},
  {Leroy}, \& {Rosolowsky}}]{Robitaille2016}
{Robitaille}, T., {Ginsburg}, A., {Beaumont}, C., {Leroy}, A., \& {Rosolowsky},
  E. 2016, {spectral-cube: Read and analyze astrophysical spectral data cubes},
  Astrophysics Source Code Library, ascl:1609.017

\bibitem[{{Rosenberg} {et~al.}(2013){Rosenberg}, {van der Werf}, \&
  {Israel}}]{Rosenberg2013}
{Rosenberg}, M.~J.~F., {van der Werf}, P.~P., \& {Israel}, F.~P. 2013, \aap,
  550, A12

\bibitem[{{Sakamoto} {et~al.}(2011){Sakamoto}, {Mao}, {Matsushita}, {Peck},
  {Sawada}, \& {Wiedner}}]{Sakamoto2011}
{Sakamoto}, K., {Mao}, R.-Q., {Matsushita}, S., {et~al.} 2011, \apj, 735, 19

\bibitem[{{Shirley}(2015)}]{Shirley2015}
{Shirley}, Y.~L. 2015, Publications of the Astronomical Society of the Pacific,
  127, 299

\bibitem[{{Sobolev}(1960)}]{Sobolev1960}
{Sobolev}, V.~V. 1960, {Moving Envelopes of Stars} (Cambridge: Harvard
  University Press, 1960)

\bibitem[{{Teuben} {et~al.}(2015){Teuben}, {Pound}, {Mundy}, {Rauch},
  {Friedel}, {Looney}, {Xu}, \& {Kern}}]{Teuben2015}
{Teuben}, P., {Pound}, M., {Mundy}, L., {et~al.} 2015, in Astronomical Society
  of the Pacific Conference Series, Vol. 495, Astronomical Data Analysis
  Software an Systems XXIV (ADASS XXIV), ed. A.~R. {Taylor} \& E.~{Rosolowsky},
  305

\bibitem[{{Tiemann}(1974)}]{Tiemann1974}
{Tiemann}, E. 1974, Journal of Physical and Chemical Reference Data, 3, 259

\bibitem[{{Turner} \& {Ho}(1985)}]{Turner1985}
{Turner}, J.~L., \& {Ho}, P.~T.~P. 1985, \apjl, 299, L77

\bibitem[{{Ulvestad} \& {Antonucci}(1997)}]{Ulvestad1997}
{Ulvestad}, J.~S., \& {Antonucci}, R.~R.~J. 1997, \apj, 488, 621

\bibitem[{{van der Tak} {et~al.}(2007){van der Tak}, {Black}, {Sch{\"o}ier},
  {Jansen}, \& {van Dishoeck}}]{vanderTak2007}
{van der Tak}, F.~F.~S., {Black}, J.~H., {Sch{\"o}ier}, F.~L., {Jansen}, D.~J.,
  \& {van Dishoeck}, E.~F. 2007, \aap, 468, 627

\bibitem[{{van der Tak} \& {van Dishoeck}(2000)}]{vanderTak2000}
{van der Tak}, F.~F.~S., \& {van Dishoeck}, E.~F. 2000, \aap, 358, L79

\bibitem[{{Weaver} {et~al.}(2002){Weaver}, {Heckman}, {Strickland}, \&
  {Dahlem}}]{Weaver2002}
{Weaver}, K.~A., {Heckman}, T.~M., {Strickland}, D.~K., \& {Dahlem}, M. 2002,
  \apjl, 576, L19

\bibitem[{{Weiland} {et~al.}(2011){Weiland}, {Odegard}, {Hill}, {Wollack},
  {Hinshaw}, {Greason}, {Jarosik}, {Page}, {Bennett}, {Dunkley}, {Gold},
  {Halpern}, {Kogut}, {Komatsu}, {Larson}, {Limon}, {Meyer}, {Nolta}, {Smith},
  {Spergel}, {Tucker}, \& {Wright}}]{Weiland2011}
{Weiland}, J.~L., {Odegard}, N., {Hill}, R.~S., {et~al.} 2011, \apjs, 192, 19

\bibitem[{{Wei{\ss}} {et~al.}(2008){Wei{\ss}}, {Kov{\'a}cs}, {G{\"u}sten},
  {Menten}, {Schuller}, {Siringo}, \& {Kreysa}}]{Weiss2008}
{Wei{\ss}}, A., {Kov{\'a}cs}, A., {G{\"u}sten}, R., {et~al.} 2008, \aap, 490,
  77

\bibitem[{{Wyrowski} {et~al.}(1999){Wyrowski}, {Schilke}, \&
  {Walmsley}}]{Wyrowski1999}
{Wyrowski}, F., {Schilke}, P., \& {Walmsley}, C.~M. 1999, \aap, 341, 882

\bibitem[{Xu \& Lovas(1997)}]{Xu1997}
Xu, L.-H., \& Lovas, F.~J. 1997, Journal of Physical and Chemical Reference
  Data, 26, 17

\end{thebibliography}

\end{document}